\documentclass[letterpaper,12pt]{JHEP3}
\usepackage{amsmath}
\usepackage{amssymb}
\usepackage{graphicx}
\usepackage{subfig}
\usepackage{bbm}
\usepackage{epsfig}
\usepackage{yfonts}
\newcommand{\labell}[1]{\label{#1}}
\def\({\left(} \def\){\right)}
\def\[{\left[} \def\]{\right]}
\def\al{\alpha}

\newcommand{\be}{\begin{equation}}
\newcommand{\ee}{\end{equation}}
\newcommand{\bea}{\begin{eqnarray}}
\newcommand{\eea}{\end{eqnarray}}
\newcommand{\ba}{\begin{eqnarray}}
\newcommand{\ea}{\end{eqnarray}}

\newcommand{\eqn}[1]{(\ref{#1})}
\newcommand{\beq}{\begin{equation}}
\newcommand{\eeq}{\end{equation}}
\newcommand{\beqa}{\begin{eqnarray}}
\newcommand{\eeqa}{\end{eqnarray}}
\newcommand{\beqar}{\begin{eqnarray*}}
\newcommand{\eeqar}{\end{eqnarray*}}

\newcommand{\reef}[1]{(\ref{#1})}
\newcommand{\ssc}{\scriptscriptstyle}
\newcommand{\eg}{{\it e.g.,}\ }
\newcommand{\ie}{{\it i.e.,}\ }

\newcommand{\mt}[1]{\textrm{\tiny #1}}

\newcommand{\veps}{\varepsilon}

\newcommand{\X}{\mathcal{X}}
\newcommand{\Z}{\mathcal{Z}}
\newcommand{\A}{\mathcal{A}}

\newcommand{\D}{\mathcal{D}}
\newcommand{\E}{\mathcal{E}}

\newcommand{\cO}{\mathcal{O}}

\newcommand{\tL}{\tilde{L}}

\newcommand{\la}{\lambda}
\newcommand{\lp}{\ell_{\mt P}}

\newcommand{\fin}{f_\infty}

\newcommand{\w}{\omega}
\newcommand{\rh}{r_\mt{H}}

\newcommand{\ads}{a_d^*}

\newcommand{\ren}{R\'enyi\ }
\newcommand{\myeq}[1]{\begin{equation} \begin{split} #1 \end{split} \end{equation}}
\newcommand{\poin}{Poincar\'e}

\newcommand{\m}{v}

\newcommand{\tr}{{\rm tr}}
\newcommand{\rhov}{\rho_{\ssc V}}
\newcommand{\Vs}{V_{\ssc \Sigma}}
\newcommand{\xq}{x_q}
\newcommand{\CT}{{\widetilde C}_T}
\newcommand{\tfor}{{t}_4}
\newcommand{\mtt}[1]{\textrm{#1}}
\newcommand{\T}{\mathcal{T}}

\preprint{arXiv:1110.nnnn [hep-th]}

\title{Holographic Calculations of \ren Entropy}

\author{Janet Hung,$^{1}$ Robert C. Myers,$^{1}$ Michael Smolkin$^{1}$
and Alexandre Yale$^{1,2}$\\
\it
$^1$Perimeter Institute for Theoretical Physics,\\
\ Waterloo, Ontario N2L 2Y5, Canada\\
$^2$Department of Physics \& Astronomy and Guelph-Waterloo Physics Institute,\\
\ University of Waterloo, Waterloo, Ontario N2L 3G1, Canada}

\abstract{We extend the approach of \cite{casini9} to a new
calculation of the \ren
entropy of a general CFT in $d$ dimensions with a spherical entangling
surface, in terms of certain thermal partition functions. We apply this
approach to calculate the \ren entropy in various holographic models.
Our results indicate that in general, the \ren entropy will be a
complicated nonlinear function of the central charges and other
parameters which characterize the CFT. We also exhibit the relation
between this new thermal calculation and a conventional calculation
of the \ren entropy where a twist operator is inserted on the spherical
entangling surface. The latter insight also allows us to calculate the
scaling dimension of the twist operators in the holographic models.}

\begin{document}

\section{Introduction}

Entanglement entropy has emerged as a useful measure of entanglement
with which one can characterize various quantum systems \eg
\cite{wenx,cardy0,cardyCFT}. Given a subsystem $V$ described by a
density matrix $\rhov$, the entanglement entropy corresponds to the von
Neumann entropy of the corresponding density matrix: $S_\mt{EE}=-{\rm
tr}\!\left[\,\rhov \log \rhov\right]$. An approach for calculating
entanglement entropy for the boundary field theory in gauge/gravity
duality was proposed some five years ago in \cite{rt1}
--- see also \cite{rt2,rt3}. Given a particular holographic framework, the
entanglement entropy in the $d$-dimensional boundary theory between a
spatial region $V$ and its complement $\bar V$ is calculated by
extremizing the following expression
 \be
S(V) = \frac{2\pi}{\lp^{d-1}}\ \mathrel{\mathop {\rm
ext}_{\scriptscriptstyle{\m\sim V}} {}\!\!} \left[A(\m)\right]
 \labell{define}
 \ee
over ($d$--1)-dimensional surfaces $\m$ in the bulk spacetime which are
homologous to the boundary region $V$.\footnote{Hence `area' $A(\m)$ to
denotes the ($d$--1)-dimensional volume of $\m$. Note that the extremal
surface only yields the minimal area if one first Wick rotates to
Euclidean signature.} In particular then, the boundary of $\m$ matches
the `entangling surface' $\partial V$ in the boundary geometry.
Implicitly, eq.~\reef{define} assumes that the bulk theory is well
approximated by classical Einstein gravity, where we have adopted the
convention: $\lp^{d-1}=8\pi G_\mt{N}$. Hence this expression
\reef{define} bears a striking resemblance to black hole entropy,
however, the surfaces $\m$ do not coincide with any event horizon in
general. While there is a good amount of evidence supporting this
proposal, \eg see \cite{rt2,head,EEGB}, a general derivation of
eq.~\reef{define} remains lacking. However, we note that a derivation
was recently constructed for the special case of a spherical entangling
surface in \cite{casini9}.

Another interesting measure of entanglement is the \ren entropy
\cite{renyi0,karol}:
 \be
 S_q=\frac{1}{1-q}\,\log{\rm tr}\!\left[\,\rhov^{\,q}\right] \,.
 \labell{rensq}
 \ee
where $q$ is a positive real number. Given this expression for the \ren
entropy, there are an number of interesting limits which one may
consider: In particular, one finds that the entanglement entropy is
recovered in the limit $q$ tends to one, \ie $S_\mt{EE} =\lim_{q\to
1}S_q$. Another interesting case to consider is $q\to\infty$, which
yields
 \be
 S_\infty=\lim_{q\to\infty} S_q = -\log\lambda_1\,,
 \labell{mins}
 \ee
where $\lambda_1$ is the largest eigenvalue of $\rhov$. This infinite
limit is sometimes referred to as the `min-entropy'. Finally we note
that
 \be
S_0=\lim_{q\to0} S_q = \log\left[\,\D \right]\,,
 \labell{snot}
 \ee
where $\D$ corresponds to the number of nonvanishing eigenvalues of
$\rhov$. For two-dimensional CFT's, there is a universal result for the
\ren entropy of an interval of length $\ell$ \cite{cardy0,cardyCFT}:
 \be
S_q(d=2)=\frac{c}{6} \(1 + \frac{1}{q} \)\log(\,\ell/ \delta)\,,
 \labell{twod}
 \ee
where $c$ is the central charge and $\delta$ is a short-distance
regulator, in the underlying CFT. Recently, \ren entropy has drawn some
attention in the condensed matter community \cite{cmt}. Essentially,
all of these results rely on an approach to calculating \ren entropy
using the `replica trick,' which requires evaluating the partition
function on a $q$-fold cover of the original background geometry. In
fact, this approach combined with the limit $q\to1$ is regarded as the
standard calculation of entanglement entropy. Now holography
geometrizes calculations in the boundary theory and so a natural
holographic implementation of the replica trick involves a singular
bulk geometry, reflecting the singular nature of the $q$-fold cover in
the boundary \cite{furry}. However, attempting to work with this
singular bulk space in a straightforward way \cite{furry} produces
incorrect results \cite{head}. Hence it seems that without a full
understanding of string theory or quantum gravity in the bulk, one will
not be able to work with such singular bulk space in a controlled way.
This problem then stands as an obstacle both to producing a derivation
of the holographic expression of entanglement entropy \reef{define} and
to performing holographic calculations of \ren entropy \reef{rensq}.
However, one can also imagine finding a new solution for the bulk
geometry which remains smooth despite the singularity in the boundary
metric \cite{head,lost}. Essentially our calculations below follow this
route.

Recently, ref.~\cite{casini9} presented a new calculation of
entanglement entropy of a CFT for a spherical entangling surface in
flat $d$-dimensional Minkowski space,\footnote{This construction is
also easily extended to a spherical entangling surface in a
`cylindrical' background with the topology $R\times S^{d-1}$
\cite{casini9}.} which has a simple holographic translation. One first
considers the causal development of the ball enclosed by the entangling
surface and then observes that this spacetime region can be conformally
mapped to a `hyperbolic cylinder,' $R\times H^{d-1}$. The curvature
scale on the hyperbolic spatial slices $H^{d-1}$ matches the radius of
the original spherical entangling surface, $R$. Further the vacuum of
the CFT in the original spacetime is mapped to a thermal bath in the
hyperbolic cylinder with temperature
 \be
 T_0=\frac{1}{2\pi R}\,.
 \labell{tnot}
 \ee
As with any other operator in the CFT, the density matrix in the new
spacetime $R\times H^{d-1}$ is related to that in the original geometry
by a unitary transformation, \ie $\rhov=U^{-1}\,\rho_\mt{therm}\,U$.
Hence the entanglement entropy for the spherical entangling surface is
equivalent to the thermal entropy of the CFT at temperature $T_0$ in
the hyperbolic cylinder $R\times H^{d-1}$. While this insight may not
be particularly useful for a generic CFT, using the AdS/CFT
correspondence for a holographic CFT, we can relate the latter thermal
bath to an appropriate black hole in the bulk AdS space. Given the
geometry of $R\times H^{d-1}$, it is natural the thermal bath in the
boundary theory would be described by a `topological' black hole, for
which the event horizon has a hyperbolic cross-section \cite{topbh}. In
fact, the dual black hole with temperature $T_0$ is simply a particular
foliation of empty AdS spacetime with slices which are topologically
equivalent to $R\times H^{d-1}$. Here the coordinate transformation in
the bulk from Poincar\'e coordinates to the hyperbolic foliation
implements the conformal transformation described above on the
boundary.\footnote{The interested reader will find the precise details
of this transformation of the bulk coordinates in \cite{casini9}.} In
any event, in this holographic framework, the entanglement entropy is
now given by the horizon entropy of this black hole, which is easily
calculated using Wald's formula \cite{WaldEnt} for any general
gravitational theory in the bulk.

Now we observe this derivation presented in \cite{casini9} also readily
lends itself to a holographic calculation of \ren entropy for a
spherical entangling surface. Following the above construction, the
density matrix is (essentially) thermal and we can write the $q$'th
power of $\rhov$ as
 \be
 \rhov^{\,q}=U^{-1}\,\frac{\exp\left[-qH/T_0\right]}{Z(T_0)^q}\,U \quad{\rm
 where}\ \ Z(T_0)\equiv \tr\left[e^{-H/T_0}\right]\,.
 \labell{powerq}
 \ee
The unitary transformation $U$ and its inverse, appearing above, will
cancel upon taking the trace of this expression. Hence we are left with
 \be
\textrm{tr}\!\left[\,\rhov^{\,q}\right] = \frac{Z(T_0/q)}{Z(T_0)^q}\,.
 \labell{trrho0x}
 \ee
Now using the usual definition of the free energy, \ie $F(T)=-T\log
Z(T)$, the corresponding \ren entropy \reef{rensq} becomes
 \be
 S_q=\frac{q}{1-q}\frac{1}{T_0}\left[ F(T_0)-F(T_0/q)\right]\,.
 \labell{form1}
 \ee
We note that this result matches that in \cite{baez}, which examined
\ren entropies for a thermal ensemble. Further using the standard
thermodynamic identity, $S=-\partial F/\partial T$, we can rewrite this
expression as
 \be
S_q = \frac{q}{q-1} \frac{1}{T_0}\, \int_{T_0/q}^{T_0}
S_\mt{therm}(T)\, dT\,,
 \labell{finalfor}
 \ee
where $S_q$ is the desired \ren entropy while $S_\mt{therm}(T)$ denotes
the thermal entropy of the CFT on $R\times H^{d-1}$.
Eq.~\reef{finalfor} will be convenient for our holographic calculations
in the following. It is clear from this expression that we recover the
desired result for the entanglement entropy
 \be
 S_\mt{EE}=\lim_{q\to 1}S_q = S_\mt{therm}(T_0)\,.
 \labell{lim1}
 \ee
with $T_0$ given in eq.~\reef{tnot}.

This discussion demonstrates that for any CFT, \ren entropies of a
spherical entangling surface can be calculated if the thermal entropy
on the hyperbolic cylinder $R\times H^{d-1}$ is known for an arbitrary
temperature. As with the entanglement entropy, this approach will now
allow us to perform holographic calculations of the \ren entropy of the
boundary CFT in the case where the entangling surface is a sphere
$S^{d-1}$. The essential new ingredient is that we must understand the
thermodynamic properties of the dual topological black holes away from
$T=T_0$. Hence our analysis of holographic \ren entropy is restricted
to bulk gravitational theories where these black hole solutions are
known. Hence the present discussion differs from that of holographic
entanglement entropy in \cite{casini9}, which applied for any covariant
theory of gravity in the bulk.

An overview of the remainder of the paper is as follows: We begin by
explicitly calculating the holographic \ren entropy as described above
for three bulk theories: Einstein gravity, Gauss-Bonnet gravity
\cite{lovel} and quasi-topological gravity \cite{old1,old2} in sections
\ref{sec:EH}, \ref{sec:GB} and \ref{sec:QT}, respectively. Our results
indicate that the \ren entropy (and in particular, the universal
contribution) is a complicated nonlinear function of the central
charges and other parameters which characterize the underlying CFT. In
section \ref{twist}, we review the calculation of \ren entropy in terms
of twist operators for two-dimensional CFT's and investigate the
holographic representation of these calculations. Section \ref{versus}
presents the connection between the `thermal calculation' of \ren
entropy discussed above and the calculation where a twist operator is
inserted on the spherical entangling surface. This discussion applies
for CFT's in any dimension and so in section \ref{twisted}, we apply
these insights to calculate the scaling dimension of the twist
operators for the holographic models introduced in section \ref{renyi}.
Finally, in section \ref{spectrum}, we begin to examine what
information can be inferred about the eigenvalue spectrum of the
reduced density matrix using our results for the \ren entropy. Then we
conclude with a discussion of our results in section \ref{discuss}. A
number of appendices are included to cover certain ancillary
discussions. in appendix \ref{unequal}, we consider various
inequalities which \ren entropy must generally satisfy and their
implications for our holographic results.. Appendix \ref{threed}
presents a nontrivial holographic model for a three-dimensional
boundary CFT. Our calculations with this model indicate that there is
no particular simplification of the \ren entropy for $d=3$. Finally, we
present some details of holographic calculations relevant for section
\ref{twist} in appendix \ref{review}.

\section{Holographic \ren Entropy} \labell{renyi}

As described in the introduction, our goal is to use the AdS/CFT
correspondence to calculate the \ren entropy for a spherical entangling
surface in the boundary conformal field theory. We do this for three
bulk gravitational theories of increasing complexity, beginning with
Einstein gravity in section \ref{sec:EH} and then considering
Gauss-Bonnet gravity in section \ref{sec:GB} and quasi-topological
gravity in section \ref{sec:QT}. In appendix \ref{threed}, we also make
a perturbative analysis of a four-dimensional bulk theory with an
interaction term which is cubic in the Weyl tensor. As we describe
below, the additional couplings appearing in these bulk gravity
theories allow us to consider a broader class of dual boundary CFT's.
The three gravitational theories in this section were chosen because,
in each case, a family of analytic solutions is known corresponding to
topological black holes with hyperbolic horizons. Using the standard
tools of black hole thermodynamics, \ie the Wald entropy
\cite{WaldEnt}, we are then able to calculate the horizon entropy for
these black holes. The standard AdS/CFT dictionary equates this horizon
entropy to the thermal entropy of the boundary CFT on the hyperbolic
cylinder $R\times H^{d-1}$. Hence using eq.~\reef{finalfor} we can
calculate the \ren entropy for a spherical entangling surface.

Before moving to specific cases, let us consider the gravitational
calculations in a little more detail. We are considering a
$d$-dimensional boundary theory and hence the bulk spacetime has $d$+1
dimensions. For the three gravity theories of interest, the metric for
the topological black holes will take the form
 \be
ds^2 = -\( \frac{r^2}{L^2} f(r) - 1 \) N^2 dt^2 +
\frac{dr^2}{\frac{r^2}{L^2} f(r) - 1 } + r^2 d\Sigma^2_{d-1}\,,
 \labell{lineelement}
 \ee
where $d\Sigma^2_{d-1}$ is the line element for the $(d-1)$-dimensional
hyperbolic plane $H^{d-1}$ with unit curvature. The function $f(r)$ is
determined by the field equations of the particular gravity theory
under consideration. We will find that asymptotically $f(r \rightarrow
\infty)\equiv\fin$ and hence, from $g_{rr}$, we can see that the AdS
curvature scale is given by $\tilde{L}^2 = {L^2}/{\fin}$. We have
included an extra constant $N^2$ in $g_{tt}$ to allow us to adjust the
normalization of the time coordinate. In particular, we will choose
$N^2={L^2}/(\fin R^2)=\tL^2/R^2$ to ensure that the boundary metric is
conformally equivalent to
 \be
ds^2_{\infty} =  \( -dt^2 + R^2 d \Sigma^2_{d-1} \)\,.
 \labell{cftmet}
 \ee
That is, we are studying the boundary CFT on $R\times H^{d-1}$ with the
curvature scale of the hyperbolic spatial slices set to $R$.

The position of the event horizon, given by vanishing of $g_{tt}$, is
defined by the expression
 \be
\frac{r_\mt{H}^2}{L^2} f(r_\mt{H}) = 1\,.
 \labell{horizon}
 \ee
The temperature of the thermal bath in the boundary CFT is given by the
Hawking temperature of the horizon, which may be written as
 \bea
T &=& \frac{N}{4\pi}\ \partial_r\!\! \[ \frac{r^2}{L^2}f(r)
\]_{r=\rh}
 \nonumber\\
&=& \frac{\tL}{4\pi R} \[ \frac{2}{\rh} + \frac{\rh^2}{L^2}\
\left.\frac{\partial f(r)}{\partial r}\right|_{r=\rh} \]~.
 \labell{temp}
 \eea
At this point, we note that setting $f(r)$ to a constant, \ie
$f(r)=\fin$, always provides a solution of the gravitational equations
of motion. This is the case where the metric \reef{lineelement}
corresponds to a pure AdS space with a hyperbolic foliation. With this
choice, eq.~\reef{horizon} yields $\rh=\tL$ and then from
eq.~\reef{temp} we recover the expected temperature as in
eq.~\reef{tnot}, \ie $T = T_0 = (2\pi R)^{-1}$. This AdS
metric was central to the discussion of holographic entanglement
entropy in \cite{casini9}.

Now one could proceed with a holographic calculation the \ren entropy
using eq.~\reef{form1}. In this approach, one would first evaluate the
Euclidean gravitational action $I_\mtt{E}$ for the topological black
holes \reef{lineelement}. Then interpreting this result in terms of the
free energy of the dual thermal ensemble, \ie $F(T)=T\,I_\mtt{E}(T)$,
the \ren entropy is easily calculated with the expression in
eq.~\reef{form1}. However, we will instead proceed with
eq.~\reef{finalfor} where the thermal entropy is determined as the
horizon entropy of the dual black holes. Hence an essential step in our
calculations will be evaluating the horizon entropy of these
topological black holes \reef{lineelement}. Since in sections
\ref{sec:GB} and \ref{sec:QT} we will be considering bulk theories with
higher curvature interactions, we calculate the horizon entropy using
Wald's formula \cite{WaldEnt}
 \beq
S = -2 \pi \int_\mt{horizon}\! d^{d-1}x\sqrt{h}\
\frac{\partial{\mathcal{L}}}{\partial R^{a b}{}_{c d}}\,\hat{\veps}^{a
b}\,\hat{\veps}_{c d}\,,
 \labell{Waldformula}
 \eeq
where $\mathcal{L}$ corresponds to the Lagrangian of the particular
gravity theory of interest. We use $\hat{\veps}_{a b}$ to denote the
binormal to the horizon.

Again, we interprete this result as the thermal entropy of the boundary
CFT at a temperature $T$ given by eq.~\reef{temp} and then we can
calculate the \ren entropy using eq.~\reef{finalfor}. However, rather
than considering the entropy as a function of the temperature, it will
be more convenient in the following to consider $S(x)$ where
$x\equiv\rh/\tL$. In this case, we can rewrite the expression for the
\ren entropy as
 \bea
S_q &=& \frac{q}{q-1} \frac{1}{T_0}\, \int_{\xq}^{1}
S(x)\,\frac{dT}{dx}\,dx
 \labell{finalfor2}\\
&=& \frac{q}{q-1} \frac{1}{T_0}\,\[\left.S(x)\,T(x)\right|^1_{\xq}
-\int_{\xq}^{1} \frac{dS}{dx}\,T(x)\,dx\]~.
 \nonumber
 \eea
The upper endpoint in the above integral is fixed using $\rh=\tL$ and
hence $x=1$ for $T=T_0$. It remains to determine the lower endpoint $x_q$
corresponding to the temperature $T=T_0/q$. We will see below that the
precise value of $\xq$ will depend on the details of the gravitational
theory under consideration.

The black hole solutions \reef{lineelement} above inherit an
$SO(1,d-1)$ symmetry from the hyperbolic plane portion of the metric.
In particular, the geometry of the horizon is homogenous and the
integrand in eq.~\reef{Waldformula} is simply a constant in the cases
of interest. Hence we may write the entropy as:
 \beq
S = -2 \pi \, \rh^{d-1} \left. \frac{\partial{\mathcal{L}}}{\partial
R^{a b}{}_{c d}}\,\hat{\veps}^{a b}\,\hat{\veps}_{c
d}\right|_{r=\rh}\,\Vs \ \,.
 \labell{Wald2}
 \eeq
Here we have introduced $\Vs$ to denote the `coordinate' volume of the
hyperbolic plane, \ie $\Vs=\int_{H^{d-1}}d\Sigma$. Of course, this
volume is divergent and must be regulated to make sense of the entropy.
As described in detail in \cite{casini9}, this divergence is related to
the UV divergences appearing in the \ren entropy.

A convenient choice of coordinates on $H^{d-1}$ is given by\footnote{In
terms of the coordinates chosen in \cite{casini9}, $y=\cosh u$ or
$y=\sqrt{x^2+1}$.}
 \be
d\Sigma^2_{d-1}= \frac{dy^2}{y^2-1} + (y^2-1)\, d\Omega_{d-2}^2~,
 \labell{hash}
 \ee
where $d\Omega_{d-2}^2$ is the line-element on a unit $(d-2)$-sphere. As
explained in \cite{casini9}, the horizon entropy \reef{Wald2} is
regulated by integrating out to a maximum radius in this hyperbolic
geometry:
 \be
y_{max}=\frac R\delta\,,
 \labell{connect3b}
 \ee
where $\delta$ is the short-distance cut-off in the boundary CFT. Hence
the hyperbolic volume becomes
 \bea
\Vs&=&\Omega_{d-2}\int_1^{y_{max}} \left(y^2-1\right)^{(d-3)/2}\,dy
 \nonumber\\
&\simeq& \frac{\Omega_{d-2}}{d-2}\left[ \frac{R^{d-2}}{\delta^{d-2}}-
\frac{(d-2)(d-3)}{2(d-4)}\frac{R^{d-4}}{\delta^{d-4}}+\cdots\right]\,,
 \labell{vtotal}
 \eea
where $\Omega_{d-2}=2\pi^{(d-1)/2}/\Gamma((d-1)/2)$ is the area of a
unit $(d-2)$-sphere. Now using the horizon entropy in eq.~\reef{Wald2}
to calculate the \ren entropy using eq.~\reef{finalfor}, it is clear
that the UV divergences appearing in the latter are simply inherited
from the above expression for $\Vs$.

An important consequence of these calculations is that the UV
divergences appearing in the \ren entropy for any $q$ --- and the
entanglement entropy given by $q=1$ --- have precisely the same
structure. In particular, if we note that the area of the spherical
entangling surface is given by ${\mathcal
A}_{d-2}=\Omega_{d-2}R^{d-2}$, then from eq.~\reef{vtotal} the leading
divergence in the \ren entropies will have the form of the `area law'
expected to appear in the entanglement entropy \cite{rt2,rt3}. Note
that the hyperbolic geometry of the horizon was essential to ensure the
leading power was $1/\delta^{d-2}$ here despite the area integral being
($d\!-\!1$)-dimensional in eq.~\reef{vtotal}. Of course, the power-law
divergences in the \ren entropies are not universal, \eg see
\cite{rt2,rt3}, however, a universal contribution can be extracted from
the subleading terms. The form of this universal contribution to the
entanglement entropy depends on whether $d$ is odd or even
\cite{casini9,cthem}:
\be V_{\ssc
\Sigma,univ}=\frac{\pi^{d/2}}{\Gamma(d/2)}\times\left\lbrace
\begin{matrix}
(-)^{\frac{d}{2}-1}\,\, \frac2\pi
\,  \log(2R/\delta)&\quad&{\rm for\ even\ }d\,,\\
(-)^{\frac{d-1}{2}}\ \ \ \ \ \ \ \ \ \ \ \ \ \ \ \ \ &\quad&{\rm for\
odd\ }d\,.
\end{matrix}\right.
\labell{unis} \ee
Here we should acknowledge a technical point with regards to odd $d$.
As is standard, we have identified here the universal contribution as
the constant term appearing in the expansion \reef{vtotal}. While the
universal character of this constant is established in the entanglement
entropy, \ie $S_1$, for a variety of $d=3$ conformal quantum critical
systems \cite{fradkin}, as well as certain three-dimensional (gapped)
topological phases \cite{wenx}. Nevertheless, in general, one may worry
whether this constant is affected by the details of scheme chosen to
regulate the underlying field theory. However, this issue should be
circumvented by considering the mutual information with an appropriate
construction \cite{casini9,cthem}, as can be explicitly demonstrated in
holographic calculations \cite{mutual}. The potential ambiguity does
not appear with the mutual information since the latter is free of any
UV divergences --- for example, see \cite{casini55,swingle}. We expect
that the same approach can be extended to \ren entropies with general
$q$. For the present purposes, however, it will suffice to focus on the
above terms \reef{unis} to identify the universal contribution to the
\ren entropies.

\subsection{Einstein gravity in any dimension \labell{sec:EH}}

To begin the detailed calculations, we consider Einstein gravity. Hence
the bulk action is simply
 \myeq{ I= \frac{1}{2\lp^{d-1}}
\int d^{d+1}x \sqrt{-g} \( \frac{d(d-1)}{L^2} + R \)\,,
 \labell{einsact} }
and the equations of motion lead to the solution
 \be
 f(r)=1-\frac{\w^d}{r^d}\,,
 \labell{sol0}
 \ee
for the line element in eq.~(\ref{lineelement}). Above, $\w^d$ is an
integration constant which characterizes the energy (density) of the
corresponding black hole. For the present purposes, it is most useful
to parameterize this constant in terms of the position of the horizon
with
 \be
\w^d=\rh^d-\tL^2\rh^{d-2}\,.
 \labell{omega0}
 \ee
Further we note that here, $\fin=f(r\to\infty)=1$ and hence $\tL=L$,
\ie the AdS curvature scale is same as the scale appearing in the
cosmological constant term in the action \reef{einsact}. The latter
equality will not hold for the higher curvature gravity theories in the
subsequent sections. Applying eq.~\reef{temp} to the present case, we
find that the temperature is given by
 \be
T = \frac{1}{4\pi R} \[ d\frac{\rh}{\tL}-(d-2)\frac{\tL}{\rh}   \]\,.
 \labell{tempEH}
 \ee
Next we would like to determine the horizon entropy. For Einstein gravity,
the Wald formula \reef{Waldformula} reduces to the expected result
$S={\mathcal A}_{hor}/(4G_\mt{N})=\frac{2\pi}{\lp^{d-1}}{\mathcal
A}_{hor}$ and hence, in the present case, eq.~\reef{Wald2} reduces to
 \be
S=\frac{2\pi}{\lp^{d-1}}\,\rh^{d-1}\Vs\,.
 \labell{EHEE}
 \ee

Now we would like to use these results to determine the \ren entropy
using eq.~\reef{finalfor2}. First we must re-express the temperature
\reef{tempEH} and the entropy \reef{EHEE} as functions of the variable
$x=\rh/\tL$:
 \be
 T(x)=\frac{T_0}{2}\[ d\,x-\frac{d-2}x   \]\,,\qquad
 S(x)=2\pi\Vs\frac{\tL^{d-1}}{\lp^{d-1}}\,x^{d-1}\,.
 \labell{xstuff0}
 \ee
Next we must determine the lower endpoints $x_q$ of the corresponding
integration, which corresponds to $T=T_0/q$. From
eq.~\reef{xstuff0}, we see that $\xq$ satisfies
 \be
 0 =
d\, \xq^2 - \frac{2}{q}\, \xq - \( d - 2 \) \,.
 \labell{EHquad}
 \ee
We can easily find the roots of this quadratic equation and choosing
the real and positive root we have
 \be
 \xq = \frac{1}{qd} \( 1 + \sqrt{1 - 2dq^2 + d^2 q^2} \)\,.
 \labell{EHxq}
 \ee
Further we may note that $\xq\le 1$ for $q\ge1$. Now it straightforward
to show that eq.~\reef{finalfor2} yields
 \be
S_q = \frac{\pi\, q}{q-1} \Vs \( \frac{\tL}{\lp} \)^{d-1} \(2-
\xq^{d-2}\(1 + \xq^2\)\)\,.
 \labell{sqeh}
 \ee
This expression is plotted in Figure \ref{fig:EH} for various values of
$d$.
\FIGURE[!ht]{
\includegraphics[width=0.8\textwidth]{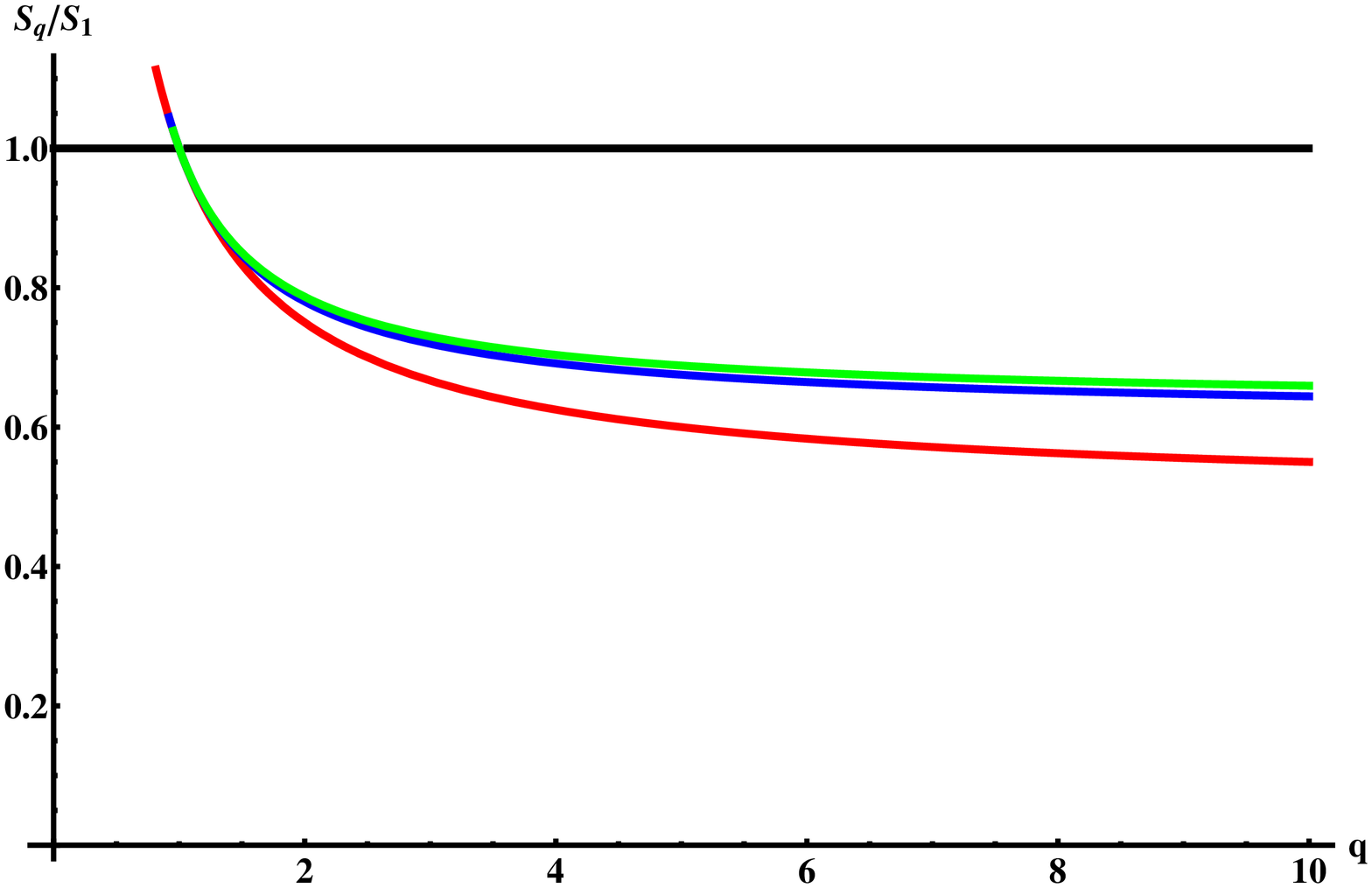}
\caption{(Colour Online) Ratio of the \ren entropy to the entanglement
entropy, $S_q/S_1$, as a function of $q$. Starting at the bottom, the three
curves correspond to $d=2$ (red), $3$ (blue) and $1000$ (green). The
dashed line with $S_q/S_1=1$ is simply a guide to the eye. Note that
$S_q/S_1<1$ for $q>1$ and $S_q/S_1>1$ for $q<1$.} \labell{fig:EH}}

As discussed in the introduction, it is interesting to consider the
limits $q\rightarrow 0$, 1 and $\infty$, which yield
 \bea
\lim_{q \rightarrow 0} S_q      &=& 2 \pi\Vs \( \frac{\tL}{\lp}
\)^{d-1}\(\frac 2d\)^d\,\frac{1}{2 q^{d-1}}\,,
 \nonumber\\
\lim_{q \rightarrow 1} S_q      &=& 2\pi\Vs \( \frac{\tL}{\lp}
\)^{d-1}\,,
 \labell{limits} \\
\lim_{q \rightarrow \infty} S_q &=& 2\pi\Vs \( \frac{\tL}{\lp} \)^{d-1}
\,\(1-\frac{d-1}d\(\frac{d-2}{d}\)^{\frac{d-2}2}\)\,.
 \nonumber
 \eea
As required, we recover the entanglement entropy in the limit $q\to1$,
as was calculated previously \cite{casini9,cthem}. This can also be
seen by substituting $\rh=\tL$ in eq.~\reef{EHEE}. We discuss various
properties of $S_0$ and $S_\infty$ later. However, let us note here
that because of the factor of $\Vs$, both of these expressions (as well
as $S_1$) are divergent in the limit $\delta\to0$.

In this general expression \reef{sqeh}, we can recognize that the
factor $(\tL/\lp)^{d-1}$ gives a count of the number of degrees of
freedom in the dual boundary CFT, \eg for even $d$, this factor is
proportional to the central charge of the boundary theory
\cite{sken}.\footnote{Recall that all of the central charges are
proportional to this same factor for any holographic CFT dual to
Einstein gravity \cite{sken}.} As discussed previously, the factor
$\Vs$ contains information about the size of the entangling surface and
the short-distance cut-off, as given in eqs.~\reef{vtotal} and
\reef{unis}. The remaining factors implicitly give some function of $q$
(and $d$) through eq.~\reef{EHxq}. Hence the qualitative form of
eq.~\reef{sqeh} is
 \be
 S_q=C\times V(R/\delta)\times f(q)\,,
 \labell{qual}
 \ee
where $C$ denotes some central charge in the boundary CFT. Hence this
expression, which applies for any $d\ge2$, has a form similar to the
\ren entropy of any two-dimensional CFT given in eq.~\reef{twod}. We
will see, however, that for holographic CFT's in higher dimensions
which are dual to a more general gravitational theory than Einstein
gravity, the general form of eq.~\reef{qual} no longer applies.

It is a non-trivial check of our result \reef{sqeh} to verify that it
yields the expected result \reef{twod} for a two-dimensional CFT when
$d=2$. In this case, eq.~\ref{EHquad} simplifies to yield $\xq=1/q$ and
then eq.~\reef{sqeh} becomes
 \be
S_q(d=2) = \frac{\pi \tL}{\lp} \(1 + \frac1q \)\Vs\,.
 \labell{holotwod}
 \ee
Now we make some observations: first, the central charge of the
boundary CFT is given by $c = 12\pi \tL/\lp $, as is well known
\cite{sken}. Second, care must be taken in evaluating the regulated
volume $\Vs$. For $d=2$, there are no power law divergences as in
eq.~\reef{vtotal}, rather the leading contribution coincides with the
universal term in eq.~\reef{unis}, \ie one finds the logarithmic
divergence $\Vs(d=2)= 2\log(2R/\delta) +\cdots$. Finally, for $d=2$,
the `spherical' entangling surface consists of two points separated by
a distance $2R$. In other words, we are calculating the \ren entropy
for an interval of length $\ell=2R$. Combining these observations,
eq.~\reef{holotwod} becomes
 \be
S_q(d=2) = \frac{c}{6} \left( 1 + \frac{1}{q} \right) \log
\frac{\ell}{\delta}\,,
 \labell{holotwod2}
 \ee
which precisely matches the required result \reef{sqeh} for two
dimensions.

\subsection{Gauss-Bonnet gravity and $d\ge4$ \labell{sec:GB}}

In this section, we examine holographic \ren entropy for higher
curvature bulk theories by considering Gauss-Bonnet (GB) gravity
\cite{lovel}. The gravitational action in $(d+1)$ dimensions can be
written:
 \be
I = \frac{1}{2\lp^{d-1}} \int d^{d+1} x \sqrt{-g} \[ \frac{d(d-1)}{L^2}
+ R + \frac{\lambda L^2}{(d-2)(d-3)} \X_4
\] ~,
 \labell{GBAction}
 \ee
where
 \be
\X_4=R_{abcd}R^{abcd}-4R_{ab}R^{ab}+R^2\,.
 \labell{GBterm}
 \ee
Of course, because this interaction is proportional to the
four-dimensional Euler density, $\X_4$ only contributes to the
equations of motion for $d\ge4$ \cite{YalePaddy}. These
higher-dimensional theories \reef{GBAction} have a variety of
interesting features but foremost amongst these is the fact that the
equations of motion are only second order in derivatives \cite{lovel},
which is related to the topological origin of the $\X_4$ interaction.
Relevant for the present analysis, asymptotically AdS black hole
solutions with hyperbolic horizons were found for GB gravity in
\cite{GBadsbh}. We might also mention that \cite{EEGB,friends} provided
a generalization of the standard prescription for holographic
entanglement entropy \reef{define} to these higher curvature theories.
However, we will not directly use the results of these studies here.

Recently, there has been renewed interest in this theory
\reef{GBAction} in the context of the AdS/CFT correspondence --- for
example, see \cite{EtasGB}. One interesting feature of the dual
boundary theory is that the different central charges have distinct
values because of the curvature-squared interaction \cite{highc}. For
example, in four dimensions, one has \cite{old2}
 \be
c = \pi^2 \frac{ \tL^3}{\lp^3} \( 1 - 2 \lambda f_\infty \)\,, \qquad a
= \pi^2 \frac{ \tL^3}{\lp^3} \( 1 - 6 \lambda f_\infty \)\,.
 \labell{fourdc}
 \ee
As before, we are using $\tL$ to denote the curvature scale of the AdS
vacuum while $\fin=L^2/\tL^2$. The precise definition of $\fin$ for GB
gravity is given below in eq.~\reef{quadf}. To facilitate our
discussion with arbitrary $d\ge4$, we would like to define two central
charges that appear any CFT for any $d$ --- including odd $d$ and hence
the trace anomaly is not a useful definition of the central charges.
One simple charge is that controlling the leading singularity of the
two-point function of the stress tensor. For GB gravity, one finds
\cite{cc}\footnote{Note that the present normalization was chosen so
that $\CT=\ads$ in the limit $\la\to0$. This choice is slightly
different from that in \cite{cc}, \ie
$C_T=\frac{d+1}{d-1}\frac{\Gamma(d+1)}{\pi^d}\CT$.}
 \be
\CT= \frac{\pi^{d/2}}{\Gamma(d/2)} \(\frac{\tL}{\lp}\)^{d-1} \left[ 1 -
2 \lambda \fin \right]\,.
 \labell{effectc}
 \ee
One can think of this central charge as playing the role of the
four-dimensional $c$ in higher dimensions. In fact, with the present
normalization, we have $\CT\big|_{d=4}=c$. As the second central
charge, we use that identified in \cite{cthem} as satisfying an
interesting holographic c-theorem:
 \be
\ads =\frac{\pi^{d/2}}{\Gamma(d/2)} \(\frac{\tL}{\lp}\)^{d-1} \left[ 1
- 2\frac{d-1}{d-3} \lambda \fin \right]\,.
 \labell{effecta}
 \ee
This central charge can be determined in the CFT from the entanglement
entropy across a spherical entangling surface \cite{cthem}. Further it
was shown \cite{cthem} that, in even dimensions, $\ads$ is the central
charge appearing in the A-type trace anomaly \cite{deser}. For example,
comparing with eq.~\reef{fourdc}, we see $\ads\big|_{d=4}=a$. In odd
dimensions, this central charge can also be identified with the
(renormalized) partition function of the boundary theory evaluated on
$S^d$ \cite{casini9}. With this holographic dictionary,
eqs.~\reef{effectc} and \reef{effecta}, we can interpret the results of
our gravitational calculations in terms of quantities defined in the
dual boundary theory.

Let us now follow the analogous steps as in the previous section to
calculate \ren entropy with GB gravity. First we must determine the
function $f(r)$ appearing in the metric \eqn{lineelement}. For GB
gravity, the equations of motion lead to the following simple quadratic
equation \cite{old1}:
 \be
f(r)-\lambda f(r)^2   = 1 - \frac{\omega^d}{r^d} \,,
 \labell{func1}
 \ee
where $\w^d$ is again an integration constant. Combining the above
expression with eq.~\reef{horizon}, we can parameterize this constant
in terms of the position of the horizon with
 \be
\w^d=\rh^d-L^2\rh^{d-2}+\la L^4 \rh^{d-4}\,.
 \labell{horiGB}
 \ee
It is straightforward to determine the roots of eq.~\reef{func1} and we
find
 \be
f(r) = \frac{1}{2\lambda} \[ 1 - \sqrt{ 1 - 4 \lambda \( 1 -
\frac{\omega^d}{r^d} \) } \]\,.
 \labell{solgb}
 \ee
We should note that, in general, eq.~\reef{func1} has two roots and we
are only considering that which yields the solution \reef{sol0} of the
Einstein theory in the limit $\la\to0$. One finds that graviton
fluctuations about the solution given by the other root are ghosts
\cite{GBghost,old1} and hence the boundary theory would not be unitary.
Demanding a reasonable holographic framework (\ie demanding that the
dual theory is causal or does not produce negative energy excitations)
produces further constraints. In particular, the gravitational coupling
must lie within \cite{EtasGB,cc}
 \be
- \frac{(3d  +2)(d - 2)}{ 4(d + 2)^2} \leq \lambda \leq \frac{(d - 2)(d
- 3)(d^2 -d+ 6)}{ 4(d^2 - 3d + 6)^2} \,.
 \labell{limitsd}
 \ee
In terms of the central charges, these bounds are conveniently written
as
 \be
\frac{d(d-3)}{d(d-2)-2} \leq \frac{\CT}{\ads} \leq \frac{d}{2} \,.
 \labell{limitsca}
 \ee
We also observe here that $\fin=(1 - \sqrt{ 1 - 4 \lambda})/(2\la)$ and
hence $\tL=L/\sqrt{\fin}\ne L$ for nonvanishing $\la$, \ie the AdS
curvature scale is distinct from the scale $L$ appearing in the action
\reef{GBAction}. Further, we note that from eq.~\reef{func1}, the
constant $\fin$ satisfies
 \be
 1-\fin+\la\fin^2=0\,,
 \labell{quadf}
 \ee
which may be used to simplify various expressions in the following.

Applying eq.~\reef{temp} to the above solution \reef{solgb} we find
that the temperature can be expressed as
 \be
T = \frac{1}{2 \pi R}\,\frac{1}{x} \( 1 + \frac{d}{2 \fin}\, \frac{ x^4
- \fin x^2 + \la\fin^2}{  x^2 - 2\la\fin } \)\,,
 \labell{tempGB}
 \ee
where, as in the previous section, we used $x=\rh/\tL$. Next the horizon
entropy is calculated to be\footnote{Note that the expression in the
first line of eq.~\reef{rensGB} is not precisely the same as the Wald
entropy \reef{Waldformula} but both expression agree when evaluated on
a Killing horizon \cite{ted1,EEGB}.}
 \beqa
S &=& \frac{2 \pi}{\lp^{d-1}} \int d^{d-1} x \sqrt{h}
 \[ 1 + \frac{2 \lambda L^2}{(d-3)(d-2)} {\cal R} \]
 \labell{rensGB} \\
&=& 2 \pi \frac{\tL^{d-1}}{\lp^{d-1}}\Vs \,x^{d-1}\( 1 - 2 \lambda\fin
\frac{d-1}{d-3}\, \frac1{x^2} \) \,.
 \nonumber
 \eea
Then applying eq.~\reef{finalfor2} we arrive at the following
expression for the \ren entropy
 \bea
S_q &=& \frac{\pi\, q}{q-1} \Vs \( \frac{\tL}{\lp} \)^{d-1}
\[ \frac{1}{f_\infty}\(1-\xq^d\) - \frac3{d-3}\(1-\xq^{d-2}\) -
\frac{d-1}{d-3}\lambda \fin\(1-\xq^{d-4}\)
 \right.
 \nonumber\\
 &&\qquad\qquad\qquad\qquad\quad +\left.
\frac{d}{d-3}\(1-4\lambda\)\(\frac{1}{1-2\la\fin}-\frac{\xq^d}{\xq^2-2\lambda
\fin}\)\]\,,
 \labell{SqGB}
 \eea
where again $\xq$ corresponds to the value of $x$ when the horizon
temperature is $T = T_0/q$. Using eq.~\reef{tempGB}, we find that the
latter is the root of a quartic equation:
 \be
0 = \frac{d}\fin\, \xq^4 - \frac{2}{q}\,x^3 - \(d-2\) x^2 + \frac{4
\lambda f_\infty}{q}\,x + \(d-4\)\lambda f_\infty\, .
 \labell{xq}
 \ee
As a check of these results, one may verify that in the limit
$\la\to0$, they reduce to the correct expressions for Einstein gravity
in eqs.~\reef{EHquad} and \reef{sqeh}.

Since these general expressions are quite complicated, in considering
the results in more detail, we focus on $d=4$. In this case, we may use
eq.~\reef{fourdc} to express the result \reef{SqGB} in terms of the
central charges of the boundary CFT:
 \be
S_q=\frac{q}{q-1}\frac{\Vs}{4\pi}(1-\xq^2) \left[
(5c-a)\xq^2-(13c-5a)+16c\,\frac{2c\,\xq^2-(c-a) }{(3c-a)\xq^2-(c-a)}
\right]
 \labell{SqGB4}
 \ee
and further eq.~\reef{xq} reduces to the following cubic equation:
 \be
0=\xq^3-\frac{3c-a}{5c-a}\left(\frac{\xq^2}{q}+\xq\right)+\frac{1}{q}
\frac{c-a}{5c-a}\,.
 \labell{xq4}
 \ee
This four-dimensional result is still quite complicated. The \ren
entropy for these theories no longer has the simple qualitative form
\reef{qual} that was found for the holographic CFT's dual to Einstein
gravity and held for any two-dimensional CFT. Here we would replace
eq.~\reef{qual} with $S_q= a \times V(R/\delta)\times f(q, c/a)$. In
particular, the \ren entropy depends on both of the two distinct
central charges in the boundary theory and the dependence on these
central charges can not be factored from the dependence on the index
$q$.

Considering the limits $q\rightarrow 0$, 1 and $\infty$ in
eq.~\reef{SqGB4}, we find\footnote{It may be useful to note the
corresponding limits for the physical root of eq.~\reef{xq4}: $\lim_{q
\rightarrow 0} \xq = \fin/(2q)$, $\lim_{q \rightarrow 1}\xq=1$ and
$\lim_{q \rightarrow \infty}\xq=\sqrt{\fin/2}$.}
 \bea
\lim_{q \rightarrow 0} S_q      &=& a \, \frac{2\Vs }{\pi}\,
\frac{1}{8q^3} \, \frac{(3(c/a)-1)^4}{(5(c/a)-1)^3} \,,
 \nonumber\\
\lim_{q \rightarrow 1} S_q      &=& a \, \frac{2\Vs }{\pi} \,,
 \labell{limitsGB} \\
\lim_{q \rightarrow \infty} S_q &=& a \, \frac{2\Vs}{\pi} \, \[ 1
-\frac{3}{2} \, \frac{(c/a)^2}{5(c/a)-1}\]\,.
 \nonumber
 \eea
As required, in the limit $q\to1$, we recover the entanglement entropy
as was calculated previously \cite{casini9,cthem}. In this case, there
is a considerable simplification and in particular this result only
depends on the central charge $a$. Of course, it is a general result
for any $d=4$ CFT that the coefficient of the universal term in $\Vs$,
given in eq.~\reef{unis}, is proportional to $a$ (and is independent of
$c$) with a spherical entangling surface, as considered in \cite{solo}.
However, no such simplification arises in either of the other limits
with $S_0$ and $S_\infty$ depending on both $a$ and $c$.

As the \ren entropy in eq.~\reef{SqGB4} is so complicated, it is useful
to analyze the result numerically. For this purpose, we consider the
ratio of $S_q/S_1$, \ie the ratio of the $q$'th \ren entropy to the
entanglement entropy. In dividing by $S_1$, the entanglement entropy,
we remove the dependence on the regulator volume $\Vs$ as well as the
overall factor of the central charge $a$, and we are left with a
function of $q$ and $c/a$. Figure \ref{fig:GB} shows a plot of
$S_q/S_1$ as a function of $c/a$ for various values of $q$. Recall that
physical constraints on the holographic framework impose constraints on
the allowed values of the gravitational coupling $\la$, as given in
eq.~\reef{limitsd}, or alternatively on the ration of the two central
charges, as given in eq.~\reef{limitsca}. For $d=4$, the latter become
\cite{diego,EtasGB}:
 \be
 \frac23\le \frac{c}{a}\le 2\,.
 \labell{physical4}
 \ee
Hence in figure \ref{fig:GB}(a), we only plot $S_q/S_1$ over this
physical range. Remarkably, the figure shows that in this range, the
resulting curves are essentially linear for any value of $q$. Further
$S_q/S_1$ is a monotonically decreasing function of $c/a$, \ie
 \be
\left.\frac{S_q}{S_1}\right|_{c_1/a_1} >
\left.\frac{S_q}{S_1}\right|_{c_2/a_2} \qquad{\rm for}\ \
\frac{c_1}{a_1}<\frac{c_2}{a_2}\,.
 \labell{ovsve}
 \ee
\FIGURE[!ht]{
\begin{tabular}{cc}
\includegraphics[width=0.5\textwidth]{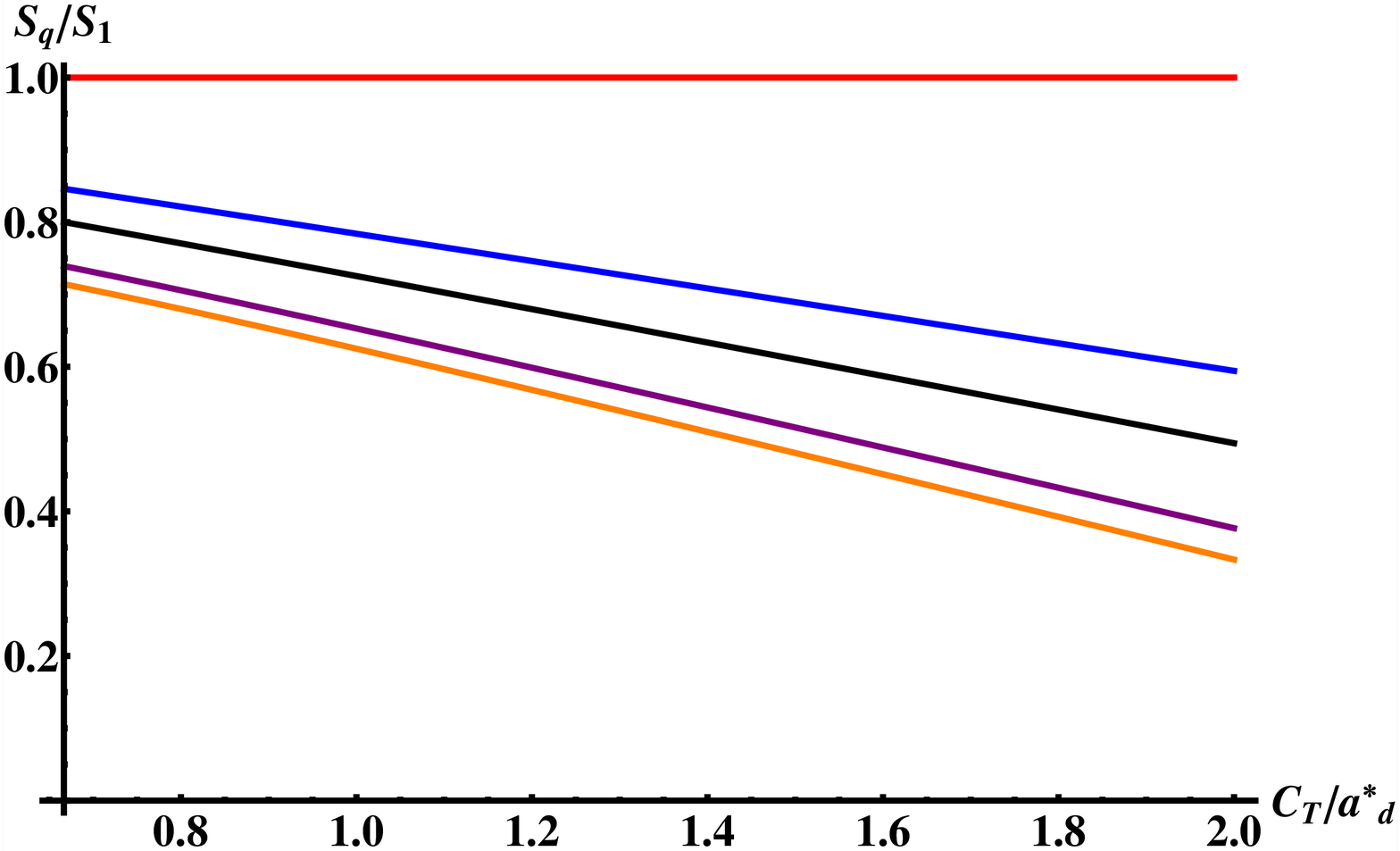}&
\includegraphics[width=0.5\textwidth]{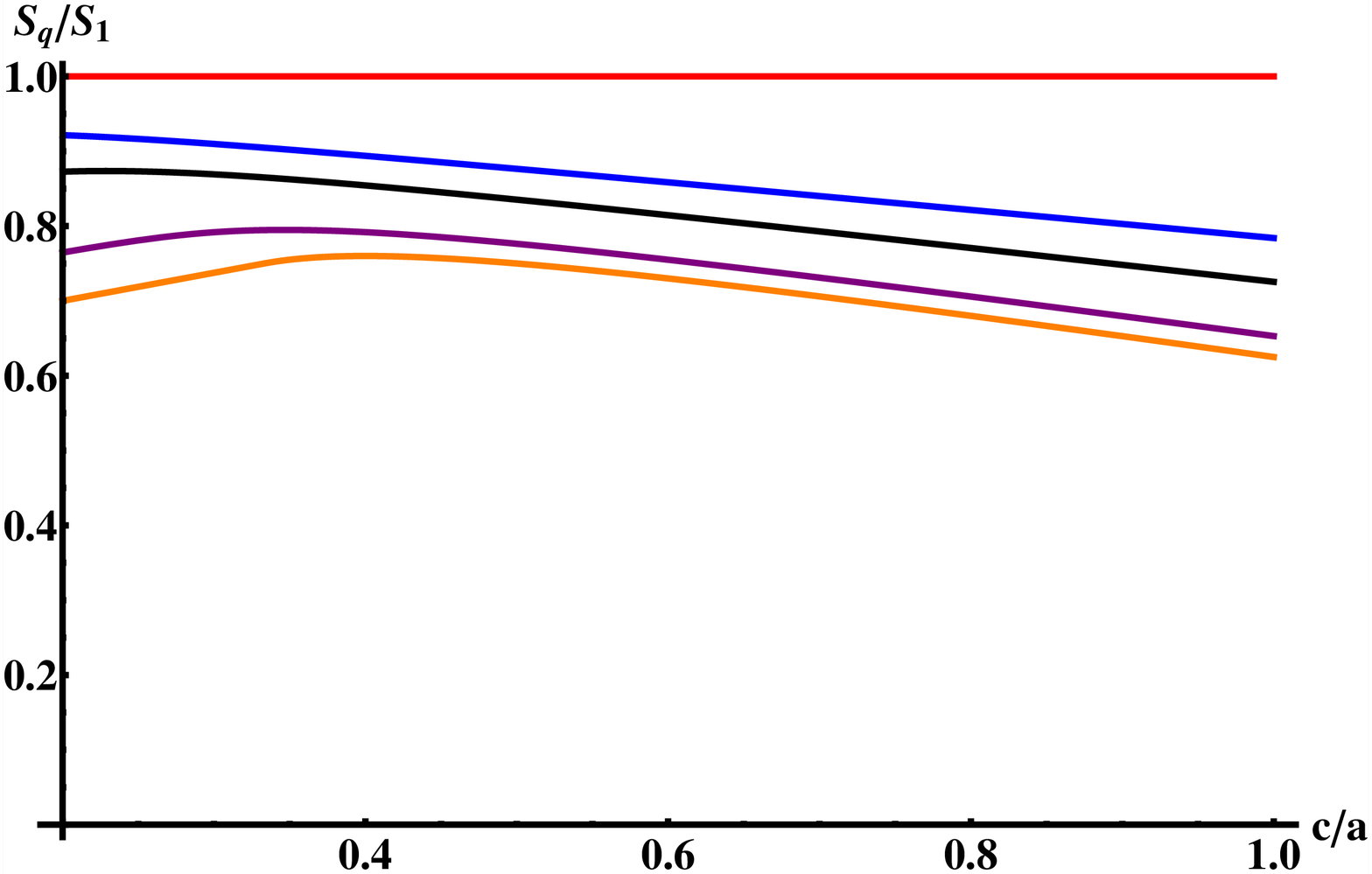}\\
(a) & (b)
\end{tabular}
\caption{\ren entropy \reef{SqGB4} divided by the entanglement entropy,
$S_q/S_1$ plotted as a function of $c/a$ for various values of
$q=\left\{1, 2, 3, 10, 100 \right\}$ with $d=4$. The corresponding
curves run from top to bottom with $q=1$ at the top. In panel (a), the
plot runs over the physical range: $2/3\le c/a \le 2$. Despite the
complicated form of eqs.~\eqn{SqGB4} and \eqn{xq4}, these curves are
all essentially linear. In panel (b), we see this nearly linear
behaviour breaks down for smaller values of $c/a$.} \labell{fig:GB}  }

Some insight into this apparent linearity comes from considering the
$q\to\infty$ limit. Using the expressions in eq.~\reef{limitsGB}, the
ratio of interest can be written as
 \bea
\frac{S_\infty}{S_1} &=& 1 -\frac{3}{10} \frac{c}{a} \frac{1}{1-\frac{1}{5}
\( \frac{c}{a} \)^{-1}}  \labell{test}\\
& \simeq&   - \frac{3}{10} \frac{c}{a} +\frac{47}{50}- \frac{3}{250}
\frac{a}{c} - \cdots \,.
 \nonumber
 \eea
In the physical regime, the term $\frac{1}{5} \( \frac{c}{a} \)^{-1}$
varies between $0.1$ and $0.3$ and hence the corresponding expression
multiplying $c/a$ in the first line above is approximately constant. In
fact, the slope varies by slightly less than $10 \%$ over the physical
region \reef{physical4}. The suppression of the higher order terms in
the expansion in the second line above essentially arises because of
the extra factor of $1/5$ and so appears to be largely a numerical
accident. Using eqs.~\reef{SqGB4} and \reef{xq4}, a similar expansion
can be made for $S_q/S_1$ with general $q$ but we do not present the
details here. These are essentially expansions for large $c/a$ and, as
evident from figure \ref{fig:GB}(a), numerical factors again strongly
suppress the higher order terms even though the ratio of central
charges is not particularly large in the physical regime
\reef{physical4}.

It is straightforward to demonstrate that the \ren entropy could not
have been a linear function of the central charges.  The latter must
obey the inequality $\partial_q S_q \leq 0$ \cite{karol} --- see also
appendix \ref{unequal} --- and hence we should find $S_q/S_1 < 1$ for
all $q>1$ and any $c/a$. Hence, $S_q/S_1$ is bounded from above and
therefore cannot be linear in $c/a$.  In figure \ref{fig:GB}(b), we
plot this ratio for smaller values of $c/a$ outside the physical
regime. In this plot, the nonlinearity becomes evident and it can be
seen that the curves for $q=10$ and 100 reach a maximum value. From
eq.~\reef{test}, one readily finds that the maximum in $S_\infty/S_1$
occurs at $c/a=2/5$. However, we note that this interesting structure
arises in a regime where the holographic model has unphysical
properties and so we do not wish to put too much weight on the
appearance of these maxima. We further note that $S_\infty/S_1$ becomes
negative for $c/a\lesssim 0.214$ or $c/a\gtrsim3.120$, which is clearly
unphysical.

Above we have focused the discussion on $d=4$ for simplicity. It is
straightforward to carry out a similar analysis for any value of $d$
--- however, recall we are assuming $d\ge4$ with this holographic model
using GB gravity in the bulk --- and the results are qualitatively
similar to those found above. In particular, one would begin using
eqs.~\reef{effectc} and \reef{effecta} to translate the \ren entropy
\reef{SqGB}, as well as eq.~\reef{xq}, to expressions in terms of the
effective central charges, $\CT$ and $\ads$, in the dual boundary
theory. In particular, one may make this translation using
 \bea
 \(\frac{\tL}{\lp}\)^{d-1}&=&\frac{\Gamma\(d/2\)}{2\pi^{d/2}}\ \ads\, \(
 (d-1)\,(\CT/\ads)-(d-3)\)\,,
 \labell{dictd}\\
 \la\fin&=&\frac{d-3}2\,\frac{(\CT/\ads)-1}{(d-1)\,(\CT/\ads)-(d-3)}\,,
 \nonumber
 \eea
from eqs.~\reef{effectc} and \reef{effecta}, as well as $1/\fin = 1 -
\la\fin$ from eq.~\reef{quadf}. One then finds expressions which are
equally complicated as eqs.~\reef{SqGB4} and \reef{xq4} were for $d=4$.
However, one may again plot $S_q/S_1$ versus $\CT/\ads$ and in the
physical regime corresponding to eq.~\reef{limitsca}, the resulting
curve is essentially linear. Figure \ref{fig:GB2} illustrates this
behaviour for $d=5$ and 6.
\FIGURE[!ht]{
\begin{tabular}{cc}
\includegraphics[width=0.5\textwidth]{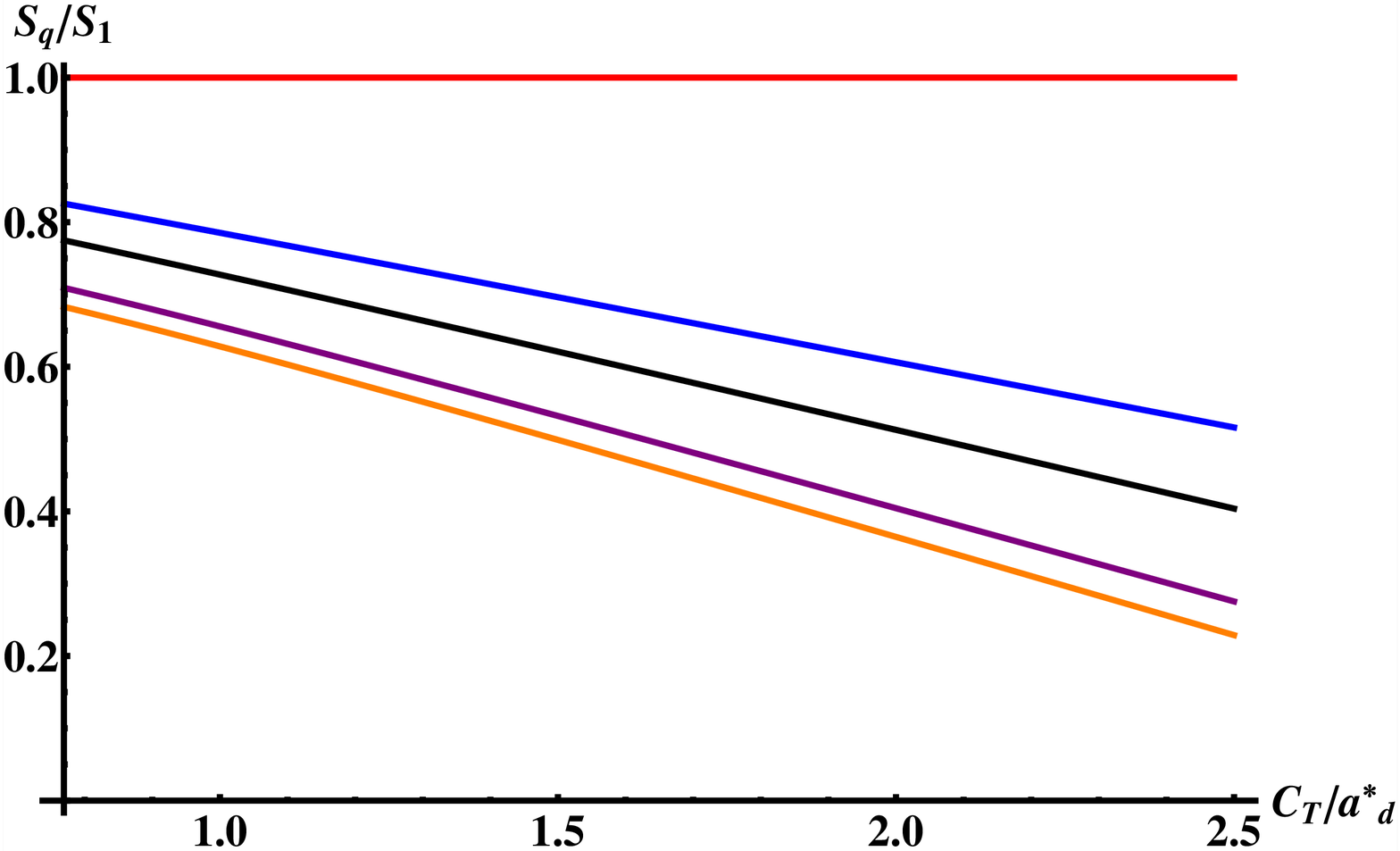}&
\includegraphics[width=0.5\textwidth]{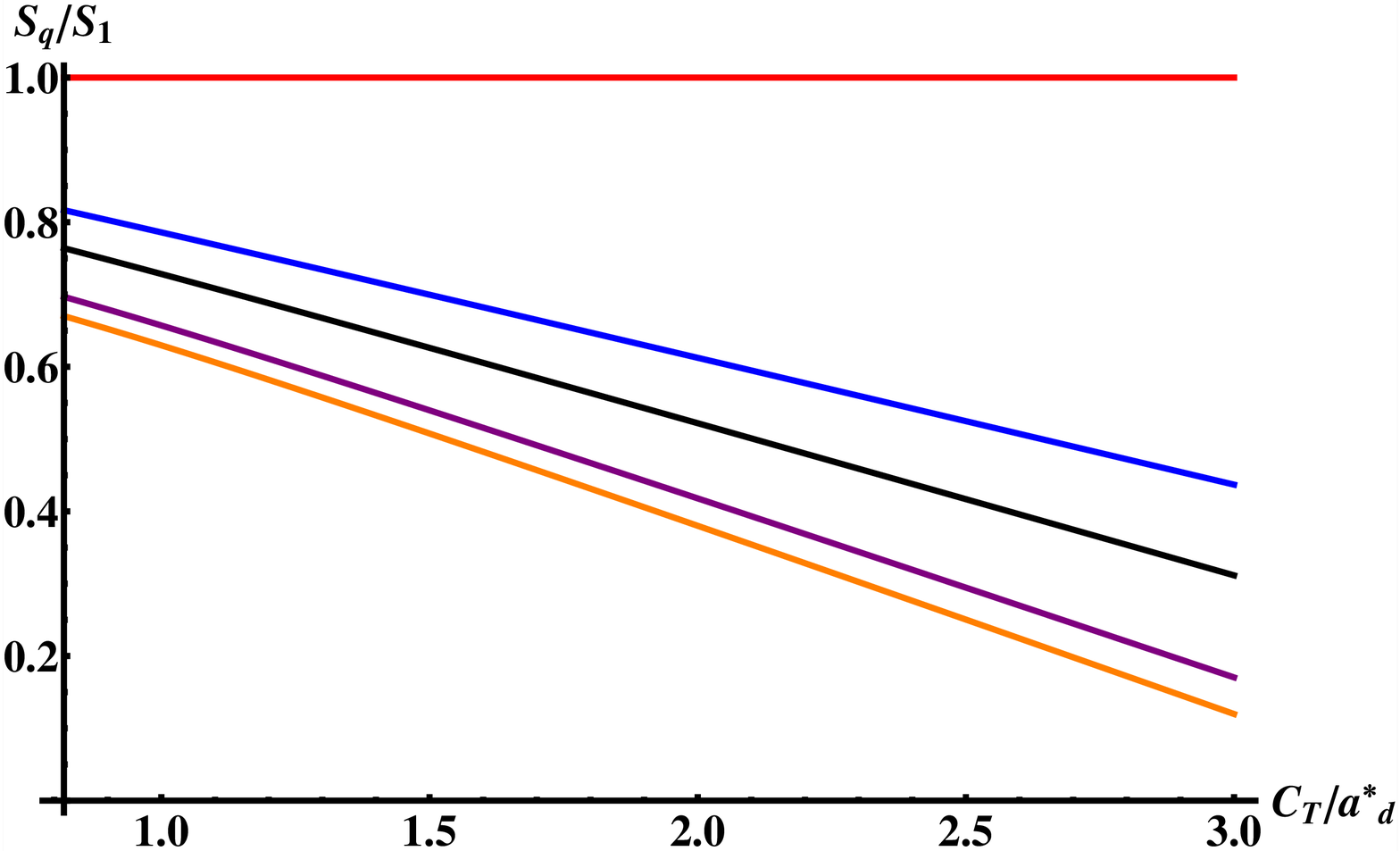}\\
 $d=5$&  $d=6$\\
\end{tabular}
\caption{$S_q/S_1$ plotted as a function of $\CT/\ads$ for $d=5,6$ and
for $q=\left\{1, 2, 3, 10, 100 \right\}$. The corresponding curves run
from top to bottom with $q=1$ at the top. Despite the complicated form
of eqs.~\eqn{SqGB} and \eqn{xq}, these curves are all essentially
linear in the physical regime of $\CT/\ads$, given in
eq.~\reef{limitsca}.} \labell{fig:GB2} }

\subsection{Quasi-topological gravity and $d=4$ \labell{sec:QT}}

It is well known that, for any even $d$, the universal contribution to
the entanglement entropy of a CFT has a logarithmic dependence on the
UV cut-off with a coefficient given by some linear combination of the
central charges appearing in the trace anomaly
\cite{rt2,cthem,solo,EEGB}. This result applies for any smooth
entangling surface and the precise linear combination of central
charges depends of the geometries of both the entangling surface and
the background in which it is embedded. In the previous section, by
extending the bulk theory to GB gravity, we were able to distinguish
various (effective) central charges in the boundary theory, as given in
eqs.~\reef{effectc} and \reef{effecta}. However, the subsequent
holographic calculations demonstrated that the \ren entropy of the
boundary CFT's has a very complicated dependence on these two charges
--- a property that holds for both even and odd $d\ge4$. In this section,
we will demonstrate that this complexity extends beyond the central
charges. In particular, we will focus on the case of even dimensions
and show that, in general, the \ren entropy for a spherical entangling
surface depends on CFT parameters beyond the central charges appearing
in the trace anomaly.

In the following, we will study a $d=4$ holographic model with
quasi-topological gravity as the bulk theory
\cite{old1,old2}.\footnote{It is straightforward to extend the
following analysis to higher values of $d$ but the qualitative
behaviour of the \ren entropy does not change. In particular, the final
result explicitly depends on the new parameter $t_4$, as well as the
two central charges.} The action is given by
 \beq
I = \frac{1}{2\lp^3} \int \mathrm{d}^5x \, \sqrt{-g}\, \left[
\frac{12}{L^2} + R + \frac{\lambda L^2}{2}\X_4 +\frac{7\mu L^4}{4}\Z_5
\right]
 \labell{QTAction}
 \eeq
where $\X_4$ is the same curvature-squared term \reef{GBterm} appearing
in GB gravity and $\Z_5$ is a new third-order interaction originally
constructed in \cite{old1,newer}:
 \beqa
\Z_5&=& R_{a\,\,b}^{\,\,c\,\,\,d} R_{c\,\,d}^{\,\,e\,\,\,f}
R_{e\,\,f}^{\,\,a\,\,\,b} + \frac{1}{56}\left(21\,R_{a b c d}R^{a b c
d} R-72\,R_{a b c d}R^{a b c}{}_{e}R^{d e}
 \right.\nonumber\\
&&\qquad\left.+ 120\,R_{a b c d} R^{a c}R^{b d}
+144\,R_a{}^{b}R_b{}^{c}R_c{}^{a} - 132\, R_a^{\,\,b}R_b^{\,\,a}R
+15\,R^3\right)\,. \labell{result5}
 \eeqa
This action \reef{QTAction} was constructed to provide a holographic
model where the three-point function of the stress tensor in the
boundary CFT had the most general possible form \cite{old1}. In any CFT
with $d\ge4$, this three-point function is fixed by conformal
invariance up to three independent constants. For $d=4$, we may
characterize these three constants as the two central charges, $a$ and
$c$, and a third parameter $t_4$ which naturally arises in describing
certain scattering experiments.\footnote{We refer the interested reader
to \cite{diego,old2} for further details. We also note that to simplify
various expressions in the following, we have changed the normalization
of this parameter with respect to these previous references, \ie
$\left(t_4\right)_\mtt{previous}=1890\,\left(t_4\right)_\mtt{present}$.}
Given the action \reef{QTAction}, the appropriate AdS/CFT dictionary
reads \cite{old2}:
 \bea
c &=& \pi^2 \frac{\tilde{L}^3}{\lp^3} \( 1 - 2 \lambda f_\infty - 3 \mu
f_\infty^2\) \,,
 \nonumber\\
a &=& \pi^2 \frac{\tilde{L}^3}{\lp^3} \( 1 - 6 \lambda f_\infty + 9 \mu
f_\infty^2\)\,,
 \labell{parms4}\\
t_4 &=& \frac{2 \mu f_\infty ^2}{1 - 2 f_\infty \lambda - 3\mu
f_\infty^2}\,.
 \nonumber
 \eea
As usual, $\tL$ denotes the AdS curvature scale while $\fin=L^2/\tL^2$.
The precise definition for $\fin$ in quasi-topological gravity is given
below in eq.~\reef{finQT}. Note that if $\mu$ is set to zero, we
recover the expressions in eq.~\reef{fourdc} for the central charges in
GB gravity and $t_4$ vanishes.

By design, the metric function $f(r)$ in eq.~\eqn{lineelement} is still
determined as the root of a simple polynomial \cite{old1}
 \be
f(r)- \lambda f(r)^2 -\mu f(r)^3  = 1 - \frac{\w^4}{r^4} \,.
 \labell{fQT}
 \ee
The integration constant $\w^4$ can again be written in terms of the
position of the horizon
 \be
\w^4=\rh^4-L^2\rh^2+\la L^4 + \mu L^6/\rh^2\,.
 \labell{horiQT}
 \ee
As noted previously, the asymptotic value of $f(r)$ fixes the AdS scale
with $\tL=L/\sqrt{\fin}$. In the present case, in the limit
$r\to\infty$, eq.~\reef{fQT} reduces to the following cubic equation
for $\fin$:
 \be
1-\fin+\la \fin^2 +\mu\fin^3=0\,.
 \labell{finQT}
 \ee
Demanding that the boundary theory does not produce negative energy
excitations constrains the two higher curvature couplings, $\la$ and
$\mu$, to satisfy \cite{old2}:
 \bea
0 & \leq & 1-10 \lambda \fin + 189 \mu \fin^2\,,
 \nonumber \\
0 & \leq &1 + 2 \lambda \fin - 855 \mu \fin^2\,,
 \labell{ineqQT} \\
0 & \leq &1 + 6 \lambda \fin + 1317 \mu \fin^2 \,.
 \nonumber
 \eea
Within this physical domain, the couplings remain relatively small.
Further, in this regime, eq.~\reef{finQT} will have three roots but
implicitly we only consider the smallest positive root, as it is the
only one to yield a physically reasonable holographic model without
ghosts \cite{GBghost,old1}.

The temperature of these black hole solutions can be calculated using
eq.~\reef{temp} and is given by
 \be
T = \frac{1}{2\pi R}\,\frac1x \( 1 + \frac2\fin\, \frac{ x^6  - \fin
x^4 + \lambda \fin^2 x^2 + \mu \fin^3}{ x^4 - 2\lambda \fin x^2 - 3\mu
\fin^2}\) \,,
 \labell{tempQT}
 \ee
where as before $x \equiv \rh/\tilde{L}$. Calculating the entropy using
the Wald formula \reef{Wald2} yields \cite{old1}:
 \be
 S = 2\pi \frac{\tL^3}{\lp^3}\Vs\, x^3 \( 1 - 6\lambda \fin\frac{1}{x^2}
+ 9\mu \fin^2 \frac{1}{x^4} \) \,.
 \labell{entQT}
 \ee
Combining the above results with eq.~\reef{finalfor2}, the \ren entropy for
a spherical entangling surface in the $d=4$ boundary CFT is
\bea
 S_q &=& \frac{\pi\,q}{q-1}\Vs \,\frac{\tL^3}{\lp^3}\,(1-\xq^2)
\Bigg\{ \frac{1+\xq^2}{f_\infty} +  \(1-16\lambda\)
-\frac{3\mu\fin^2}{\xq^2}
-16 \fin^2
 \labell{SqQT}\\
&&\ \   \times\left[
\frac{((1-4\la)\la^2+3\mu\la-2\mu)(\xq^2+3\mu\fin^2)
+3\mu\fin((1-2\la)\la+3\mu)(1+\xq^2-2\la\fin)}{(1-2 \lambda \fin - 3
\mu \fin^2)(\xq^4-2 \lambda \fin \xq^2 - 3 \mu \fin^2)}
 \right]\Bigg\} \,.
 \nonumber
 \eea
As before, $\xq$ is defined by the relation $T =T_0/q$. In this case,
eq.~\reef{tempQT} yields the following sixth order polynomial to solve
for $\xq$:
 \be
 0 = \frac2\fin\,\xq^6 - \frac{1}{q}\xq^5 - \xq^4 + \frac{2
\lambda f_\infty}{q}\, \xq^3 + \frac{3 \mu f_\infty^2}{q}\,\xq - \mu
f_\infty^2\,.
 \labell{xQT}
 \ee
Although it is not immediately obvious, the $\mu \rightarrow 0$ limit
correctly reproduces the corresponding expressions for GB gravity in
eqs.~\reef{SqGB4} and \reef{xq}, respectively.

The above expressions are readily rewritten in terms of the CFT
parameters using eq.~\reef{parms4}, which yields
 \bea
 \frac{\tL^3}{\lp^3}&=&\frac{a}{2\pi^2} \( 3 \frac{c}{a}\left(1 + 3 \tfor \right) - 1\)
 \,,
 \nonumber\\
 \la\fin&=&\frac{1}{2}\, \frac{\frac{c}{a}\left(1 + 3 \tfor \right)-1}{3
 \frac{c}{a}\left(1 + 3 \tfor \right) -1}
 \,,
 \labell{dict4x}\\
 \mu\fin^2&=&\frac{\frac{c}{a} \tfor}{3 \frac{c}{a}\left(1 + 3 \tfor \right) - 1 }
 \,,
 \nonumber
 \eea
as well as eq.~\reef{finQT}, which yields
 \bea
 \frac1\fin&=&1-\la\fin-\mu\fin^2
 \labell{housefire}\\
 &=&\frac{1}{2}\, \frac{5 \frac{c}{a} \( 1 - 2 \tfor \)-1}{3 \frac{c}{a} \( 1 + 3 \tfor \)-1}
 \,.
 \nonumber
 \eea
Also note that the constraints in eq.~\reef{ineqQT} become
 \bea
2 & \geq & \frac{c}{a} \( 1 - 84\, \tfor \)\,,
 \nonumber \\
\frac{1}{2} & \leq & \frac{c}{a} \( 1 - 210\, \tfor\) \,,
 \labell{ineqx} \\
\frac{2}{3} & \leq & \frac{c}{a} \( 1 + 224\, \tfor\)  \,.
 \nonumber
 \eea

We will not explicitly display eqs.~\reef{SqQT} and \reef{xQT} after
their conversion to CFT parameters since they are rather cumbersome and
not particularly enlightening. Of course, one important feature which
they do demonstrate is that the \ren entropies have a very complicated
dependence on these parameters characterizing the boundary CFT. As in
the previous section with GB gravity in the bulk, the \ren entropy no
longer has the simple qualitative form \reef{qual} that appeared for
CFT's dual to Einstein gravity. Instead the qualitative form of the
results is $S_q= a \times V(R/\delta)\times f(q, c/a, \tfor)$. In
particular then, the \ren entropy depends on the new parameter $\tfor$,
as well as both of the central charges. Demonstrating this additional
dependence of the \ren entropy on parameters beyond the central charges
in an even-dimensional CFT was the main goal of the present section.

For a fixed value of $\tfor$, the \ren entropy here has the same
qualitative shape as with GB gravity in the previous section (for which
$\tfor=0$). In particular, within the physical regime, $S_q/S_1$ is
nearly linear with a negative slope, similar to the dependence shown in
figure \ref{fig:GB}. Figure \ref{fig:QT} illustrates the new dependence
on $\tfor$. In figure \ref{fig:QT}(a), we see that within the
physically allowed regime fixed by eq.~\reef{ineqx}, the dependence of
$S_q/S_1$ on $t_4$ (at fixed $c/a$) is nearly linear, but in fact the
slope is very small. Figure \ref{fig:QT}(b) shows that the dependence
on $t_4$ and $c/a$ only becomes significant well outside of the
physical regime.
\FIGURE{
\begin{tabular}{cc}
\includegraphics[width=0.45\textwidth]{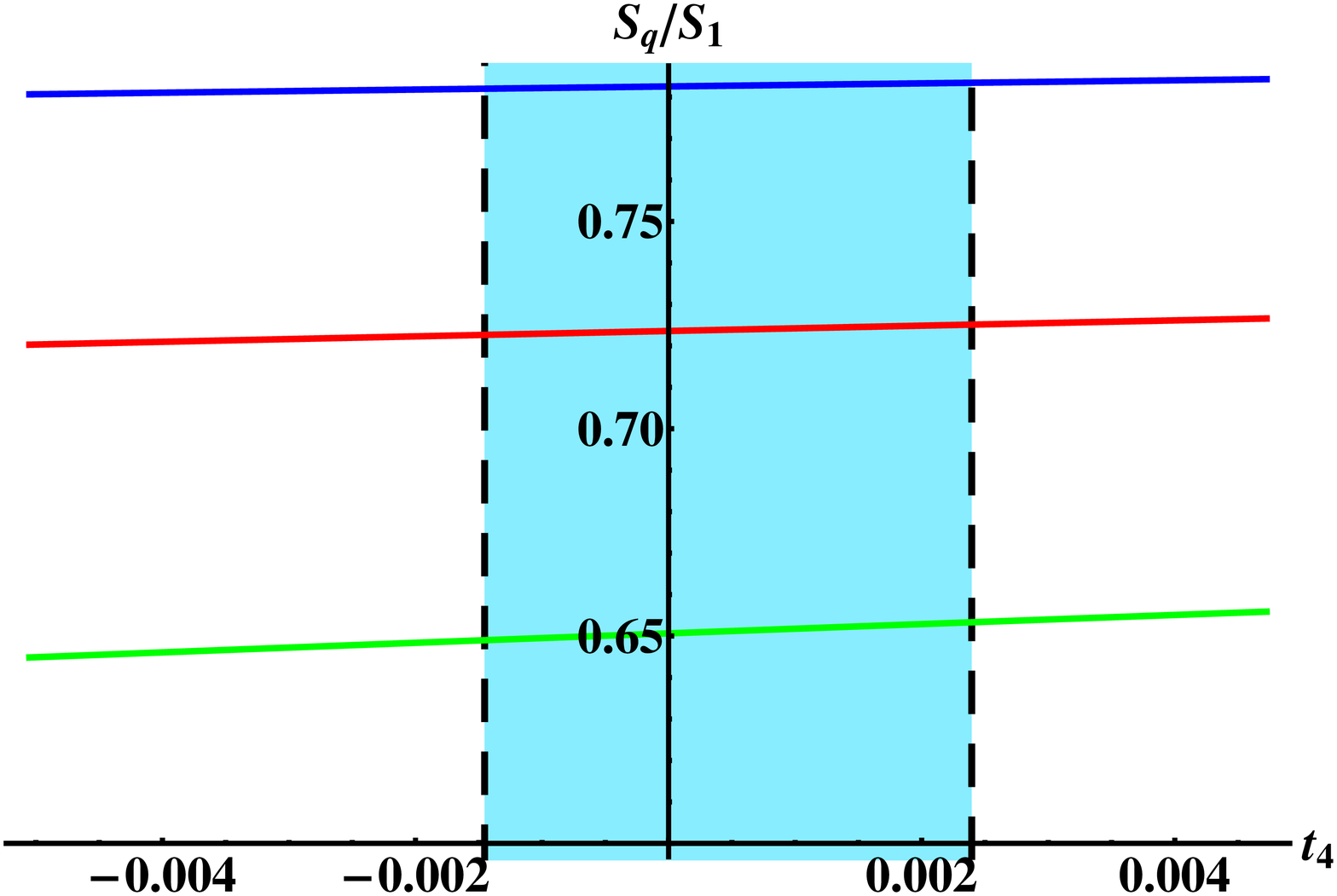} &
\includegraphics[width=0.50\textwidth]{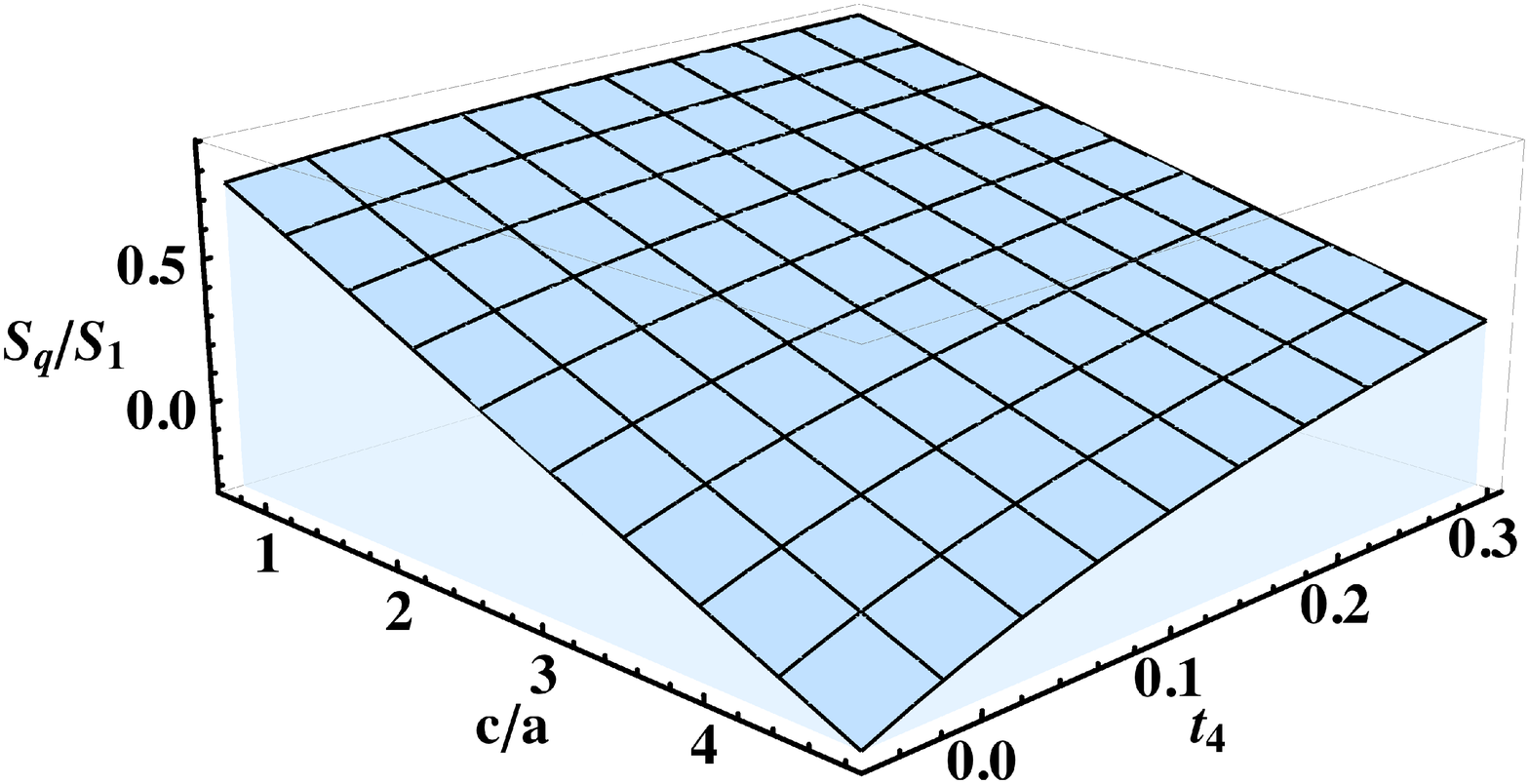} \\
(a) & (b)
\end{tabular}
\caption{(Colour Online) $S_q/S_1$ as a function of $\tfor$ with fixed
$c/a=1$.  Starting from the top, the three curves correspond to $q=2$
(blue), 3 (red) and 10 (green).  The shaded blue region between the
dashed vertical lines corresponds to the allowed physical regime for
$\tfor$ (with $c/a=1$) dictated by eq.~\reef{ineqx}. (b) $S_q/S_1$ as a
function of both $\tfor$ and $c/a$.  Note that in this case, the plot
extends beyond the physical regime.} \labell{fig:QT} }

Recall that in the previous section, there was a certain simplification
in the limits $q\to0,1,\infty$. Considering the $q\to1$ limit, \ie the
entanglement entropy, first here, we find
 \be
 S_\mt{EE}=\lim_{q\to1}S_q=a\,\frac{2\Vs}\pi\,.
 \labell{QTEE}
 \ee
The fact that, for a spherical entangling surface, the entanglement
entropy only depends on the central charge $a$ could be anticipated
from the holographic calculations of \cite{casini9,cthem} or from the
CFT calculations of \cite{solo}. Note that this expression matches that
given for $S_1$ in eq.~\reef{limitsGB} for GB gravity.

Next turning to the limit $q \rightarrow 0$, we find a certain
simplification in the \ren entropy
 \bea
 \lim_{q\to 0} S_q &=& \Vs \frac{\tL^3}{\lp^3}
\frac{\pi f_\infty^3}{16q^3}
 \nonumber\\
&=& \frac{a \Vs}{4 \pi q^3}
\frac{ \( 3 \frac{c}{a} \( 1 + 3 \tfor \)-1\)^4}{\(5 \frac{c}{a} \(1+2 \tfor\)-1\)^3}
\,.
 \labell{soqt}
 \eea
As in the previous section, $\xq = f_\infty/2q$ in this limit. We note
that the expression in the first line, which expresses $S_0$ in terms
of the gravitational parameters, takes the same form here for
quasi-topological gravity as it did for GB gravity in the previous
section. Therefore there is no dependence on the dimensionless
couplings for the higher curvature terms except implicitly through the
factor of $\tL^3f_\infty^3=L^3\fin^{3/2}$. However, despite this
simplification, this expression still depends on $t_4$, as well as the
central charges.

Finally, we consider the limit $q \rightarrow \infty$. In this case,
there is no particular simplification of the final result and so
$S_\infty$ again depends on all three parameters, \ie $t_4$ as well as
the two central charges, $a$ and $c$. To explicitly exhibit the $t_4$
dependence, we first observe that eq.~\reef{ineqx} constrains the
physical values of $\tfor$ to be rather small --- see, for example,
figure \ref{fig:QT}(a). Hence we construct a Taylor series expansion of
$s_\infty$ for small $\tfor$ which yields
 \myeq{
\lim_{q \rightarrow \infty} S_q & \simeq \frac{2 a \Vs}{\pi} \Bigg[ 1 + \frac{3}{2}
\, \frac{ \(\frac{c}{a}\)^2}{1-5\frac{c}{a}} \\
& \qquad + \tfor \frac{1-17\,\frac{c}{a} + 98 \(\frac{c}{a}\)^2 - 194
\(\frac{c}{a}\) ^3 - 17 \(\frac{c}{a}\)^4 + 215 \(\frac{c}{a}\)^5}{2\,
\frac{c}{a}\(1-3\frac{c}{a}\) \(1-5\frac{c}{a}\)^2 } + {\cal O}
(\tfor^2) \Bigg] \,. \labell{sinfqt} }


\section{Twist Operators and Regulator Surfaces} \labell{twist}

In the previous section, the \ren entropy for a spherical entangling
surface was computed by mapping the calculation to a thermodynamic one
\cite{casini9}. This approach contrasts with `standard' field theoretic
computations, particularly in two dimensions \cite{cardy0,cardyCFT},
which make use of the \emph{replica trick}. The replica trick
essentially replaces the problem of computing the $q$-th power of the
density matrix with that of another density matrix of a different
theory that comprises $q$ copies of the original theory. The various
copies only talk to each other along the $q$-fold branch-cuts
introduced along the region of interest. In two dimensions, the latter
is generically a collection of line segments and so this construction
can be realized with the insertion of twist operators in the path
integral over the replicated theory, which are responsible for opening
and closing the branch cuts at the end-points of the various intervals.
Let us denote the twist operator at the end-point $\w_i$ as
$\sigma_{\eta_i q}(\omega_i)$ where $\eta_i=+1$ (--1) to open (close) a
$q$-fold branch-cut. Then the $q$-th power of the reduced density
matrix is written
 \be
\textrm{tr}\left[\rhov^{\,q}\right] = \frac{\int
[\prod_{j=1}^q\textrm{D}\phi_j]\,\prod_k \sigma_{\eta_k q}(\omega_k)\,
\exp\left[-\sum_{i=1}^q I_i[\phi]\right]}{\int
[\prod_{j=1}^q\textrm{D}\phi_j]\, \exp\left[-\sum_{i=1}^q
I_i[\phi]\right]}\equiv\frac{Z_q}{Z_1^q}\,.
 \labell{trrho0}
 \ee
which corresponds to the correlation function of these twist operators.
Here $Z_q$ denotes the path integral of the replicated theory on the
orbifold, and $Z_1$ that of the original theory on the original
background. In the case of a single interval, the \ren entropy is given
by the correlation function of just two twist operators and the
conformal dimension of these operators then completely fixes the form
of this correlator.

The idea above of introducing branch-cuts immediately generalizes to
higher dimensional theories. Hence applying the replica trick to
calculate \ren entropies in higher dimensions should also be a
straightforward extension of the previous discussion. However, in
practice the construction and properties of the corresponding twist
operators are not well understood for $d>2$
--- however, see \cite{swingle,casini6}. Twist operators in a
$d$-dimensional theory would be some ($d-2$)-dimensional \emph{surface}
operators. From previous work on various surface operators and their
holographic duals \cite{surfop}, it seems that Ryu and Takayanagi's
holographic prescription \cite{rt1} for entanglement entropy is in
agreement with this interpretation. For a three-dimensional CFT, for
example, the holographic entanglement entropy computation is
essentially the same as that of the vacuum expectation value of a
spatial `Wilson-line' operator. As so little is known about these
operators, it would of interest to use holography to study their
general properties, \eg the conformal dimension, as this could shed
light on the explicit construction of these objects. We return to these
considerations in section \ref{versus}. In the present section, we will
begin with a discussion of two-dimensional CFT's and the corresponding
holographic description of calculations using twist operators.

\subsection{Twist operators in $d=2$} \labell{twistor}

Before considering the holographic calculation, let us review the
computation the \ren entropy $S_q$ in a two-dimensional CFT using the
replica trick in more detail. In particular, we calculate $S_q$ for a
single interval $V$ bounded by two points $\omega=v_1$ and $\omega=v_2$
in the complex $\omega$-plane. Without loss of generality we can put
both $v_1$ and $v_2$ on the real line, with $v_2>v_1$. Recall the general
definition \reef{rensq} of the entanglement entropy in terms of the
reduced density matrix $\rhov$:
 \be
 S_q=\frac{1}{1-q}\,\log{\rm tr}\!\left[\,\rhov^{\,q}\right] \,.
 \labell{defsn}
 \ee
As explained above, using the replica trick, one computes instead the
path integral of a theory where $q$ copies of all the fields are
introduced accompanied by the insertion of appropriate twist operators
at the end-points, $v_1$ and $v_2$. Explicitly, eq.~\reef{trrho0}
reduces to
 \be \labell{trrho1}
\textrm{tr}\left[\rhov^{\,q}\right] = \frac{Z_q}{Z_1^q}=\langle\,
\sigma_q(v_1)\, \sigma_{-q}(v_2)\,\rangle\,.
 \ee
The branch points at which the twist operators are inserted are
singular points in the universal cover of the $\omega$-plane, suffering
an angular excess of $2\pi(q-1)$. However, in $d=2$, all metrics are
conformally flat. Hence the singular metric on the universal cover can
be mapped to a flat metric by a conformal transformation. In
particular, the transformation \cite{cardy0}
 \be
z=\left(\frac{\omega-v_1}{\omega-v_2}\right)^{\frac{1}{q}}\,,
 \labell{ctrans}
 \ee
maps the singular $\w$-plane to the smooth $z$-plane. From
eq.~\reef{ctrans}, we easily see that $\w=v_1\to z=0$ and $\w=v_2\to
z=\infty$ and hence, this transformation maps the interval between
$v_1$ and $v_2$ in the $w$-plane onto the half-line in the $z$-plane.
The power $1/q$ ensures that the singularity in the universal cover is
smoothed out by patching together $q$ copies of $\omega$-planes. In
terms of the $z$ coordinate, each $\omega$-plane is a wedge subtending
an angle of $2\pi/q$. The $q$ wedges are sewn together along edges
which correspond to the interval between $v_1$ and $v_2$ on either side
of the branch-cut, as illustrated in figure \ref{fig:beachball}. We
also observe that $\w=\infty$ is mapped to the $q$-th roots of unity in
the $z$-plane. The above transformation \reef{ctrans} implies that
 \be
 dz d\bar{z} =
\frac{(v_2-v_1)^2}{q^2}\frac{d\omega
d\bar{\omega}}{\vert(\omega-v_1)^{(1-1/q)}\,(\omega-v_2)^{(1+1/q)}\vert^2}\,.
 \labell{2dmetric}
 \ee
This line element explicitly illustrates that the curvature
singularities in the $\omega$-plane can be completely absorbed by a
conformal factor. Finally, let us note that with $q=1$ the
transformation \reef{ctrans} is in fact the Euclidean version of the
map to the Rindler wedge for $d=2$ discussed in \cite{casini9}.
\FIGURE[!ht]{
\includegraphics[width=0.6\textwidth]{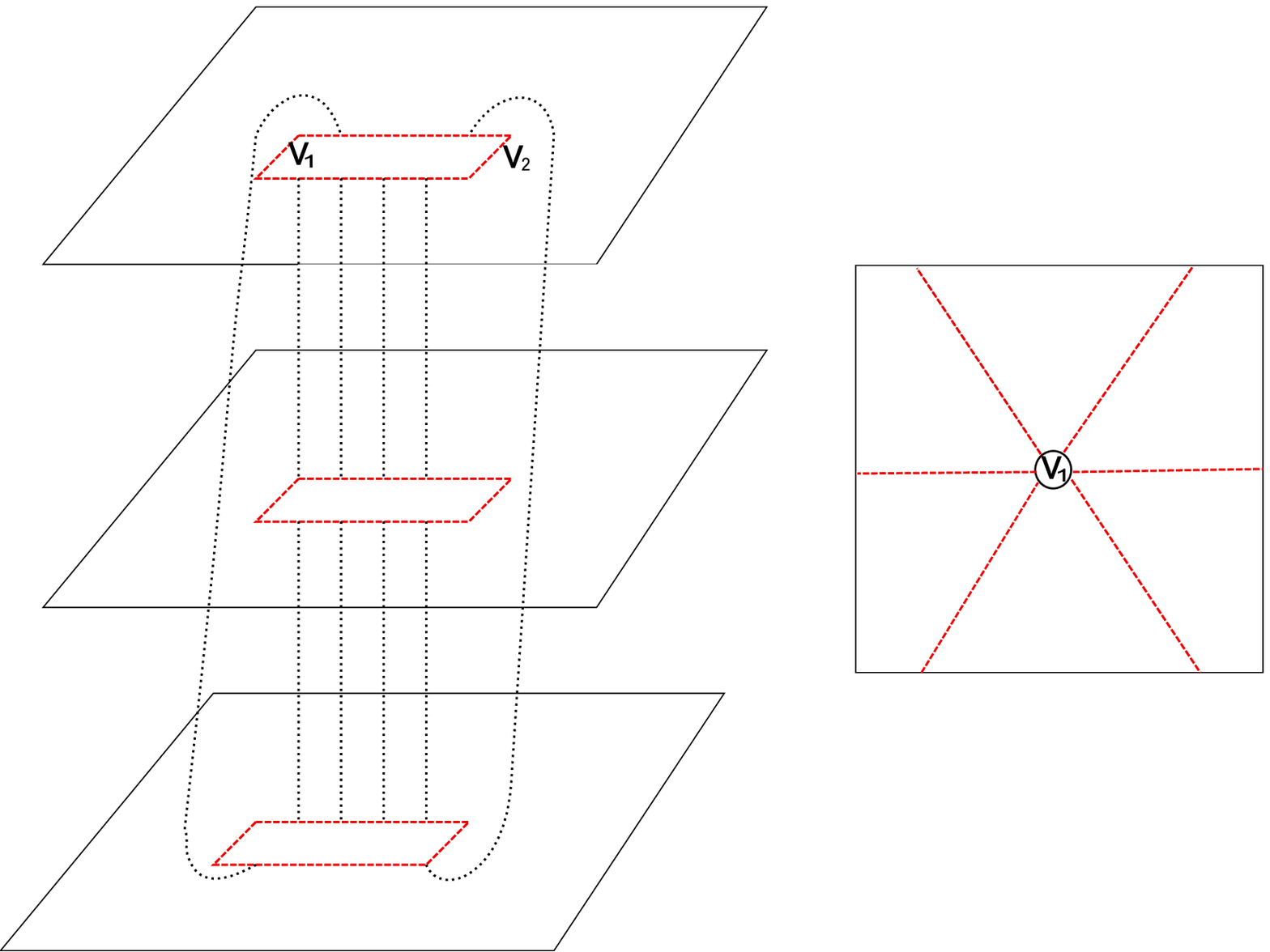}
\caption{The picture on the left depicts the universal cover of the
$\omega$-orbifold. The dashed lines indicate the identifications
between planes across the branch cut. The right picture depicts the
image of the universal cover in the $z$-plane under the conformal
mapping given in eq.~\reef{ctrans}. } \labell{fig:beachball}}

Since the $z$-plane is flat and preserves translation invariance,
$\langle T_{zz}(z)\rangle=0$, where $T_{zz}(z)$ is the holomorphic part
of the stress tensor of the two-dimensional CFT. Hence, transforming
back to the $\omega$-plane, we have\footnote{Note that the present
conventions are such that $T_{\mu\nu}=\frac2{\sqrt{g}}\frac{\delta
I_E}{\delta g^{\mu\nu}}$. We refer the interested reader to
\cite{cthem} for a full discussion of our conventions but let us note
here that the present conventions disagree with those in
\cite{cardy0,cardyCFT}. Comparing to these original references,
$\left(T_{\mu\nu}\right)_\mtt{original}=-2\pi
\,\left(T_{\mu\nu}\right)_\mtt{present}$. However, in defining the
scaling dimensions below, we write the OPE of (the holomorphic part of)
the stress tensor with a primary operator as
 \be
T_{\w\w}(\w)\,\cO(z,\bar{z})\simeq -\frac{\Delta}{2\pi}\,
\frac{\cO(z,\bar{z})}{(\w-z)^2} +\cdots
 \labell{footy}
 \ee
and hence our result in eq.~\reef{dim} is identical to that presented
in \cite{cardyCFT}. Of course, the total scaling dimension there is
given by $h=\Delta+\bar{\Delta}$. \labell{musical}}
 \be
\langle T_{\omega\omega}(\omega)\rangle = -\frac{c}{24\pi}\{z,\omega\}
= -\frac{c}{48\pi}\(1-\frac{1}{q^2}\)\,\frac{(v_2-v_1)^2}{(\omega
-v_2)^2(\omega -v_1)^2}\,,
 \labell{Tvev}
 \ee
where $c$ is the central charge, and $\{z,\omega\}$ is the Schwarzian
derivative. That is,
 \be
\{z,\omega\} =
\frac{z'''(\omega)}{z'(\omega)}-\frac{3}{2}\left(\frac{z''(\omega)}{z'(
\omega)}\right)^2\,.
 \labell{green}
 \ee

As explained in \cite{cardyCFT}, this expectation value \reef{Tvev} is
actually for the stress tensor in the presence of the twist operators,
\ie
 \be
\langle T_{\omega\omega}(\omega)\rangle =\frac{\langle\,
T_{\omega\omega}(\omega)\,\sigma_q(v_1)\,
\sigma_{-q}(v_2)\,\rangle}{\langle\, \sigma_q(v_1)\,
\sigma_{-q}(v_2)\,\rangle}\,.
 \labell{Tvev2}
 \ee
This is however, the result from the insertion of $T_{\omega\omega}$
for a single copy of the field theory, \ie on one sheet of the
universal cover \cite{cardyCFT}. If we insert $T_{\omega\omega}$ on all
$q$ sheets, the right-hand side of eq.~\reef{Tvev} acquires an extra
factor of $q$ to become
 \be
\langle T_{\omega\omega}(\omega)\rangle =-
\frac{q\,c}{48\pi}\(1-\frac{1}{q^2}\)\,\frac{(v_2-v_1)^2}{(\omega
-v_2)^2(\omega -v_1)^2}\,.
 \labell{Tvevall}
 \ee
Keeping in mind the correlator in eq.~\reef{Tvev2}, we can easily read
off the conformal dimension of the twist operators by considering the
limits $\w\to v_2,\,v_1$ in eq.~\reef{Tvevall} and its complex
conjugate. One finds that both $\sigma_{\pm q}$ are (spinless) primary
operators, where holomorphic and anti-holomorphic scaling dimensions,
$\Delta_q$ and $\bar\Delta_q$, are both given by the pre-factor in the
expression above. Adding these two results, the total scaling dimension
of both twist operators is given by
 \be h_q=2\Delta_q =
\frac{q\, c}{12} \(1-\frac{1}{q^2}\) = \frac{c}{12} \(q -
\frac{1}{q}\)\,.
 \labell{dim}
 \ee
Hence we thus have
 \be
\textrm{tr}\left[\rhov^{\,q}\right] = \langle\, \sigma_q(v_1)\,
\sigma_{-q}(v_2)\,\rangle = s_q
\left\vert\frac{v_2-v_1}{\delta}\right\vert^{-2h_q}\,,
 \labell{twocorrelate}
 \ee
where $\delta$ is the short-distance UV cut-off and $s_q$ is some
constant. Since $\textrm{tr}[\rhov]=1$, we know $s_1=1$. Otherwise, it
is not determined purely from conformal symmetry and depends on the
details of the CFT. However, $s_q$ will not effect the universal
contribution to the \ren entropy. Substituting the above expression
into eq.~\reef{defsn}, we reproduce the well known result \reef{twod}
for the \ren entropy of a single interval for any two-dimensional CFT.

\subsection{Holographic calculation for $d=2$}\labell{2dholo}

Let us now turn to the holographic description of the above calculation
with twist operators in $d=2$. In the previous section, we saw that the
action of the twist operators was captured by the conformal
transformation \reef{ctrans} and hence the expectation value of the
stress tensor was simply given by the Schwarzian in eq.~\reef{Tvev}. As
a result, the conformal dimension of the twist operators in a
two-dimensional CFT is a universal quantity depending only on the
central charge of the theory. We expect that we should be able to
recover precisely this dimension \reef{dim} and the corresponding
two-point correlator \reef{twocorrelate} with our holographic
calculations.

Related holographic calculations of \ren entropies based on the replica
trick first appeared in \cite{furry}. However, the approach presented
there yields incorrect results because of singularities in the bulk
geometry \cite{head}. An alternative approach avoiding such
singularities was explored in \cite{bulkcal}, although there is no
natural way there to obtain the finite contribution to the partition
function. The discussion there is closely related to our analysis
below. In the following, we will demonstrate a bulk calculation that
closely parallels that presented in the previous section for the
boundary CFT. Unlike the derivation reviewed above, we will directly
evaluate the path integral in the presence of the twist operators from
the classical gravity action. The extra factor of $q$ acquired from the
$q$-sheets in the final answer would emerge naturally from the
geometry. In many respects, our bulk calculations closely resemble the
CFT calculations found in \cite{oleg}, which considered correlation
functions of twist operators in a symmetric orbifold CFT, which are
relevant for the microscopic description of certain black holes in
string theory.

The standard holographic prescription dictates that we should work with
an asymptotically AdS bulk geometry that extremizes the gravity action,
and that the AdS boundary is chosen to coincide with the geometry in
which the CFT correlator of the twist operators is evaluated. From the
discussion above, the boundary should be the universal cover of the
$\omega$-plane. One could alternatively work with the $z$ coordinate,
in which case these curvature singularities are taken into account by
the conformal factor in eq.~\reef{2dmetric}. However, in the following,
we find it more convenient to work with the $\omega$ coordinate.
Nevertheless one should bear in mind that $\w$-plane only corresponds
to a single patch of the universal cover.

At this point, it is most systematic to work in the Fefferman-Graham
(FG) gauge \cite{feffer} --- see also \cite{construct} --- where the
asymptotically AdS$_3$ metric reads
 \be
ds^2 =\frac{L^2d\rho^2}{4\rho^2} + \frac{g_{a
b}(\omega,\bar{\omega},\rho)\, dx^a dx^b}{\rho}\,,
 \labell{metric2}
 \ee
with
 \be
g_{a b}(\omega,\bar{\omega},\rho) = g_{(0)\,\,a b} + \rho\, g_{(2)\,\,a
b}+ \rho^2\, g_{(4)\,\,a b}+\cdots\,.
 \labell{expand}
 \ee
The relevant boundary metric $g_{(0)\,\,a b}$ in the present case is
simply
 \be
ds_{(0)}^2 =g_{(0)\,\,a b}\,dx^a dx^b = d\omega \,d\bar{\omega}\,.
 \labell{tentmetric}
 \ee
Despite appearances, we again remind ourselves that this metric
\reef{tentmetric} is not everywhere flat. Rather we are thinking of
this as a single patch of the universal cover which has curvature
singularities at $\w=v_2$ and $\w=v_1$. Again, the above metric is
related to the flat metric in the $z$-plane by the conformal factor
shown in eq.~\reef{2dmetric}. In any event, since the boundary is
conformally flat, the bulk solution is still simply pure AdS$_3$ space.

A procedure to generate pure AdS$_3$ solutions for given conformally
flat boundary metric defined in the FG gauge was introduced in
\cite{skenderis}. We will make use of these techniques in what follows
to illustrate some features of the holographic calculation. However, as
it turns out, the computation of \ren entropy does not require the full
machinery of \cite{skenderis}. Hence we relegate most of the details to
appendix \ref{review}, quoting only the salient results here. In
particular, following this analysis for a pure AdS$_3$ with an
arbitrary boundary metric $g_{(0)}$ in the FG gauge, the on-shell
gravity action \reef{action} reduces to
  \be
I_\mtt{tot}=-\frac{c}{12\pi}\int d\omega d\bar{\omega}
\sqrt{g_{(0)}}\left[L^2 \frac{R_{(0)}}{4}\ln \rho_\textrm{crit}^- +
\frac{1}{2}\sqrt{|\T_{\omega\omega}|^2} + \frac{3}{8}L^2
R_{(0)}\right]\,.
 \labell{baction1}
 \ee
where $R_{(0)}$ is the Ricci scalar for $g_{(0)}$ and
$\rho^-_{\textrm{crit}}$ is the upper limit of the radial integral. As
explained in the appendix, the latter is determined by where the
measure of the bulk metric first reaches zero. Further $\T$ is a tensor
appearing in evaluating $g_{(2)}$ and is proportional to the stress
tensor of the boundary CFT, as shown in eq.~\reef{cftstresst}. Recall
that $c=12\pi L/\lp$ is the central charge of (a single copy of) the
boundary CFT. Explicit expressions for $\rho^-_{\textrm{crit}}$ and
$\T_{\omega\omega}$ are given in eqs.~\reef{rhocrit} and \reef{Tww},
respectively.

Now, the boundary curvature (\ie $R_{(0)}$ for the universal cover) and
$\T_{\omega\omega}$ are both singular at $\w=v_2$ and $v_1$. Hence to
evaluate the above action \reef{baction1}, we cut-off the integration
in the $\w$-plane with
 \be
|\w-v_2|> \delta\quad{\rm and} \quad |\w-v_1|> \delta\,,
 \labell{regulation}
 \ee
where $\delta$ is again the short-distance cut-off in the boundary
theory. That is, we are regulating this integral with the same cut-off
that appears, for example, in evaluating $\Vs$ in section \ref{renyi}
--- see eqs.~\reef{vtotal} and \reef{unis}. This regulator simplifies the
integrand since $R_{(0)}$ vanishes away from the two singular points.
Hence upon substituting eq.~\reef{Tww} for $\T_{\omega\omega}$, the
gravity action \reef{baction1} becomes
 \bea
I_\mtt{tot}&&= - \frac{c}{24\pi }\int d\omega d\bar{\omega}\,\frac12
\(1-\frac{1}{q^2}\)\,
\frac{(v_2-v_1)^2}{|\omega-v_2|^2\,|\omega-v_1|^2}\nonumber \\
&&= -\frac{c}{48\pi}\(1-\frac{1}{q^2}\) \int d\omega d\bar{\omega}
\left\vert\,\partial_\omega\!
\(\ln\left\vert\frac{\omega-v_1}{\omega-v_2}\right\vert^2\)
\right\vert^2 \nonumber \\
&&= -\frac{c}{96\pi}\(1-\frac{1}{q^2}\)\int d\omega
d\bar{\omega}\,\sqrt{g_{(0)}}\,g_{(0)}^{ab}\,\partial_a \chi\,
\partial_b\chi \,,
 \eea
where we have introduced a convenient auxiliary field
  \be
\chi = \ln\left \vert\frac{\omega-v_1}{\omega-v_2}\right\vert^2\,.
 \ee
Now we integrate by parts in the above integral to find
 \be
I_\mtt{tot}= -\frac{c}{96\pi}\(1-\frac{1}{q^2}\)\int d\omega
d\bar{\omega}\,\Big[ \sqrt{g_{(0)}}\,\nabla_a\! \(\chi\,
g_{(0)}^{ab}\,\nabla_b \chi\)- 2\,\chi\,\partial_\omega
\partial_{\bar{\omega}}\chi\Big]\,.
 \ee
Again since the integrand is evaluated away from the singular points as
in eq.~\reef{regulation}, the second term above vanishes. Now the
integral reduces to two boundary terms integrated around the edges of
the two cut-off disks \reef{regulation} at $\omega = v_2$ and $v_1$ in
the universal cover. Adopting polar coordinates in the vicinity of the
singular points, \eg $\w-v_2\simeq r e^{i\theta}$, these boundary terms
yield
 \bea
  I_\mtt{tot}&=&
-\frac{c}{96\pi}\(1-\frac{1}{q^2}\)\ \(\oint_{\w=v_2}+ \oint_{\w=v_1}\)
d\theta \, \Big[\delta\times \chi
\partial_r \chi\Big]
 \nonumber\\
  &=& \frac{q\,c
}{6}\(1-\frac{1}{q^2}\)\,\ln\(\frac{v_2-v_1}{\delta}\)\,.
 \labell{result1}
 \eea
Note that the overall factor of $q$ appears in the final result above
because the total range of the angular coordinate around the singular
points on the universal cover is $\Delta\theta=2\pi q$. We should also
comment that implicitly we have used the fact that the integrand dies
off sufficiently rapidly as $|\w|\to\infty$ so that the potential
boundary contribution coming from large $\omega$ actually vanishes.

Now according to the AdS/CFT dictionary, the above on-shell action can
be interpreted as giving the leading saddle-point approximation of the
partition function of the boundary theory evaluated on the universal
cover of the orbifold, \ie $ Z_q =\exp(- I_\mtt{tot})$. Now combining
eqs.~\reef{defsn} and \reef{trrho1}, we find
 \be
S_q=\frac{1}{q-1}\,\(q \log Z_1 - \log Z_q\)
 \labell{useful}
 \ee
Since $Z_1$ is the partition function of the original (un-replicated)
CFT in flat space, the corresponding holographic calculation would
compute the on-shell gravity action \reef{action} evaluated on empty
AdS$_3$ space in the usual \poin \, coordinates. However, using the
prescription outlined here and in appendix \ref{review}, this gravity
action evaluates to zero and so $Z_1=1$. Therefore our leading
semi-classical result for the \ren entropy follows simply from
eq.~\reef{result1}:
 \be
S_q=\frac{1}{q-1}\,I_\mtt{tot}=\frac{c
}{6}\(1+\frac{1}{q}\)\,\ln\(\frac{v_2-v_1}{\delta}\)
 \labell{result1x}
 \ee
which again matches precisely with the expected result \reef{twod} for
the \ren entropy of a two-dimensional CFT.

\subsection{UV regulator geometry} \labell{submarine}

In appendix \ref{review}, we give a detailed discussion of the UV
regulator surface that is implicit in the previous holographic
calculations. Of course, one component of this surface is given as
usual by the radial cut-off $\rho=\rho_\mtt{min}=\delta^2/L^2$.
However, there are two additional caps which keep the bulk integration
away from the singularities in the boundary metric at $\w=v_2$ and
$\w=v_1$. Hence we can think of the full regulator surface as being
composed of these three components. This choice of an `unusual'
regulator surface is in fact the key to our gravitational calculation
which gives a holographic translation of the determination of the \ren
entropy using twist operators discussed in section \ref{twistor}. That
is, with $d=2$, the bulk geometry is simply empty AdS$_3$ space. Hence
the main effect of choosing coordinates with FG gauge \reef{metric2} is
to alter the UV cut-off surface near the AdS$_3$ boundary relative to
the `constant radius' cut-off surface in Poincar\'e coordinates. Hence
the interesting physical results in the previous section are coming
from this `unusual' regulator surface whose choice is motivated by the
problem of calculating the \ren entropy. In particular, this choice is
made so that (to leading order in $\delta/L$) the induced metric of the
cut-off surface coincides with the boundary CFT metric in the conformal
frame of interest.

This observation implies that we could have obtained the same result
\reef{result1x} in section \ref{2dholo} through the following route: We
describe the AdS$_3$ bulk with conventional Poincar\'e coordinates,
 \be
ds^2 =\frac{L^2}{\xi^2}\left( d\xi^2 + dz \,d\bar{z}\)\,,
 \labell{lucky}
 \ee
but choose a nontrivial cut-off surface
 \be
\xi = \frac{\delta}{L}\, e^{-\phi(z,\bar{z})}\,, \qquad {\rm where}\ \
\ e^{\phi}\equiv q\,(v_2-v_1) \frac{|z|^{q-1}}{|z^q-1|^2}\,.
 \labell{cutsurf}
 \ee
Here we are following the analysis of \cite{Krasnov} --- see also
appendix \ref{review}. With this choice, the induced metric on this
cut-off surface has the same conformal factor appearing in
eq.~\reef{2dmetric}. Note that the factor $e^{-\phi}$ diverges as $z\to
0,\,\infty$. Hence we would introduce additional components to
regulator surface which impose $|z|>\delta/L$ and $|z|<L/\delta$. These
additional surfaces would correspond to the two caps which keep the
bulk integration away from the twist operators in the boundary metric.

In fact, using the above approach, the on-shell gravity action
evaluates to
 \be
I_\mtt{tot}=-\frac{c}{48\pi}\int dz d\bar{z} \(
\partial\phi\,\bar{\partial}\phi -4\partial\bar{\partial}\phi\)
 \labell{baction2}
 \ee
In the limit $\delta\to 0$, the above result is robust against changes
in the choice of cut-off surface \reef{cutsurf} which are sub-leading
in $\delta/L$. This result is in keeping with the idea that a Weyl
rescaling will shift the action of any $d=2$ CFT action by an
expression proportional to the Liouville action, where the Liouville
field takes the value of the Weyl factor \cite{friedan}. In the above
expression, the first term is precisely the kinetic term of the
Liouville field $\phi$. This result was also noted in \cite{Krasnov},
despite some small discrepancy in the choice of surface terms. At first
sight, eq.~\reef{baction2} does not appear identical to the computation
in FG coordinates, unless the second order derivatives of $\phi$
vanishes. However, as is noted in \cite{oleg}, the key non-vanishing
contribution that is responsible for the final answer is in fact the
kinetic term, whereas the rest drops out on careful regularization. For
our purpose, $R_{(0)}=2\partial\bar\partial\phi=0$ within the
integration region, and indeed inserting the explicit form of $\phi$
into \reef{baction2}, we find agreement with \reef{result1}, as is
required by conformal symmetry.

In passing, let us comment on the black hole geometries appearing in
section \ref{renyi}. The analog of the action calculation applied here
would be to first evaluate the Euclidean gravitational action
$I_\mtt{E}$ for the topological black holes \reef{lineelement}. Then
interpreting this result in terms of the free energy of the dual
thermal ensemble, \ie $F(T)=T\,I_\mtt{E}(T)$, the \ren entropy would be
calculated using eq.~\reef{form1}. Implicitly the regulator surface in
this action calculation has distinct components. Of course, there would
be the standard radial cut-off surface at $r=L^2/\delta$. However, as
discussed after eq.~\reef{hash}, we must also impose an `IR cut-off' in
the radial integration across the hyperbolic geometry foliating the AdS
geometry. That is, we will have a new cut-off surface near where these
surfaces approach the asymptotic AdS boundary. As we will see from the
discussion in the following section, these new surfaces are analogous
to the caps above which restrict the bulk integration from approaching
the position of the twist operators in the boundary metric.

\section{Twist operators and Thermal ensembles} \labell{versus}

In the previous sections, we presented two holographic calculations of
\ren entropy, each of which was motivated by a distinct approach to
calculating in the boundary CFT. We would now like to compare the
latter two approaches and in fact, with closer examination, we find
that they are closely related. To begin let us consider
eq.~\reef{trrho1} where one evaluates
$\textrm{tr}\left[\rhov^{\,q}\right] = Z_q/Z_1^q$. Here $Z_q$ denotes
the path integral of the $q$-fold replicated theory on the orbifold
generated by the insertion of twist operators. The factor of $Z_1$ is
simply the vacuum partition function of the original theory, which is
used to normalize the result. Now we consider eq.~\reef{trrho0x} where
one instead evaluates $\textrm{tr}\!\left[\,\rhov^{\,q}\right] =
Z(T_0/q)/Z(T_0)^q$. Here the factor $Z(T_0)$ corresponds again to the
vacuum partition function, however, the conformal transformation
introduced in \cite{casini9} allows it to be interpreted as a thermal
partition function. Hence it is natural to also associate the
numerators in both of these expressions. That is, the thermal partition
function $Z(T_0/q)$ should be equivalent to $Z_q$ in the previous
calculation. Further just as the path integrals which define $Z_1$ and
$Z(T_0)$ are related by a conformal transformation, it is natural to
expect that the same should hold for $Z_q$ and $Z(T_0/q)$.

In fact, it is straightforward to establish these relations. Since the
calculation with twist operators in section \ref{twistor} explicitly
referred to $d=2$, we start our discussion there. The thermal partition
functions with be defined by a path integral over the CFT on a
Euclidean background
 \be
 ds^2=dt_\mt{E}^2+du^2\,,
 \labell{twodxx}
 \ee
where the Euclidean time coordinate is periodic with period $\Delta
t_\mt{E} = q/T_0 = 2\pi R\, q$ and $u\in\lbrace -\infty,\infty
\rbrace$. Now for convenience, let us introduce the complex coordinate
$\sigma=(u+ i t_\mt{E})/R$, which is then periodic along the imaginary
axis with $\sigma = \sigma + 2\pi i\,q$. We can map the
$\sigma$-cylinder to the flat $z$-plane with the conformal
transformation: $z=\exp[\sigma/q]$. Note that we have included the
factor of $1/q$ in the exponent to ensure that this mapping is
one-to-one between the $\sigma$-cylinder and the $z$-plane. However,
let us now combine this transformation with eq.~\reef{ctrans} which
relates $z$ and $\w$ in the twist operator calculation. The combined
transformation becomes
 \be
e^\sigma=\frac{\omega-v_1}{\omega-v_2}
 \labell{ctranszz}
 \ee
Naively, this mapping is independent of $q$, however, we must remember
that there is an implicit $q$ dependence in the periodicity of
$\sigma$. Therefore, this transformation \reef{ctranszz} provides a
conformal mapping from the $({u},t_\mt{E})$-cylinder, which defines the
thermal partition function at temperature $T=T_0/q$, to the $q$-fold
universal cover of $\w$-plane, which defines the partition function
with twist operators at $\w=v_2$ and $\w=v_1$. Of course, with the
choice $q=1$, we simply have a one-to-one mapping from
$({u},t_\mt{E})$-cylinder to the $\w$-plane. Hence in general, we may
equate $Z_q=Z(T_0/q)$ and $Z_1=Z(T_0)$ because the underlying CFT path
integrals are simply related by conformal transformations.

Having made the desired identifications in $d=2$, let us now turn our
attention to higher dimensions. As noted above, it seems
straightforward to extend the replica trick to higher dimensional
theories. However, this yields a rather formal definition of twist
operators for higher dimensions and in practice beyond $d=2$, the
construction (and properties) of these operators is not well understood
--- however, see \cite{swingle,casini6}. In $d$ dimensions, a twist
operator $\sigma_q$ would be some ($d-2$)-dimensional \emph{surface}
operator which again introduces a branch cut in the path integral over
a $q$-fold replicated theory. Note that for $d>2$, we would not have
distinct operators $\sigma_{\pm q}$ to open and close the branch cuts.
For example, in the problem of interest, the entangling surface is an
$S^{d-2}$ and so there would be a single operator $\sigma_q$ which
resides on this surface to open a branch cut over the ball on the
interior.

Having introduced the notion of a twist operator for $d>2$, let us
proceed with the consideration of $Z(T_0/q)$. In fact, the conformal
mapping analogous to eq.~\reef{ctranszz} is discussed in
\cite{casini9}. However, the discussion is presented for Minkowski
signature backgrounds and so we adapt their analysis to Euclidean
signature. With $Z(T_0/q)$ in $d$ dimensions, we are considering a
thermal ensemble at temperature $T=T_0/q$ on the background $R\times
H^{d-1}$. The partition function can be evaluated by path integral on
the Euclidean background $S^1\times H^{d-1}$
 \be
ds^2=d\tau_\mt{E}^2+ R^2\,\left(du^2+\sinh^2\!
u\,d\Omega^2_{d-2}\right)\,,
 \labell{hyper2}
 \ee
where the Euclidean time coordinate has period $\Delta \tau_\mt{E} =
q/T_0 = 2\pi R\, q$. In the following, it will be convenient to
introduce complex coordinates:
 \be
\sigma=u+i \tau_\mt{E}/R \qquad {\rm and}\qquad \w = r+i t_\mt{E}\,.
 \labell{ccord}
 \ee
The latter will be used to describe a conformally mapped geometry
below. Note that both $u$ and $r$ are radial coordinates, we must have
Re$(\sigma)=u>0$ and Re$(\w)=r>0$. Now, with the first of these new
coordinates, the above metric \reef{hyper2} can be written
 \be
ds^2=R^2\( d\sigma\,d\bar{\sigma}+
\sinh^2\!\(\frac{\sigma+\bar{\sigma}}{2}\)\,d\Omega^2_{d-2}\)\,.
 \labell{hyper2a}
 \ee
Now we make the coordinate transformation
 \be
 e^{-\sigma}=\frac{R-\w}{R+\w}\,.
 \labell{magic}
 \ee
Since we are considering $d\ge3$ there is no guarantee that this
holomorphic change of coordinates will result in a conformal
transformation. However, one can readily verify the above metric
\reef{hyper2a} becomes
 \bea
ds^2&=&\Omega^2\, \left[ d\w\,d\bar{\w}+ \(\frac{\w+\bar{\w}}{2}\)^2
d\Omega^2_{d-2}\right]
 \nonumber\\
&=&\Omega^2\, \left[\, dt_\mt{E}^2 + dr^2+
r^2\,d\Omega^2_{d-2}\,\right]
 \labell{flat}
 \eea
where
 \be
 \Omega=\frac{2R^2}{|R^2-\w^2|}\,.
 \labell{confact}
 \ee
Hence, after eliminating the conformal factor $\Omega^2$ in the second
line of eq.~\reef{flat}, we recognize that the final line element is
simply the metric on $d$-dimensional flat space $R^d$. However, we must
again pay special attention to the identification of the $\sigma$. In
particular, since $\Delta \tau_\mt{E} = q/T_0 = 2\pi R\, q$, we must
identify $\sigma = \sigma + 2\pi i\,q$. Therefore the transformation
\reef{magic} actually maps the original background $S^1\times H^{d-1}$
to a $q$-fold cover of $R^d$ with an orbifold singularity on precisely
the ($d-2$)-dimensional sphere given by $r=R$. Hence evaluating the
path integral on the new geometry would yield precisely the partition
function $Z_q$ of a $q$-fold replicated theory with an $S^{d-2}$ twist
operator inserted at $r=R$. Hence as long as we are considering a
conformal field theory, we may equate $Z_q=Z(T_0/q)$ because the path
integrals are simply related by a conformal transformation. Of course,
with the choice $q=1$, we again have a simple one-to-one mapping from
$S^1\times H^{d-1}$ to $R^d$.

Turning now to our holographic calculations, we evaluate the partition
functions above in the semi-classical approximation by evaluating the
on-shell gravitational action for a particular saddle-point. For both
$Z_1$ and $Z(T_0)$, the corresponding saddle-point is simply the AdS
vacuum evaluated with a conventional UV regulator. The latter
description applies equally for $d\ge3$ or for $d=2$. In the case of
$d=2$, the saddle-point corresponding to $Z_q$ and $Z(T_0/q)$ is again
AdS$_3$ but an unconventional choice of regulator surface is chosen, as
discussed in section \ref{submarine}. Now in higher dimensions (\ie
with $d\ge3$), the saddle-point dual to $Z(T_0/q)$ is a topological
black hole \reef{lineelement} with an appropriately chosen horizon
temperature. This highlights a basic difference for higher dimensions.
In $d=2$ any boundary metric is (locally) conformally flat and so the
corresponding bulk is always just AdS$_3$. In higher dimensions, we can
rarely transform to a flat metric and so generically our bulk will not
just be the AdS$_{d+1}$ vacuum. Rather in higher dimensions, we have to
find new smooth solutions of the gravitational equations with
appropriate boundary conditions.  In particular, to evaluate $Z_q$, we
would want a new bulk solution where the conical singularity introduced
by the twist operator arises only at the boundary, \ie it would only
appear in the induced metric on UV regulator surface, as discussed in
section \ref{submarine}. Now in principle, the black hole geometries
provide precisely the desired bulk space and one is only required to
make a coordinate transformation in the bulk which would implement the
conformal transformation \reef{magic} on the boundary. Unfortunately,
this still seems to be quite a challenging exercise. However, it does
appear to be less formidable than constructing the bulk geometry
without the insight of equating $Z_q=Z(T_0/q)$.

\subsection{Twist operators and Black holes} \labell{twisted}

In this section, we wish to apply the insights above to our holographic
models in section \ref{renyi}. In particular, we will use our
holographic calculations to evaluate the scaling dimension of the twist
operators for the boundary theory in general dimensions.

As a warm-up exercise, we will apply holography to reproduce the
expression \reef{dim} for the total scaling dimension of twist
operators in a two-dimensional CFT. In the case of $d=2$,
eqs.~\reef{lineelement} and \reef{sol0} describe a black hole solution
for three-dimensional Einstein gravity
 \be
ds^2 =- \frac{r^2 - \rh^2}{R^2}\, dt^2+  \frac{L^2\,dr^2}{r^2-\rh^2} +
\frac{r^2}{R^2}\, d\hat{u}^2\,,
 \labell{btz}
 \ee
where we have modified slightly the normalization of the spatial
coordinates.  With $d=2$, rather than a hyperbolic geometry, the
spatial CFT geometry is simply a line and so we have chosen a
normalization such that the boundary metric \reef{cftmet} becomes
simply $ds^2_\infty=-dt^2+d\hat{u}^2$. Note that with $d=2$, the
formula for the temperature \reef{xstuff0} simplifies to $T=T_0\,x$
where $T_0=1/(2\pi R)$ and $x=\rh/L$ as was used throughout section
\ref{renyi}. As is already evident from these expressions, the solution
of eq.~\reef{EHxq} is simply $\xq=1/q$. Now we should note that this
solution \reef{btz} is nothing other than the spinless BTZ black hole
\cite{btz} but without any orbifolding along the spatial direction to
preserve the noncompact boundary. Without the latter orbifolding, it is
also true that this solution is simply AdS$_3$ space written in
unconventional coordinates. Before proceeding further, it is convenient
to go to Euclidean signature with which the three-dimensional black
hole metric becomes
 \be
ds^2 = L^2 \frac{dr^2}{r^2-\rh^2} + \frac{r^2 - \rh^2}{R^2}\,
dt_\mt{E}^2 + \frac{r^2}{R^2}\, d\hat{u}^2\,.
 \labell{btzE}
 \ee
Now adapting well-known results for the BTZ black hole \cite{carlip},
we use the following bulk coordinate transformation to take this
Euclidean black hole metric back into the Poincar\'e metric
\reef{lucky}:
 \be\labell{btzpoin}
z= \frac{\sqrt{r^2 - \rh^2}}{r}\exp\left[x\(\frac{\hat{u}+ i
t_\mt{E}}{R}\)\right]\,, \qquad \xi
=\frac{\rh}{r}\,\exp\left[x\,\frac{\hat{u}}{R}\right]\,,
 \ee
where as before $x=\rh/L$. The following is simplified by using a the
complex coordinate $\sigma=(u+ i t_\mt{E})/R$, which was introduced in
the previous section. Note that if we substitute $x=1/q$ in
eq.~\reef{btzpoin}, then in the boundary limit, \ie with $r\to \infty$,
$z\simeq \exp[\sigma/q]$ and so this bulk coordinate transformation
\reef{btzpoin} implements the first conformal mapping considered in the
previous section.

For our present purposes, the essential information provided by the
black hole solution is the thermal energy  density, which can be read
off with the conventional holographic map
 \be
T_{\sigma\sigma}(T) = \frac{\pi}{12}\,c\,T^2 \,.
 \ee
Now the conformal map between the $\sigma$ and $\w$ coordinates is
given above by eq.~\reef{ctranszz}. Hence the stress tensor is mapped
between these two conformal frames as
 \bea
T_{\omega\omega} &=& \sigma'(\omega)^2\ T_{\sigma\sigma}(T_0/q) -
\frac{c}{24\pi}\{\sigma,\omega\}
 \nonumber\\
&=&-\frac{c}{48\pi}\(1-\frac{1}{q^2}\)\,\frac{(v_2-v_1)^2}{(\omega
-v_2)^2(\omega -v_1)^2}\,.
 \labell{stressz}
 \eea
Hence we have reproduced the expected result for the correlator of the
stress tensor with a pair of twist operators, as given in
eq.~\reef{Tvev}. Hence following the same logic presented in section
\ref{twistor}, we can infer that the scaling dimension of the twist
operators is given by eq.~\reef{dim}.

Before concluding, let us comment on the anomalous contribution given
the Schwarzian in eq.~\reef{stressz}. Recall that in general the
Schwarzian is given by eq.~\reef{green} and so depends only on the form
of the coordinate transformation. If we evaluate this expression for
the present transformation \reef{ctranszz}, we note that it can be
expressed as
 \be
\{\sigma,\omega\}={24\pi}\,{c}\,\sigma'(\omega)^2\
T_{\sigma\sigma}(T_0)\,.
 \labell{okay}
 \ee
One can also infer this relation because if we insert $q=1$ into
eq.~\reef{stressz}, the result vanishes.

Now we would now like extend the previous discussion, which focussed on
$d=2$ and used the three-dimensional bulk black hole \reef{btz}, to
higher dimensions and determine $h_q$ for the general holographic
models of section \ref{renyi}. As background for this analysis in
higher dimensions, we first consider the following result which would
apply for any CFT: Imagine making an insertion of the stress tensor in
the presence of a {\it planar} twist operator $\sigma_q$. Let us
describe the $R^d$ background with Cartesian coordinates $x^\mu=\{x^i,\
i=0,1;\ x^a,\ a=2,\ldots,d-1\}$. The ($d-2$)-dimensional twist operator
will be positioned at $x^0=0=x^1$ while it extends throughout all of
the $x^a$ directions. We insert the stress tensor at
$x^\mu=\{\al^i,\beta^a\}$ and define
$\alpha=\sqrt{(\alpha^1)^2+(\alpha^2)^2}$. Then symmetry\footnote{The
translation and rotation symmetries dictate the basic geometric
structure in the following and the relative normalization of various
contributions is fixed by imposing $\langle T^\mu{}_{\mu}\, \sigma_q
\rangle=0=\nabla^\mu \langle T_{\mu\nu}\, \sigma_q \rangle$. Let us
also note here that the correlators in eq.~\reef{weight} are implicitly
normalized by dividing out by $\langle \sigma_q \rangle$ but we left
this implicit to avoid the clutter.} dictates that the singularity in
the corresponding correlator takes the following form
 \bea
\langle T_{a b}\, \sigma_q \rangle&=& -\frac{h_q}{2\pi}
\,\frac{\delta_{a b}}{\al^d} \,,\qquad \langle T_{a i}\, \sigma_q
\rangle=0\,,\nonumber \\
\langle T_{ij}\, \sigma_q \rangle&=& \frac{h_q}{2\pi}
\,\frac{(d-1)\delta_{i j}-d\,n_i\,n_j}{\al^d}\,.
 \labell{weight}
 \eea
where $n^i=\al^i/\al$ is the unit vector point directed orthogonally
from the twist operator to the $T_{\mu\nu}$ insertion. These
correlators \reef{weight} define the coefficient $h_q$ which is
commonly called the scaling dimension of $\sigma_q$, since its
appearance here is analogous to that of the scaling dimension of a
local primary operator. We are assuming that $T_{\mu\nu}$ corresponds
to the total stress tensor for the entire $q$-fold replicated CFT, \ie
$T_{\mu\nu}$ is inserted on all $q$ sheets of the universal cover. Then
if one reduces these expressions to $d=2$, one finds that the present
definition for $h_q$ matches precisely with those given in section
\ref{twistor}. Hence for $d=2$, $h_q$ is the total scaling dimension
given in eq.~\reef{dim}.

As noted above, the essential information that we must extract from the
topological black holes is the stress-energy of the thermal ensemble on
$R\times H^{d-1}$ or rather the Euclidean CFT lives on the background
$S^1\times H^{d-1}$. However, on general grounds, we know the
expectation value of the stress tensor is restricted to take the form
 \begin{equation}
 T^\mu{}_\nu=\textrm{diag}(-\E,\,p,\,\cdots,\,p)\,,
 \labell{none5}
 \end{equation}
with $\E$ and $p$ constants --- this form is independent of the choice
of either Minkowski or Euclidean signature. Further for a conformal
field theory the trace of this expression must vanish\footnote{One may
worry that for even $d$ that the trace anomaly will lead to a
nonvanishing trace. However, one can readily verify that in fact the
trace anomaly vanishes for the background geometry $R\times H^{d-1}$.
In particular, note that the Euler density vanishes because the
background is the direct product of two lower dimensional geometries.
Further this background is conformally flat and so any conformal
invariants also vanish.} and so in eq.~\reef{none5}, we will find
\begin{equation}
p=\E/(d-1)\,. \labell{none6}
\end{equation}
In particular then, we will use the black hole solution in higher
dimensions to determine the energy density $\E(T)$.

To connect this thermal energy density to the correlator with the twist
operator, we must conformally map the $S^1\times H^{d-1}$ background to
flat metric on $R^d$ using the transformation in eq.~\reef{magic}. As
discussed above, when the temperature is tuned to $T=T_0/q$, this
mapping actually takes the original background to a $q$-fold cover of
$R^d$ with an orbifold singularity on the sphere given by $r=R$. Now in
higher dimensions, a conformal mapping transforms the stress tensor as
 \be
T_{\al\beta} =\Omega^{d-2}\,\frac{\partial X^\mu}{\partial
x^\al}\,\frac{\partial X^\nu}{\partial x^\beta}\,\Big(T_{\mu\nu}(T_0/q)
- {\A}_{\mu\nu}\Big)\, ,
 \labell{stressd}
 \ee
where $\mu,\nu$ and $\alpha,\beta$ denote indices on the flat geometry
and $S^1\times H^{d-1}$, respectively. Since the stress tensor is not a
primary operator, in general, the transformation \reef{stressd}
includes an anomalous contribution, which we denoted $\A_{\mu\nu}$.
This contribution is the higher dimensional analog of the Schwarzian
appearing in eq.~\reef{stressz}. Again we note that this anomalous term
depends entirely on the form of the transformation. Hence it can be
fixed by observing that with $q=1$, eq.~\reef{magic} actually yields a
one-to-one mapping from $S^1\times H^{d-1}$ to $R^d$. In the absence of
any singularities, \ie without the insertion of a twist operator,
$T_{\al\beta}$ is simply the vacuum expectation value of the stress
energy in $R^d$ and hence it must vanish. Since the left-hand side of
eq.~\reef{stressd} vanishes with $q=1$, we must have ${\A}_{\mu\nu}=
T_{\mu\nu}(T_0)$. Hence the transformation \reef{stressd} becomes
 \be
T_{\al\beta} =\Omega^{d-2}\,\frac{\partial X^\mu}{\partial
x^\al}\,\frac{\partial X^\nu}{\partial x^\beta}\,\Big(T_{\mu\nu}(T_0/q)
-T_{\mu\nu}(T_0)\Big)\, ,
 \labell{stressfin}
 \ee
Recall that the required conformal factor $\Omega$ is given in
eq.~\reef{confact}.

Now the conformal mapping \reef{magic} generates a spherical twist
operator while scaling dimension $h_q$ was defined by the correlator
with a planar twist operator. However, if the insertion of the stress
tensor approaches very close to the spherical twist operator, the
leading singularity will take the form given in eq.~\reef{weight}.
Hence we can still easily identify $h_q$ in this limit. To simplify the
analysis, we will consider evaluating $T_{\al\beta}$ at $t_\mt{E}=0$
and $r = R-\alpha$ with $\alpha \ll R$ (as well as some fixed angles).
With this choice, one finds that various relevant expressions simplify.
For example, from eq.~\reef{confact}, we find $\Omega\simeq R/\alpha$
and using eq.~\reef{magic}, we can show that $\frac{\partial
u}{\partial t_\mt{E}}\simeq 0$ and $\frac{\partial
\tau_\mt{E}}{\partial t_\mt{E}}\simeq R/\alpha$. Given the last two
expressions, one finds that evaluating eq.~\reef{stressfin} for
$\al=t_\mt{E}=\beta$ yields
 \bea
 T_{t_\mt{E}t_\mt{E}} &=&\Omega^{d-2}\,\(\frac{\partial
\tau_\mt{E}}{\partial
t_\mt{E}}\)^2\,\Big(T_{\tau_\mt{E}\tau_\mt{E}}(T_0/q)
-T_{\tau_\mt{E}\tau_\mt{E}}(T_0)\Big)+\cdots
 \nonumber\\
&=& \(\frac{R}{\al}\)^d\,\Big(\E(T_0)-\E(T_0/q) \Big) +\cdots\,.
 \labell{last}
 \eea
Now this result should be compared to the $i=t_\mt{E}=j$ component in
eq.~\reef{weight}, \ie $\langle T_{t_\mt{E}t_\mt{E}}\, \sigma_q
\rangle= -h_q\,(d-1)/(2\pi\,\al^d)$. However, we note that the latter
expectation value involves the total stress tensor for the entire
$q$-fold replicated CFT. However, it is clear that eq.~\reef{last}
corresponds to an insertion of $T_{\mu\nu}$ in a single sheet of the
universal cover and hence we must multiply by an extra factor of $q$ to
compare the two expressions. The final result is
 \be
h_q=2\pi\frac{q\,R^d}{d-1}\,\Big(\E(T_0)-\E(T_0/q) \Big)
 \labell{heavy}
 \ee

We are now almost ready to evaluate these weights for the holographic
models considered in section \ref{renyi}. The simplest way to connect
the above expression to the results in this section is to use the first
law of thermodynamics, \ie $dE = T\,dS$. Then one may write
 \bea
\E(T_0)-\E(T_0/q)  &=&  \frac1{R^{d-1}\Vs}\,\int_{\xq}^{1} T(x)
\,\frac{dS}{dx}\,dx
 \labell{differ2}\\
&=& \frac1{R^{d-1}\Vs}\,\(T_0\,S(T_0) - \frac{T_0}{q}\,S(T_0/q) -
\frac{q-1}q\,T_0\,S_q\)\,.
 \nonumber
 \eea
Here we have used $\E=E/(R^{d-1}\Vs)$ to define the energy density on
the hyperbolic plane $H^{d-1}$. The second expression is derived by
integrating by parts in the first line and substituting
eq.~\reef{finalfor2} for the remaining integral. Combining
eqs.~\reef{heavy} and \reef{heavy} then yields
 \be
h_q=\frac1{(d-1)\,\Vs}\,\(\,q\,S(T_0) - S(T_0/q) - (q-1)\,S_q\)\,,
 \labell{heavy2}
 \ee
where we have simplified the pre-factor above using $T_0=1/(2\pi R)$.

Hence we may now apply our results in section \ref{renyi} to calculate
the scaling dimension $h_q$. First using the expressions of Einstein
gravity in section \ref{sec:EH}, we find
 \be
h_q =  \pi \left( \frac{L}{\lp}\right)^{d-1}\,q\, \xq^{d-2}
\,(1-\xq^2)\,,
 \labell{hqresult1}
 \ee
where $\xq$ is related to $q$ through eq.~\reef{EHxq}. Next using the
results of section \ref{sec:GB}, we have for the boundary CFT's dual to
GB gravity in arbitrary dimensions
 \be h_q = \frac{1}{4}
\Gamma\(\frac{d}{2}\)\pi^{1-\frac{d}{2}}\,q\,\xq^{d-4}(\xq^2-1)
\left((d-3) (\xq^2-1)\,\ads+ (d-3 -(d+1)\xq^2)\,\CT\right)
 \labell{hqresult2}
 \ee
where $\xq$ is now given by eq.~\reef{xq}. Interestingly, we find
tremendous simplicity in the $q\to 1$ limit:
 \be
 \partial_q h_q|_{q=1} = \frac{2}{d-1}\, \pi^{1-\frac{d}{2}}\,
 \Gamma \( {d}/{2}\)\ \CT \,.
 \labell{interest1}
 \ee
Repeating the calculation once more for quasi-topological gravity with
$d=4$, using the formulae from section \ref{sec:QT}, we find that $h_q$
has a complicated dependence both central charges, $a$ and $c$, as well
as $\tfor$:
 \be h_{q} = \frac{q \,a}{4\pi
\xq^2}(\xq^2-1)\left(\xq^4\(1-5\frac{c}{a} -10 \frac{c}{a} \,\tfor\) -
\xq^2 \(1-\frac{c}{a}-8 \frac{c}{a}\, \tfor\) + 2 \frac{c}{a}\, \tfor
\right )
 \labell{hqresult3}
  \ee
where eq.~\reef{xQT} determines the value of $\xq$. Nevertheless, we
once again notice that the derivative $\partial_q h_q|_{q=1}$ is simply
proportional to the central charge $c$ and independent of $a$ and
$\tfor$:
 \myeq{
\partial_q h_q|_{q=1} = \frac{2\, c }{3 \pi}\,,
 \labell{interest2}}
which matche precisely with eq.~\reef{interest1} for GB gravity with
$d=4$.

Hence in general, these scaling weights have a complicated nonlinear
dependence on the various parameters characterizing the boundary
theory, including ${t}_4$ in the last example. Note that although the
result in, \eg eq.~\reef{hqresult2} appears linear in both $\CT$ and
$\ads$, since $\xq$ is determined as the root of eq.~\reef{xq}, it
introduces a complicated nonlinear dependence on these parameters.
However, we find a remarkable simplification in evaluating $\partial_q
h_q$ in the limit $q \rightarrow 1$ with the result being proportional
to the central charge $\CT$.\footnote{Of course, by itself, the scaling
dimension vanishes in the limit $q\to1$, \ie $h_1=0$, since there is no
twist operator at $q=1$.} This is somewhat analogous to the
simplification found for the \ren entropy in this limit, \ie the
entanglement entropy $S_1$ has a simple form being proportional to the
central charge $a$ and independent of any other CFT parameters.


\section{Eigenvalue Spectrum} \labell{spectrum}

In this section, we begin to explore what information can be inferred
about the eigenvalue spectrum of the reduced density matrix using our
holographic results for the \ren entropy. Of course, having access to
all of the \ren entropies $S_q$ for arbitrary $q$ gives us far more
information about this spectrum than is contained in the entanglement
entropy $S_1$ alone. In fact, it was shown in \cite{calab8} that the
full spectrum of eigenvalues can be calculated by knowing ${\rm
tr}\!\left[\,\rhov^{\,q}\right]$ for only integer $q$.

First let us recall that the limit of infinite $q$ yields
 \be
 S_\infty=\lim_{q\to\infty} S_q = -\log\lambda_1\,,
 \labell{mins2}
 \ee
where $\lambda_1$ is the largest eigenvalue of $\rhov$. This result is
usually discussed in the context of a quantum mechanical system with a
discrete spectrum. In the present context, we are studying holographic
CFT's and so can expect there may be various complications. First of
all, we are evaluating the entire expression $S_q$ above, \ie this
expression does not apply for the universal contribution to $S_q$.
Hence we must think of $S_q$ for the boundary CFT in the explicit
presence of the short distance cut-off $\delta$, Hence $S_q$ is
dominated by the leading contribution in $\Vs$, which for $d>2$ is
proportional to ${\A}/\delta^{d-2}$ where $\A$ is the area of the
entangling surface. Therefore eq.~\reef{mins2} indicates that the
{largest} eigenvalue (and hence all the eigenvalues) are approaching
zero with
 \be
\lambda_1\propto \exp[-\text{const.} \times \ads\times
{\A}/\delta^{d-2}]
 \labell{suggest}
 \ee
as $\delta\to0$. Of course, our results indicate that the (positive)
constant prefactor in the exponent above will in general have a
complicated dependence on various CFT parameters. However, we have
explicitly included the factor $\ads$ to remind ourselves that even
with a fixed cut-off $\delta$, the exponent is large in the classical
gravity limit since it grows with the central charge $\ads$. For
example, in the well-known case where the boundary CFT is $d=4$
super-Yang-Mills, this factor grows like $N_c^2$. While this
holographic result may seem rather counter-intuitive, we find an
analogous result in computing $S_q$ for free fields living on a
hyperbolic waveguide geometry \cite{unpub}.

To proceed further, we will consider expanding the \ren entropy for
large $q$. As a warm-up exercise, let us assume that the eigenvalue
spectrum is discrete, in which case we have
 \bea
S_q&=& \frac{1}{1-q}\,\log{\rm tr}\!\left[\,\rhov^{\,q}\right]
 \nonumber\\
 &=& \frac{1}{1-q}\,\log\left[d_1\lambda_1^q+d_2\lambda_2^q
 +\cdots\right]\,.
 \labell{rens}
 \eea
In the second line, we have an explicit sum over the eigenvalues
$\lambda_i$ of the density matrix but we also allow the various
eigenvalues to be degenerate with multiplicity $d_i$. Now with large
but finite $q$, we can expand the last expression as
 \bea
S_q &\simeq& -\frac1q \left(1+\frac1q\right)\left[q\,\log\lambda_1
+\log d_1 +
\log\left(1+\frac{d_2}{d_1}\left(\frac{\lambda_2}{\lambda_1}\right)^q
\right)\right]
 \nonumber\\
 &\simeq& -\log\lambda_1 -
\frac{1}{q}\log\left(d_1\lambda_1\right)+\mathcal{O}
\left[\frac{1}{q^2},\frac1q\left(
\frac{\lambda_2}{\lambda_1}\right)^q\right]
 \labell{expands}
 \eea
In this case, the next-to-leading term at order $1/q$ tells us about
the degeneracy $d_1$ of the largest eigenvalue. Note also, that this
large-$q$ expansion contains contributions of the form $\frac1q\left(
\frac{\lambda_2}{\lambda_1}\right)^q$, which are nonanalytic in $1/q$.

Let us now evaluate $\lambda_1$ and $d_1$ from large-$q$ expansion of
the holographic \ren entropy in the simplest case of Einstein gravity
discussed in section \ref{sec:EH}. From eqs.~\reef{EHxq} and
\reef{sqeh}, we have
 \bea
 \xq&=&x_\infty+{1\over d\, q}+\mathcal{O}\Big({1\over
 q^2}\Big)\,,\nonumber
\\
 S_q&=&2\pi \Vs \(\frac{L}{\lp} \)^{d-1} \[\(1- {d-1\over d-2}\,x_{\infty}^d \)
 +{1\over q}\(1- {d-1\over d-2}\,x_{\infty}^d - x_{\infty}^{d-1} \)
 +\mathcal{O}\Big({1\over q^2}\Big)\]\,,
 \labell{deeep}
 \eea
where $x_\infty=\sqrt{d(d-2)}/d$. In particular, this means that
 \be
 \lambda_1=\exp\[-2\pi \Vs\( \frac{L}{\lp} \)^{d-1}\Big(1- {d-1\over d-2}\,x_{\infty}^d
   \Big)\]\,,\quad
 d_1=\exp\[2\pi \Vs \( \frac{L}{\lp} \)^{d-1} \,x_{\infty}^{d-1}\]\,.
 \ee
Note that the factor of $(L/\lp)^{d-1}$ in the second exponent
indicates the degeneracy $d_1$ is also growing as an exponential of the
central charge of the boundary CFT. In the following, we show that the
value of $d_1$ is related to the degeneracy of the ground state of CFT
on $R\times H^{d-1}$. It is interesting to note that for $d=2$, the
above holographic result states that $d_1=1$. The same result follows
from the general form \reef{twod} of the \ren entropy for $d=2$ CFT's
since the leading coefficient precisely matches that for the $1/q$
term. We also note that the latter result implies that the expansion of
the \ren entropy \reef{expands} for $d=2$ terminates at order $1/q$.
This will imply certain constraints on the structure of the spectral
function, which in turn will allow us to evaluate it.

A further observation is that our holographic result yields an
expansion \reef{deeep} which appears to be completely analytic in the
limit of large $q$.\footnote{Note that this may be an artifact of the
limit of large central charge, \ie the large $N_c$ limit, in the
boundary CFT.} In contrast, the general expansion in eq.~\reef{expands}
contains nonanalytic terms. However, it seems that the latter are a
consequence of our simplifying assumption that the spectrum is
discrete. Hence, instead one can consider the density matrix to have a
smooth distribution $d(\lambda)$ of eigenvalues and then the \ren
entropy would become
 \be
S_q= \frac{1}{1-q}\,\log\left[\int_0^{\lambda_1}d\lambda\ d(\lambda)\,
\lambda^q \right]\,.
 \labell{contins}
 \ee
While we leave the details as an exercise to the reader, one can easily
show that this ansatz also leads to nonanalytic contributions in the
large-$q$ expansion. Hence we are led to conclude that the eigenvalue
spectrum must have both smooth and discrete components, \ie
$d(\lambda)$ must contain delta-functions supported on a discrete set
of eigenvalues.

To provide evidence for the above structure and to facilitate further
analysis, we continue by examining the spectral structure of the
density matrix $\rhov$ from the thermodynamic perspective, which is
succinctly encoded in eq.~\reef{form1}. We will assume that there are
no phase transitions in the system under consideration and so the free
energy and its derivatives are continuous functions of the temperature
$T$. Expanding eq.~\reef{form1} near $q=\infty$ and using the
thermodynamic identity $F'(T)=-S(T)$ then yields
 \begin{multline}
 S_q=-{F(T_0)-F(0) \over T_0}-{1\over q}\( {F(T_0)-F(0) \over T_0}+S(0)\)
 \\
 -{1\over q^2}\( {F(T_0)-F(0) \over T_0}+S(0)-{1\over 2}T_0F''(0)\)+\mathcal{O}(1/q^3)
 \labell{exprenyi1}
 \end{multline}
Again the leading term yields $-\log\lambda_1$ and so we have
 \be
\lambda_1=\exp\[\frac{F(T_0)-F(0) }{ T_0}\]=\frac{e^{-
E_1/T_0}}{Z(T_0)}~,
 \labell{Espec}
 \ee
where $E_1$ is the energy of the ground state.  That is, from the
thermal perspective, we see that $\lambda_1$ is indeed the largest
eigenvalue of the thermal density matrix of the CFT at temperature
$T_0$ on $R\times H^{d-1}$. Of course, this result is in complete
agreement with the statement that there is a one-to-one correspondence
between the eigenvalues and their degeneracies of the reduced density
matrix $\rhov$ and the thermal density matrix. Indeed, comparing to
eq.~\reef{expands}, we next extract
 \be
 \log d_1=S(0)
 \ee
from the term of order $1/q$ in eq.~\reef{exprenyi1}. That is, the
degeneracy $d_1$ of the largest eigenvalue $\lambda_1$ equals to the
degeneracy of the ground state of the thermal CFT. Substituting these
identifications in eq.~\reef{exprenyi1}, we find entropy. Hence,
 \be
S_q=-\log\lambda_1-{1\over q}\log(d_1\lambda_1)-{1\over q^2}\(
\log(d_1\lambda_1)-\frac12 T_0F''(0)\)+\mathcal{O}(1/q^3)~.
  \labell{exprenyi2}
 \ee

Again we observe that the expansions in eqs.~\reef{exprenyi1} and
\reef{exprenyi2} will be analytic in $1/\beta=T_0/q$ as long as the
free energy is analytic in the vicinity of vanishing temperature. Given
our previous analysis, this again points to a distribution of
eigenvalues that has both smooth and discrete components. Note that
from the thermal perspective, it is natural to think in terms of energy
eigenvalues and the thermodynamic spectral function $\rho(E)$. Hence
let us consider an ansatz where $\rho(E)$ contains delta functions
supported on some discrete set of energies $\{E_i, i=1,...,i_{max}\}$
as well as a smooth continuous distribution, \ie
 \be
 Z(\beta)=\sum_{i=1}^{i_{max}}d_i \, e^{-\beta E_i}+
 \int_{ E_0}^\infty  \overline\rho(E) e^{-\beta E}dE~,
 \labell{partition1}
 \ee
where $\overline\rho(E)$ is the smooth component of the spectral
function and $\{d_i,i=1,\ldots,i_{max}\}$ are the degeneracies of the
discrete eigenvalues $E_i$. Without loss of generality, we are assuming
the smooth component has support down to some $E_0$ with $E_0\ge E_1$.
Note also that $\overline\rho(E_0)\neq0$, since otherwise the free
energy will be nonanalytic in the neighbourhood of $T=0$ and thus
contradicts our holographic result.

Expanding in $1/\beta$, yields to leading order
 \be
 F(\beta)=-\frac1{\beta}\log Z(\beta)=E_1-\frac1{\beta}\log d_1+...
 \ee
This expansion agrees with \reef{exprenyi1} and in particular, as
claimed before $S(T=0)=\log d_1$. Furthermore, if we expand to higher
orders in the vicinity of $T=0$, we will encounter corrections
proportional to $e^{-\beta(E_i-E_0)}/\beta$ with $i>1$, which are again
nonanalytic in $1/\beta$. While this sort of correction does not
contradict any thermodynamic reasoning and it would seem to indicate
that the free energy is nonanalytic in the neighbourhood of $T=0$ (or
equivalently $q=\infty$). However, this would conflict with our
holographic result, where only inverse powers of $q$, or alternatively
$\beta$, are present in the expansion. Hence we proceed by
investigating the consequences of demanding that these corrections are
absent in keeping with holography. In particular, one immediate
consequence is that $i_{max}=1$, namely there is only one delta
function supported on the ground state of the CFT. Similarly, one can
show that analyticity of $S_q$ near $q=\infty$ (\ie our holographic
restriction) requires $E_0=E_1$ as well as analyticity of
$\overline\rho(E)$ in the vicinity of the ground state $E=E_1$.

Combining altogether, we have for the density of states
 \be
 \rho(E)=d_1\, \delta(E-E_1)+\overline\rho(E)~, \labell{bong}
 \ee
with $\overline\rho(E)$ being analytic at $E_1$. Then expanding the
resulting free energy for large $\beta$ then yields
 \be
F(\beta)=-\frac1{\beta}\log Z(\beta)=E_1-\frac1{\beta}\log d_1-
\frac1{\beta}\log\(1+\frac1{d_1\beta}\sum_{n=0}^\infty
{\overline\rho^{(n)}(E_1) \over \beta^{n} }\)~. \labell{freengy}
 \ee
Comparing this result to eq.~\reef{exprenyi2}, we find by comparing the
terms of order $1/q^2$:
  \be
 F''(0) = -2\,\frac{\overline\rho(E_1)}{d_1}~.
  \ee
Recall the thermodynamic identity $F''(T)=-{C_v}/{T}$ where $C_v$ is
the heat capacity. Hence the this implies a finite limit $\lim_{T\to0}
C_v/T$ if $\rho(E_1)$ is nonvanishing.  Substituting this result for
the free energy \reef{freengy} into eq.~\reef{form1}, we would find the
large-$\beta$ expansion of the \ren entropy. Then we could expand our
holographic results for $S_q$ in \eg eq.~\reef{SqGB} around $q=\infty$
and by comparing with the two expansions, we can reconstruct the
spectral function $\rho(E)$ in the neighbourhood of the ground state
$E=E_1$.

Let us consider the case of $d=2$ (holographic) CFT's. In this case,
the hyperbolic plane reduces to an infinite line and the free energy of
the field theory is given by
 \be
 F(T)=-{c \over 6} \, {T^2 \over T_0} \log(2R/\delta)\,.
 \ee
To get this result, one merely notes that for $d=2$, the Euclidean
background $S^1\times R$ can be conformally mapped to the infinite flat
plane, as discussed in section \ref{versus}. Therefore bearing in mind
that the field theory is conformal, we deduce $F\sim T^2 \Vs (d=2) \sim
T^2 \log (2R/\delta)$ and the proportionality constant is fixed by
requiring the thermal entropy derived from this expression to be equal
the entanglement entropy, $S_1$. Comparing this expression with
\reef{freengy}, yields
 \be
 d_1=1, \quad \overline\rho^{(n)} (E_1)= {\overline\rho_0^{\, n+1}
  \over (n+1)!}~ \text{for} ~ n \geq 0\, ,
 \ee
where
 \be
 \overline\rho_0= {c \over 6 T_0} \log(2R/\delta)~.
 \ee
Combining these results yields a simple spectral function:
 \be
  \rho(E)=\, \delta(E-E_1)+\theta(E-E_1){ \overline\rho_0 \over \sqrt{\overline\rho_0(E-E_1)}}
  \,I_1(2\sqrt{\overline\rho_0(E-E_1)})\,.
  \labell{rhoE}
 \ee
where $I_1$ is a modified Bessel function.

We observe that eq.~\reef{bong} exhibits  the same form as found in
\cite{calab8} from general considerations of the entanglement spectrum
for two-dimensional CFT's. Let us review their analysis. In equation
\reef{contins}, one could consider the \ren entropy as related to the
Laplace transform of the spectral function $d(\lambda)$, with $q$
acting as the Laplace parameter. \ie
\begin{equation}
e^{(1-q)S_{q}} = \int_{0}^{\lambda_1} d\lambda \, d(\lambda) \lambda^q =
\int_{\mathcal{E}_1}^{\infty} d\mathcal{E} \,e^{-\mathcal{E} (q+1)} \,d(e^{-\mathcal{E}}),
\end{equation}
where we have made the change of variables $\lambda= e^{-\mathcal{E}}$.
Note that comparing with \reef{Espec}, $\mathcal{E}$ is related to
energy measured in the hyperbolic space by $\mathcal{E} = E/T_0 + \ln
Z(T_0)$, and $\mathcal{E}_1 = E_1/T_0 + \ln Z(T_0) $. Given $S_{q}$,
therefore, one could consider inverting the expression via an inverse
Laplace transform: \be \labell{inlap} d(e^{-\mathcal{E}}) =
\frac{1}{2\pi i} \int_{\gamma - i\infty}^{\gamma + i\infty}  dq\,
e^{(1-q) S_q } e^{q \mathcal{E}}, \ee where $\gamma$ is suitably chosen
for the convergence of the integral. This has been considered in
\cite{calab8}. The complete spectral function for the \ren entropy
\reef{holotwod2} is given by
\begin{eqnarray}
d(\lambda) &&= \delta(\lambda_{1}- \lambda) +
b \frac{\theta(\lambda_{1}- \lambda)}{\lambda \phi}I_1(2\phi)\,, \nonumber \\
\phi &&= \sqrt{b\ln(\lambda_{1}/\lambda)} \,, \qquad b= -\ln \lambda_{1}.
\end{eqnarray}
Hence we can see that there is full agreement between this result and
eq.~\reef{rhoE}.

In higher dimensions, one can consider pursuing the same exercise.
However, given the increased complexity of the expression for $S_q$, we
would satisfy ourselves with an asymptotic expression, where $\lambda =
\exp(-\mathcal{E})\ll 1$ or $\mathcal{E}\gg 1$. In which case, we can
approximate the integral via the saddle-point approximation. The
effective Lagrangian in the integral \reef{inlap} for Einstein gravity
is given by \be L=  q \mathcal{E} - (q-1) S_q=  q \mathcal{E} -  \pi q
\Vs \( \frac{\tL}{\lp} \)^{d-1} (2 - x_q^{d - 2}(x_q^2 + 1)) . \ee The
saddle point is difficult to solve for general $d$, but can be worked
out easily for explicit values of $d$.

Checking the results for $d=4$, we find \be d(\lambda =
e^{-\mathcal{E}}) \sim \sqrt{\pi 3^{1/8}} \frac{e^{2
\Lambda^{\frac{1}{4}} (\frac{\mathcal{E}}{3})^\frac{3}{4}}}{2
\mathcal{E}^{\frac{5}{8}} \Lambda^{\frac{1}{8}}}. \ee where we have
$\Lambda= \pi \Vs \( \frac{\tL}{\lp} \)^{d-1}$. Combining the results
at $d=2$, this fits into the pattern of \be d(\lambda =
e^{-\mathcal{E}}) \sim
\frac{\exp(\mathcal{E}^{(d-1)/d})}{\mathcal{E}^{(d+1)/2d}}. \ee The
power dependence of $\mathcal{E}$ in the exponential is exactly what is
expected of the spectral density of a conformal field theory in flat
space. Since we are in some high energy limit, the background curvature
becomes irrelevant which thus leads to the same behavior. It would be
interesting to find out whether field theories in higher dimensions
display the same degree of universality as in two
dimensions\cite{calab8}.


\section{Discussion} \labell{discuss}

We have presented several new results in this paper. The first was an
extension the approach of \cite{casini9} for calculating entanglement
entropy to a new approach to calculate the \ren entropy of a general
CFT in $d$ dimensions with a spherical entangling surface. This new
calculation evaluates the \ren entropy in terms of certain thermal
partition functions of the CFT on the background $R\times H^{d-1}$.
While this result is independent of holography, this calculation has a
simple holographic translation to a gravitational calculation in the
context of the AdS/CFT correspondence. Hence we applied this approach
to calculate the \ren entropy for variety of holographic models in
section \ref{renyi}. Our results, \eg in eqs.~\reef{SqGB} and
\reef{SqQT}, indicate that the \ren entropy in higher dimensions is a
complicated nonlinear function of the central charges and other
parameters which characterize the CFT. We emphasize that the latter
result should be expected to apply beyond holographic CFT's. That is,
since no simple universal formula appears in our holographic results,
we should not expect any simplification in general. The fact that more
data than the central charges enters the \ren entropy for a spherical
entangling surface in higher dimensions is perhaps not so surprising.
Even in two dimensions, if we consider two separate intervals
calculating the \ren entropy requires detailed information about the
full spectrum of the corresponding CFT \cite{joint}.

In section \ref{versus}, we elucidated the relation between this new
thermal calculation and a conventional calculation of the \ren entropy
where a twist operator is inserted on the spherical entangling surface.
Here one begins with the thermal path integral on the Euclidean
background $S^1\times H^{d-1}$ where the period of the circle is given
by $q/T_0$. If we choose $q$ to be a positive integer, the Euclidean
version of the conformal mapping employed in \cite{casini9} takes this
background to a $q$-fold cover of flat space $R^d$ with an orbifold
singularity on a sphere of radius $R$ with an angular excess of
$2\pi(q-1)$. This relation also allowed us to calculate the conformal
scaling dimension of the twist operator. The latter was expressed in
terms of the energy density of the thermal ensemble in eq.~\reef{heavy}
or in terms of the thermal entropy and the corresponding \ren entropy
in eq.~\reef{heavy2}. Again these results do not depend on holography,
however, the analysis is easily translated to a gravitational
calculation using the AdS/CFT correspondence. Hence we calculated $h_q$
for the various holographic models introduced in section \ref{renyi}.
Again, in general, the result was found to be a complicated nonlinear
expression written in terms of the central charges, as well as other
parameters in the CFT. However, we observed a remarkable simplification
upon considering the expression $\partial_qh_q|_{q=1}$, which is given
by simply the central charge $\CT$ appearing in the two-point function
of the stress tensor. We should add that our calculations here have a
strong similarity to those appearing in \cite{donm}. The latter
analyzed holographic conformal field theory in de Sitter space but at
an arbitrary temperature which need not match that of the cosmological
horizon.

An interesting consequence of eq.~\reef{finalfor} is that all of the
\ren entropies calculated for a spherical entangling surface in a
$d$-dimensional CFT exhibit the same divergence structure as the
corresponding entanglement entropy. These UV divergences arise from the
integral over the same (infinite) hyperbolic geometry in all cases and
were encapsulated by the factor $\Vs$ given in eq.~\reef{vtotal}. While
in general the individual power law divergences are sensitive to the
details of the regulator in the CFT, we also identified a universal
contribution \reef{unis} in this expression. However, if one calculates
all of the \ren entropies with a certain fixed regulator, it should be
that one can consider the ratio $S_q/S_1$ which remains finite in the
limit $\delta\to0$. In this limit, the ratio is determined by the
coefficients of the leading power law divergent terms and hence their
ratio yields universal information characterizing the CFT. However,
since the entire regulator dependence of $S_q$ factors out as $\Vs$,
this ratio will not contain any new information aside from the
coefficients already identified in the universal contributions in the
individual $S_q$.

While our results indicate that the \ren entropy in higher dimensions
is a complicated nonlinear function of the various parameters
characterizing the CFT, it is interesting that when we explicitly plot
$S_q$ as a function of, \eg $c/a$ in figure \ref{fig:GB}a or $t_4$ in
figure \ref{fig:QT}a, the results appear to be essentially linear in
these parameters over the physical regime. We examined the behaviour of
$S_q$ in more detail for the holographic model with $d=4$ and GB
gravity in section \ref{sec:GB}. There we observed that  the linearity
cannot continue to arbitrarily towards small $c/a$ because we must have
$S_q/S_1<1$. However, it seems that the nonlinear terms are suppressed
for large $c/a$ and in this regime, the linear approximation of
$S_q/S_1$ is even better. One consequence is that eventually $S_q/S_1$
becomes negative for sufficiently large $c/a$, which is clearly
unphysical! However, as we emphasized in our discussion in section
\ref{sec:GB}, this unusal behaviour arises in a regime where we already
know that the holographic model has unphysical properties, \eg
violations of causality. Behaviour similar to that observed for $d=4$
also appears for $d>4$ as well, as shown in figure \ref{fig:GB2}.
However, we note that for sufficiently large $d$ (\ie $d>7$) we find
that $S_q/S_1$ can become negative within the physical regime defined
in eq.~\reef{limitsca}. As we will discuss elsewhere \cite{tobe}, it
seems that this unusual behaviour can be used to impose additional
physical constraints on these holographic models with interesting
consequences.

Another remarkable simplification was observed in section
\ref{twisted}. There again, the scaling dimension $h_q$ in higher
dimensions is found to be a complicated nonlinear function of the
central charges and other parameters characterizing the CFT, however,
in the limit $q\to1$, $\partial_qh_q$ is given by a very simple
expression \reef{interest1} for all of our holographic models and it is
precisely proportional to $\CT$. In fact, the simplicity of this result
strongly suggests that it should be a general result that applies
beyond a holographic setting. That is, we conjecture that
eq.~\reef{interest1} in fact holds for any CFT. Here we comment on a
possible physical consequence of this result. Generally, any even
dimensional surface operator will suffers from a Graham-Witten anomaly
\cite{grwit} and it has been observed that the anomaly controls how the
vacuum expectation value of the operator scales under scaling
transformations \cite{surfop}. It is therefore a natural question to
ask whether the Graham-Witten anomaly is related to the generalized
scaling dimension $h_q$, when $d$ is even, for the twist operators
appearing in the present discussion. In this context, the universal
coefficient $\mathcal{C}$ of the logarithmic contribution to the
entanglement entropy is given by precisely the Graham-Witten anomaly
\cite{cthem,solo}. As explained in the sections \ref{twist} and
\ref{versus}, the partition function $Z_q$ is given by the expectation
value of a twist operator $\sigma_q$ inserted on the entangling
surface. Hence we may write the \ren entropy as $S_q= \log \langle\,
\sigma_{q}\,\rangle/(1-q)$. Hence in the limit $q\to 1$, we identify
the entanglement entropy as $S_1=-\partial_q \log \langle\,
\sigma_{q}\,\rangle|_{q=1} $. Therefore, the universal coefficient
$\mathcal{C}$ is equal to the derivative with respect to $q$ of the
corresponding coefficient in the expectation value of the twist
operator in the same limit. At $d=2$, we can compare $\mathcal{C}$ with
$\partial_qh_q$ at $q=1$ and they coincide. In higher even dimensions,
for a spherical entangling surface in flat space, it is well known that
$\mathcal{C}$ is proportional to the $A$-type central charge of the
conformal field theory \cite{cthem,solo}. However, our holographic
results \reef{interest1} suggests that $\partial_q h_q\vert_{q\to 1}$
is given instead by $\CT$, which controls the two-point function of the
stress tensor, which is a distinct from the $A$-type central charge in
higher dimensions. Hence it seems that there is no simple relation
between the generalized conformal dimension and the Graham-Witten
anomaly for twist operators in higher dimensions.

One important aspect of our holographic calculations of \ren entropy
was that the bulk geometries of which we make use are nonsingular. As
described in sections \ref{twist} and \ref{versus}, the standard field
theory calculation of \ren entropy involves evaluating the partition
function on a singular orbifold background. The naive holographic
translation of this boundary calculation extends the orbifold
singularity into the bulk spacetime \cite{furry}. However, the
resulting bulk geometry does not satisfy the classical gravitational
equations of motion precisely because of the presence of this
singularity and so it seems that without a full understanding of string
theory or quantum gravity in the bulk, we will not understand how to
properly calculate with this bulk geometry. More importantly, it can be
explicitly shown that the naive prescription suggested in \cite{furry}
leads to incorrect results for the \ren entropy in certain instances
\cite{head}. The key step in our approach was to find an alternative
field theory calculation of the \ren entropy involving a thermal
partition function which had a simple holographic translation involving
hyperbolic black holes in the bulk spacetime and of course, the
Euclidean versions of these geometries are completely nonsingular. As
stressed in section \ref{submarine}, one should be able to perform a
holographic calculation of the \ren entropy by choosing the boundary
metric to match that on the orbifold appearing in the replica trick
calculation and finding a new smooth bulk geometry which properly
solves the gravity equations. In this approach, the conical
singularities introduced by the twist operators would appear only in
the induced metric on UV regulator surface. Of course, as shown in
section \ref{versus}, our thermal approach and the standard replica
trick calculation are simply related by a conformal transformation.
Hence in principle, the black hole geometries provide precisely the
desired nonsingular bulk solution for the problem where the boundary
metric matches that on the relevant orbifold. Of course, our
holographic calculations only strictly apply for a spherical entangling
surface and so calculating the \ren entropy with more general
entangling surfaces still presents a formidable challenge in
constructing the appropriate smooth bulk geometry.

An interesting observation in \cite{head} was that although
prescription of \cite{furry} using singular bulk geometries to
calculate $S_q$ led to incorrect results for general $q$, it still
seems that the correct entanglement entropy results in the limit
$q\to1$. This correct result still relies on extending the boundary
singularity into the bulk and so strictly speaking the bulk geometry
does not satisfy the correct equations of motion. However we note that,
in the $q\to1$ limit, the deviation from the on-shell geometry is
infinitesimal. A similar holographic calculation of the entanglement
entropy in the case of a spherical entangling surface was presented in
\cite{cthem} where again an infinitesimal singularity appears in the
bulk. However, following the analysis of \cite{solo1}, it was
emphasized that the leading off-shell behaviour of this singularity is
universal and hence the result for the entanglement entropy is
reliable. In particular, the entanglement entropy requires expanding
the gravity action only to linear order in the deviation from the
on-shell value. However, this linear deviation is proportional to the
gravitation equations of motion in the bulk. Hence the bulk term does
not contribute and the non-trivial contribution comes exclusively from
the boundary. It seems that similar reasoning may apply for the more
general calculations in \cite{furry} so that the detailed extension of
the singularity into the bulk is irrelevant to the final result for the
entanglement entropy. However, it seems that one must still invoke some
additional rotational symmetry about the bulk singularity to properly
justify this approach. As noted in \cite{cthem}, an analogous result
applies in calculations of black hole entropy where it is understood
that the off-shell approach \cite{frank1} will generally yields the
correct Wald entropy \cite{WaldEnt} for stationary black hole
solutions.

\acknowledgments

We would like to thank Matt Headrick, Patrick Hayden, Andreas Karch,
Brian Swingle and Tadashi Takayanagi for useful conversations. Research
at Perimeter Institute is supported by the Government of Canada through
Industry Canada and by the Province of Ontario through the Ministry of
Research \& Innovation. RCM also acknowledges support from an NSERC
Discovery grant and funding from the Canadian Institute for Advanced
Research. AY is also supported by a fellowship from the Natural
Sciences and Engineering Research Council of Canada.

\appendix

\section{\ren entropy inequalities} \labell{unequal}

In information theory, the \ren entropy refers to a probability
distribution with $p_i>0$ and $\sum_i p_i =1$ \cite{renyi0}. It follows
that $S_q$ must satisfy a variety of different inequalities
\cite{karol}. In the following, we consider four such inequalities:
 \bea
\frac{\partial S_q}{\partial q} &\le &0 \,, \labell{ineq1}\\
\frac{\partial}{\partial q}\( \frac{q-1}{q} S_q\) &\ge& 0 \,, \labell{ineq2} \\
\frac{\partial}{\partial q}\( (q-1)S_q\) &\ge &0\,,\labell{ineq3}\\
\frac{\partial^2}{\partial q^2}\((q-1)S_q\) &\le& 0\,.\labell{ineq4}
 \eea
In particular, we will show that these inequalities are naturally
satisfied by the entropies found in our holographic calculations.
Actually, these results do not rely on the holographic framework but
rather they follow from the relation which we established in the
introduction between the \ren entropies for a spherical entangling
surface and the thermal ensemble on the hyperbolic cylinder $R\times
H^{d-1}$. Hence it follows that these inequalities are satisfied for
any CFT, as long as the thermal ensemble is stable, as we will see
below. First, let us recall the formula \reef{finalfor} relating the
\ren entropy and the thermal entropy,
 \be
S_q = \frac{q}{q-1} \frac{1}{T_0} \int^{T_0}_{T_0/q} S_\mt{th}(T)\,
dT\,,
 \labell{holoren}
 \ee
where $T_0 = 1/(2\pi R)$ as usual and $S_\mt{th}(T)$ denotes the
thermal entropy on $R\times H^{d-1}$.

If we begin by considering eq.~(\ref{ineq2}), the expression on the
right-hand side yields
 \be
\frac{\partial}{\partial q}\( \frac{q-1}{q} S_q\) = \frac{1}{q^2}\,
S_\mt{th}(T_0/q)\,.
 \labell{idenS}
 \ee
Hence the inequality \reef{ineq2} follows simply because the thermal
entropy is a positive quantity. Given that eq.~(\ref{ineq2}) is
satisfied, simple corollary that follows is $S_q \le q/(q - 1)\,
S_\infty$ for $q> 1$ \cite{head}. However, we note that it seems that
the basic physical property that $S_\mt{th}(T)\ge0$ can provide a
nontrivial constraint on some of the holographic models with higher
curvatures --- see the discussion in section \ref{discuss}.

Next turning to eq.~(\ref{ineq4}), the right-hand side yields
 \be
\frac{\partial^2}{\partial q^2}\((q-1)S_q\) = - \frac{T_0}{q^3}\,
\left.\frac{\partial S_\mt{th}}{\partial
T}\right|_{T_0/q}=-\frac1{q^2}\, \left.\frac{\partial
E_\mt{th}}{\partial T}\right|_{T_0/q}\,. \labell{holoineq4}
 \ee
Hence this expression is proportional to the specific heat and it
follows from the stability of the thermal ensemble that the specific
heat must be positive. This stability then ensures that the inequality
\reef{ineq3} is satisfied by the corresponding \ren entropy.

Using eq.~\reef{holoren}, the relevant expression in the first
inequality \reef{ineq1} becomes
 \be
\frac{\partial S_q}{\partial q} =
-\frac1{(q-1)^2}\,\frac1{T_0}\int_{T_0/q}^{T_0}\Big[ S_\mt{th}(T) -
S_\mt{th}(T_0/q) \Big]\, dT \,.
 \labell{holoineq1}
 \ee
The fact that the integrand here is positive for $q>1$ again follows
from the stability requirement above. That is, a positive specific heat
implies $\partial_T S_\mt{th}>0$ and hence the finite difference in the
integrand above is also positive. Hence the desired inequality
\reef{ineq1} is again satisfied for $q>1$. Examining the case $0<q<1$
is facilitated by rewriting the above expression \reef{holoineq1} as
 \be
\frac{\partial S_q}{\partial q} =
-\frac1{(1-q)^2}\,\frac1{T_0}\int^{T_0/q}_{T_0}\Big[S_\mt{th}(T_0/q) -
S_\mt{th}(T) \Big] \,dT \,.
 \labell{holoineq1b}
 \ee
Now it is clear that the same reasoning as above can be used to
establish that eq.~\reef{ineq1} is still satisfied for this range of
$q$. Again a simple corollary that follows from eq.~\reef{ineq1} is
that $S_{n+m}\le S_n$ for positive integers, $n$ and $m$.

Finally considering eq.~\reef{ineq3}, the right-hand side can be
written as
 \be
\frac{\partial}{\partial q}\( (q-1)S_q\) =
\frac1{T_0}\int_{T_0/q}^{T_0} S_\mt{th}(T)\, dT
+\frac1q\,S_\mt{th}(T_0/q) \,.
 \labell{holoineq3}
 \ee
For $q>1$, the positivity of this expression readily follows simply
because the thermal entropy is always a positive quantity and hence the
corresponding inequality \reef{ineq3} holds. In considering the range
$0<q<1$, it is better to write this expression as
 \be
\frac{\partial}{\partial q}\( (q-1)S_q\) =
\frac1{T_0}\int^{T_0/q}_{T_0}\Big[S_\mt{th}(T_0/q) - S_\mt{th}(T) \Big]
\, dT + S_\mt{th}(T_0/q) \,.
 \labell{holoineq3b}
 \ee
In this case, the integral being positive follows from the stability of
the thermal ensemble which implies $\partial_T S_\mt{th}>0$, as
discussed above with eq.~\reef{holoineq1}. Of course, the positivity of
the last term in this expression again follows because the thermal
entropy is positive. Hence the desired inequality also holds in the
range $0<q<1$.

\section{\ren entropy in $d=3$} \labell{threed}

The results in section \ref{renyi} indicate that the \ren entropies
with a spherical entangling surface for CFT's in higher dimensions do
not have a simple form analogous to the two-dimensional result
\reef{twod}. In fact, for even dimensions, the \ren entropies depend on
CFT parameters beyond simply the central charges. The latter contrasts
with the known result for the entanglement entropy in any dimension
where the coefficient of the universal term is simply a linear
combination of the central charges appearing in the trace anomaly
\cite{rt2,cthem,solo,EEGB}. The precise linear combination depends on
the geometry and for a spherical entangling surface, it is simply
proportional to $A$, the central charge in the A-type anomaly
\cite{deser}.

However, the holographic calculations leading to the above conclusion
only refer to $d\ge4$. Hence one might still hope for some kind of
simplification for the form of the \ren entropy in $d=3$. We would like
to examine this question with a nontrivial holographic model with a
four-dimensional bulk. The theory we analyze here is a higher curvature
theory with an interaction which is cubic in the Weyl
tensor.\footnote{Note that in four dimensions, the Weyl tensor is given
by
\beq
 C_{abcd}=R_{abcd}-  g_{a[c}\,R_{d]b}+g_{b[c}\,R_{d]a}
+\frac13\,R\,g_{a[c}\,g_{d]b}\ .
 \label{weyl}
\eeq} To be concrete, we consider the following action
 \be
I_{C^3}=\frac{1}{2\lp^2}\int d^4x \sqrt{-g}\(\frac{6}{L^2} +R + \gamma
L^4 C_{abcd}C^{cdef}C_{ef}{}^{ab}\)\,,
 \labell{act3d}
 \ee
While no exact analytic black hole solutions are known for this theory,
we can readily construct solutions perturbatively in the
(dimensionless) coupling $\gamma$. However, before doing so, let us
consider the role of $\gamma$ in the dual boundary theory. Since the
boundary is three-dimensional, there is no trace anomaly, however, we
can still consider the effective central charges $C_T$ and $\ads$, as
introduced in eq.~\reef{effectc}. However, it turns out that because
the cubic curvature interaction above is constructed with the Weyl
tensor, neither of these parameters in the dual CFT is effected by
$\gamma$. For example, $C_T$ is essentially determined by the graviton
propagator in AdS$_4$ space. However, since the Weyl curvature vanishes
in AdS space, the cubic term in eq.~\reef{act3d} can not contribute to
the graviton kinetic term in this background. As this reasoning
suggests, the effect of the cubic interaction only begins to be felt
with the introduction of a new interaction cubic in the gravitons.
Therefore the new coupling is first seen in a modification of the
parameters appearing in the three-point function of the boundary stress
tensor. In fact, if we are working perturbatively in this coupling, one
would find that to leading order $t_4\propto\gamma$ \cite{diego}, where
$t_4$ is the parameter introduced in eq.~\reef{parms4}. Hence as a
measure of simplicity, we wish to determine whether or not the \ren
entropies are independent of this additional CFT parameter $t_4$.

In the following analysis of the holographic \ren entropy, we are
working perturbatively in $\gamma$ and we will only present the
relevant results to first order in this coupling. A useful check of any
of the following expressions is that in the limit $\gamma\to0$, they
must reduce to the corresponding result for Einstein gravity with
$d=3$, as given in section \ref{sec:EH}. Further note that since the
Weyl tensor vanishes in AdS space, the new gravitational interaction
does not effect the AdS curvature, \ie as with Einstein gravity,
$\tL=L$

In constructing the perturbative black hole solutions, we extend the
metric ansatz \reef{lineelement} slightly to
 \be
ds^2 = - \(\frac{r^2}{L^2} f(r)-1\)
N^2(r)\,dt^2+\frac{dr^2}{\frac{r^2}{L^2} f(r)-1} + r^2
d\Sigma_{\,2}^2\,,
 \ee
\ie now $g_{tt}$ now contains an independent function $N(r)$. As
before, $d\Sigma_{\,2}^2$ denotes the metric of a two-dimensional
hyperbolic plane with unit curvature. As noted above, to zeroth order
in $\gamma$, the solution is just the hyperbolic black hole solution
for Einstein gravity presented in section \reef{sec:EH} with $d=3$. The
perturbative solution to linear order in $\gamma$ is then given by
 \bea
f(r) &=& 1 - \frac{\w^3}{r^3} - \gamma \frac{2\, \w^6}{L^2r^9} \(8 \w^3
+ 9 L^2r - 12r^3\)
\,,\nonumber \\
N(r) &=& N_0\(1 - \gamma \frac{6 \w^6 }{r^6}\)\,,\qquad N_0 =
\frac{L}{R}\,.\labell{persol}
 \eea
The constant $N_0$ is chosen so that the boundary metric matches that
given in eq.~\reef{cftmet}. It is convenient to again parameterize the
constant $\w^3$ in terms of the horizon radius $\rh$ which is still
determined by eq.~\reef{horizon}. Hence to first order in $\gamma$, we
have
 \be
 \w^3 = \rh^3 - L^2\rh + \gamma \frac{2(4 \rh^2-L^2) (
\rh^2 - L^2)^2}{\rh^3}\,. \labell{const3}
 \ee

The temperature is given by eq.~\reef{temp} except that we should
replace the constant $N$ with the function $N(r)$ (evaluated at
$r=\rh$). That is,
 \be
  T = \frac{N(\rh)}{4\pi } \[ \frac{2}{\rh} + \frac{\rh^2}{L^2}\
\left.\frac{\partial f(r)}{\partial r}\right|_{r=\rh} \]\,.
 \labell{temp3}
 \ee
To leading order in $\gamma$, we may write the temperature as
 \be
T= \frac{1}{4\pi R}\left(3x - \frac{1}{ x }+ \gamma \frac{6(1 - x^2)^2
}{x^3}\right)\,.
 \labell{temp3a}
 \ee
where as before $x=\rh/L$.

Next we determine the horizon entropy using Wald's formula
\reef{Waldformula}, which yields
 \be
S=\frac{2\pi}{\lp^2}\int_{\mt{horizon}} d^{2}x\, \sqrt{h}\left[1+
\frac{3}{2}\left( C_{ab}{}^{cd}C_{cd}{}^{ef} \hat{\veps}^{ab}
\hat{\veps}_{ef} -2\, C_{ab}{}^{cd}C_{cd}{}^{eb}\, \hat{\veps}^{af}
\hat{\veps}_{ef}- \frac{2}{3}\,C_{abcd}C^{abcd} \right) \right]\,.
 \labell{entt}
 \ee
Recall that $\hat{\veps}_{ab}$ is the binormal to the horizon and in
Lorentzian signature, $\hat{\veps}_{ab}\, \hat{\veps}^{ab}=-2$.
Substituting the above solution into this expression and expressing the
result in terms of $x$ yields
 \be
S(x) = 2\pi\Vs \frac{L^2}{\lp^2} \left(\,x^{d-1}+ 6 \gamma
\frac{(x^2-1)^2}{x^2}\right)  \,,
 \labell{enttb}
 \ee
to first order in $\gamma$.

Finally we apply eq.~\reef{finalfor2} to calculate the \ren entropy and
we find
 \be
S_q = \frac{\pi\,q}{q-1}\Vs \frac{L^2}{\lp^2}  \left( 2- \xq(1+\xq^2)
+2\gamma\,\frac{(1 - 4 \xq^2) (1 - \xq^2)^2}{\xq^3} \right)\,,
 \labell{result3}
 \ee
where again $\xq$ is the value of $x= \rh/L$ when the temperature
\reef{temp3a} is $T=T_0/q$. Working perturbatively in $\gamma$, we find
 \be
\xq = \frac{1 + \sqrt{1 + 3 q^2}}{3q} - 4\gamma \frac{ \left( \sqrt{1 +
3 q^2}-2\right)^2}{ 3q\sqrt{1 + 3 q^2} }\,.
 \labell{root3}
 \ee
Considering the limits $q\rightarrow 0$, 1 and $\infty$, we find to
leading order in $\gamma$
 \bea
\lim_{q \rightarrow 0} S_q      &=& \pi\Vs \frac{L^2}{\lp^2}
\,\frac8{27}\(1+2\gamma\)\,\frac{1}{q^{2}}\,,
 \nonumber\\
\lim_{q \rightarrow 1} S_q      &=& 2\pi\Vs \frac{L^2}{\lp^2}\,\,,
 \labell{limits3} \\
\lim_{q \rightarrow \infty} S_q &=& 2\pi\Vs \frac{L^2}{\lp^2}
\,\(1-\frac{2}{3\sqrt{3}}+\frac{8}{3\sqrt{3}}\gamma\)\,.
 \nonumber
 \eea
Here we simply note that our perturbative expression \reef{result3} of
$S_q$ depends on $\gamma$ and further this dependence persists above in
the limits $q\to0,\,\infty$. Hence, it does not seem that there is any
simplification in the form of the \ren entropy in $d=3$, compared to
our results for $d\ge4$.

The scaling weight $h_{q}$ of the twist operators in the dual boundary
theory can again be calculated using eq.~\reef{heavy2} which yields
\myeq{
h_q =   \frac{-2 \pi}{x^2(1-3x^2)^2} \frac{L^2}{\lp^2} \left(-2 \gamma +10 x^2 \gamma - (1+26 \gamma )x^4 +3 (1+6 \gamma )x^6\right) \,.
\labell{weight3}
}
The $q \rightarrow 1$ limit of course has the same simple form as that found in section \reef{twisted}:
\myeq{
\partial_q h_q |_{q=1} = \pi \frac{L^2}{\lp^2}
\labell{interestB}
}

\section{The on-shell bulk action for $d=2$}\labell{review}

In this appendix, we review some of the details of the holographic
construction presented in \cite{skenderis} when applied in our
holographic calculation of the \ren entropy in section \ref{2dholo}.
First, we write the three-dimensional bulk metric in FG gauge as given
in eq.~\reef{metric2}, which we reproduce here
 \be
ds^2 =\frac{L^2d\rho^2}{4\rho^2} + \frac{g_{a
b}(\omega,\bar{\omega},\rho) dx^a dx^b}{\rho}\,.
 \labell{repeatmetric}
 \ee
As discussed in the main text, the bulk geometry is the pure AdS$_3$
space for $d=2$. In this case, the curvature tensor can be written in
terms of the bulk metrc $G$ as
 \be
R_{\mu\nu\rho\sigma} =-
\frac{1}{L^2}\(G_{\mu\rho}G_{\nu\sigma}-G_{\mu\sigma}G_{\nu\rho} \)\,,
 \labell{eom}
 \ee
where $L$ is the AdS curvature scale. Substituting the metric $G$ in
eq.~\reef{repeatmetric} into this expression implies that
 \be
\partial^3_{\rho}\,g_{a b}(\omega,\bar{\omega},\rho)=0\,,
 \labell{short}
 \ee
and so the expansion \reef{expand} of
$g_{ab}(\omega,\bar{\omega},\rho)$ terminates at order $\rho^2$. In
fact, the full result may be written as
 \be g =
(1+ \frac{\rho}{2} g_{(2)}g_{(0)}^{-1})\,g_{(0)}\,(1+ \frac{\rho}{2}
g_{(0)}^{-1}g_{(2)})\,.
 \labell{gsol}
 \ee
In $d=2$,  the conformal symmetry at the boundary alone does not
completely determine $g_{(2)}$ since this is the order at which the
expansion contains information about the state of the CFT, \eg the
vacuum expectation value of the energy-momentum tensor. Of course, in
higher dimensions, this date enters the FG expansion at order $d$. In
any event, the boundary conformal symmetry does fix the form of
$g_{(2)}$ as follows
 \be
g_{(2)\,\, a b} = -\frac{L^2}{2}\left(R_{(0)}\, g_{(0)\,\,a b} + \T_{a
b}\right)\,,
 \labell{g2eom0}
 \ee
where
 \be
\nabla^a\,\T_{ab}=0\,,\quad{\rm and}\quad g_{(0)}^{a b}\,\T_{ab}=
-R_{(0)}\,,
 \labell{g2eom}
 \ee
with $R_{(0)}$, the Ricci scalar for the boundary metric $g_{(0)}$.

The symmetric tensor $\T_{ab}$ above behaves very much like a stress
tensor. Hence, to generate solutions, \cite{skenderis} proposed to take
the analogy more seriously, by introducing an auxiliary scalar
Liouville field $\phi$ on the AdS boundary, whose action is
 \be
 I_{\phi}=\frac{1}{48\pi}\int d\omega
d\bar{\omega}\sqrt{g_{(0)}}\left(\frac{1}{2}g_{(0)}^{a
b}\,\partial_a\phi\,\partial_b\phi - \phi R_{(0)}\right)\,.
 \labell{louis}
 \ee
The resulting energy-momentum tensor, which we equate with the unknown
tensor in eq.~\reef{g2eom0}, is given by
 \be
\T_{ab}=-\frac{1}{2}\partial_a\phi\,\partial_b\phi
-\nabla_a\partial_b\phi +g_{(0)\,\,
ab}\left(\frac{1}{4}g_{(0)}^{ab}\,\partial_a\phi\,\partial_b\phi +
 g_{(0)}^{ab}\,\nabla_a \partial_b\phi\right)\,,
 \labell{tab}
 \ee
where covariant derivatives $\nabla$ are again evaluated with
$g_{(0)}$. With the introduction of $\phi$, the second expression in
eq.~\reef{g2eom} becomes
 \be
  g_{(0)}^{ab}\,\nabla_a \partial_b\phi= -R_{(0)}\,.
 \labell{boxphi}
 \ee

We now focus on the specific case where the boundary is the universal
cover of the orbifold, in which case $g_{(0)}$ is given by
eq.~\reef{tentmetric}. We know that the universal cover is not flat
everywhere and its curvature can be deduced from the conformal
transformation \reef{2dmetric} that takes us to the flat $z$-plane, \ie
 \be
 R_{(0)}[g_{z\bar z}] =\Omega^{-2}
\left(\,R_{(0)}[g_{\omega\bar\omega}] - 4 g^{\omega \bar\omega}\,
\partial_\omega\partial_{\bar\omega}\ln \Omega\, \right)=0 \,,
 \labell{Rw}
 \ee
where
 \be
\Omega =
\bigg\vert(\omega-v_1)^{-(1-1/q)}(\omega-v_2)^{-(1+1/q)}\bigg\vert\,.
 \labell{Rw2}
 \ee
Substituting this expression into eq.~\reef{boxphi} then yields
 \be
\overline{\partial}_\omega\partial_\omega \phi =
-2\,\overline{\partial}_\omega\partial_\omega \ln \Omega\,,
 \labell{bang}
 \ee
which is readily solved with
 \be
 \phi = -2\ln \Omega + F(\omega) + \overline{F(\omega)}\,,
 \labell{sol}
 \ee
where  $F(\omega)$ is some general holomorphic function. For
simplicity, let us set $F(\omega)=0$ and as we will see this is the
correct choice given the conformal transformation \reef{ctrans} which
we performed at the boundary. We may simplify eq.~\reef{tab} for the
boundary metric $g_{(0)}$ given by eq.~\reef{tentmetric} to find
 \be
\T_{\omega\omega}= -\partial\partial\phi -\frac{1}{2}(\partial\phi)^2
=\overline{\(\T_{\bar{\omega}\bar{\omega}}\)}\,, \qquad \T_{\omega
\bar{\omega}} = \partial\bar{\partial}\phi\,.
 \labell{bang2}
 \ee
Substituting the explicit solution \reef{sol} (with $F=0$) into the
above expression then yields
 \be
\T_{\omega\omega} = \frac12\,\(1-\frac{1}{q^2}\)
\,\frac{(v_2-v_1)^2}{(\omega-v_2)^2\,(\omega-v_1)^2}\,.
 \labell{Tww}
 \ee

Of course, the tensor $\T_{ab}$ is simply to the expectation value of
the stress tensor $T_{ab}$ in the boundary theory. With $d=2$, the
relation is found to be\footnote{We comment here that the definitions
of the energy-momentum tensor in \cite{cardy0} and \cite{skenderis}
differ by the same factor of $-2\pi$ that was noted in footnote
\ref{musical}.}
 \be
\langle\, {T}_{a b}\,\rangle = \frac{1}{ L\,\lp} \(g_{(2)\,\, ab}-
g_{(0)\,\,ab}\,\textrm{Tr}(g_{(0)}^{-1}\,g_{(2)})\)
=-\frac12\,\frac{L}{\lp}\,\T_{ab}\,.
 \labell{cftstresst}
 \ee
Here the final expression comes from substituting eqs.~\reef{g2eom0}
and \reef{g2eom} for $g_{(2)}$. Hence combining eq.~\reef{Tww} with the
usual result for the central charge $c = 12\pi L/\lp$ (for Einstein
gravity in the bulk), we arrive at
 \be
\langle T_{\omega\omega}\rangle = -\frac{c}{48\pi}\(1-\frac{1}{q^2}\)
\, \frac{(v_2-v_1)^2}{(\omega-v_2)^2\,(\omega-v_1)^2}\,,
 \labell{holovev}
 \ee
which precisely matches the CFT result \reef{Tvev}.

Perhaps this result should not be surprising since it is well known
that a Weyl rescaling will shift the action of any $d=2$ CFT action by
an expression proportional to the Liouville action, where the Liouville
field takes the value of the Weyl factor \cite{friedan}. Further, in
the case where the Weyl factor is generated by a conformal
transformation, the stress tensor of the Liouville action evaluates
precisely to the Schwarzian. The holographic result suggests that the
bulk gravitational action should also be shifted by the same Liouville
action (as a surface term) relative to action evaluated in Poincar\'e
coordinates.\footnote{This need not seem obvious at this point.
However, we return to addressing this issue in section
\ref{submarine}.}

Nevertheless, to proceed with our holographic calculation of the \ren
entropy, we begin by substituting the metric \reef{gsol} into the
gravitational action and evaluating it on-shell. The latter action
consists of the Einstein-Hilbert action $I_{EH}$ and two boundary
contributions, the Gibbons-Hawking term $I_{GH}$ and a counter-term
action $I_{ct}$ \cite{ct}:
 \bea
&I_\mtt{tot} = \frac{1}{2\lp}\(I_{EH} + I_{GH}+ I_{ct}\)\,,\qquad
&I_{EH}
= -\int d^{3}x \sqrt{G} \(R+ \frac{2}{L^2}\)\,,\nonumber \\
&I_{GH} = - 2\,\int_{UV} d^2x\sqrt{ h}\,
 K\,,&I_{ct} = \frac{2}{L}\int_{UV}
  d^2x\sqrt{ h}\,.
 \labell{action}
 \eea
Above the two boundary contributions are evaluated on the UV regulator
surface $\rho=\rho_\mtt{min}=\delta^2/L^2$, where $\delta$ is the short
distance cut-off in the boundary theory which was introduced in the
main text. In particular then, $h(\omega,\bar{\omega})$ and $K$ are
respectively the induced metric and the extrinsic curvature on this
regulator surface.

Let us begin with $I_{EH}$. Since the bulk geometry is pure AdS$_3$
space, this contribution reduces to
 \be
I_{EH}=\int d\rho\, d\omega d\bar{\omega} \sqrt{\det
g(\rho,\omega,\bar{\omega})}\,\frac{2}{\rho^2 L^2}\,.
 \ee
Now using eq.~\reef{gsol}, we may evaluate $\det g$ as
 \be
\det\!\big[\, g\, \big]= \det\!\left[1+ \frac{\rho}{2}\,
g_{(2)}g_{(0)}^{-1}\right]\,\det[g_{(0)}]\,\det\!\left[1+
\frac{\rho}{2}\, g_{(0)}^{-1}g_{(2)}\right]\,.
 \ee
Since the matrix $(1+ \frac{\rho}{2}\, g_{(2)}g_{(0)}^{-1})$ is the
transpose of $(1+ \frac{\rho}{2}\,g_{(0)}^{-1}g_{(2)})$, the
corresponding determinants are the same. Therefore we have
 \be
\sqrt{\det g} = \sqrt{\det g_{(0)}}\,\det\!\(1+ \frac{\rho}{2}\,
g_{(2)}g_{(0)}^{-1}\)\,.
 \ee
In the bulk action, $\rho$ is integrated from the UV cut-off
$\rho_\mtt{min}$ to some upper bound $\rho_\mtt{crit}$. In
\cite{skenderis}, it was suggested that a natural upper bound would be
$\det[ g]$ becomes zero, which can be interpreted as the center of bulk
space. Taking this approach, we must solve
 \be
\det\!\left.\(1+ \frac{\rho}{2} g_{(2)}g_{(0)}^{-1}\)
\right|_{\rho_\mtt{min}}=0\,,
 \ee
which, using eq.~\reef{g2eom}, yields
 \be\labell{rhocrit}
\rho^{\pm}_\textrm{crit}= \frac{8}{L^2(R_{(0)}\mp
4\sqrt{\Delta})}\,,\qquad{\rm with}\ \  \Delta =
\frac{1}{8}\(\textrm{tr}(\T^2)-\frac{1}{2}(\textrm{tr}\T)^2\)\,.
 \ee
Above, the traces are again taken using the boundary metric $g_{(0)}$.
The appropriate physical solution is the smaller solution
$\rho^{-}_\textrm{crit}$. Finally, let us note that in these
coordinates, the surface terms $I_{GH}+I_{ct}$ cancel out all of the
power law divergences in the bulk term $I_{EH}$. Explicitly, we have
\begin{eqnarray}
I_{ct} &=& 2L \int d\omega d\bar\omega  \frac{\sqrt{\det g(\delta)}}{\delta^2} =
\frac{2L}{\delta^2}\int d\omega d\bar\omega  \sqrt{\det g_{(0)}}\left(1 -
\frac{\delta^2}{4}R_{(0)}\right) \,,\labell{gamma9}  \\
I_{GH} &=&   L\int  d\omega d\bar\omega  \frac{\sqrt{\det g(\delta)}}{\delta^2}
\(-4 + 2 \delta^2\, g^{\mu\nu}\!(\delta)\,g'_{\mu\nu}\!(\delta) \)
= -\frac{4L}{\delta^2}\int  d\omega d\bar\omega \sqrt{\det g_{(0)}}\,.
 \nonumber
\end{eqnarray}
Hence by combining the above expressions, the on-shell gravity action
\reef{action} becomes
 \be
I_\mtt{tot}= -\frac{c}{12\pi L^2}\int d\omega
d\bar{\omega}\sqrt{g_{(0)}} \left[ \frac{L^2}{4}R_{(0)}\ln
\rho^-_\textrm{crit} + \frac{1}{\rho^-_\textrm{crit}} +
\frac{L^2}{4}R_{(0)}\right]\,,
 \labell{integrated}
 \ee
One further simplification follows if we substitute the expression for
$\rho^-_{\textrm{crit}}$ given in eq.~\reef{rhocrit} into the second
term in the integrand above:
 \be
I_\mtt{tot}=-\frac{c}{12\pi}\int d\omega d\bar{\omega}
\sqrt{g_{(0)}}\left[ L^2\frac{R_{(0)}}{4}\ln \rho_\textrm{crit}^- +
\frac{1}{2}\sqrt{|\T_{\omega\omega}|^2} + \frac{3}{8}L^2
R_{(0)}\right]\,.
 \labell{baction1x}
 \ee
We use this result to evaluate the \ren entropy of the boundary CFT in
section \ref{2dholo}.

As noted in section \ref{submarine}, the primary role of the FG
coordinates is to select an interesting UV regulator surface in our
holographic calculation of the \ren entropy. That is, with $d=2$, the
bulk geometry is empty AdS$_3$ space and so the interesting physics
comes from the `unusual' asymptotic regulator surface whose choice is
motivated by the problem of calculating the \ren entropy. Let us
examine the geometry of this regulator surface in more detail here.
First we note that the coordinate transformation that takes us from
general FG coordinates \reef{repeatmetric} to the standard \poin \,
coordinates coordinates,
 \be
ds^2 =\frac{L^2}{\xi^2}\left( d\xi^2 + dz d\bar{z}\)\,, \labell{point}
 \ee
for an AdS$_3$ bulk was found by \cite{Krasnov}. Applying this result
to the present case, we find
 \be
 \labell{transform}
\xi = \frac{\rho^{1/2}\, e^{-\hat\phi}}{1+ L^2 \rho\,
e^{-2\hat\phi}|\partial_y \hat\phi|^2}\, \qquad z= y +
 \frac{L^2\rho\, e^{-2\hat\phi}\,\partial_{\bar{y}}\hat\phi}{1+
L^2\rho\, e^{-2\hat\phi}|\partial_y \hat\phi|^2}\,,
 \ee
where
 \be
y\equiv\left(\frac{\omega-v_1}{\omega-v_2}\right)^{\frac{1}{q}}\,.
 \labell{gamma8}
 \ee
Further $\hat\phi$ is related to the conformal factor appearing in
eq.~\reef{2dmetric},
 \be \labell{conformalfactz}
e^{\hat\phi}\equiv
\frac{q}{v_2-v_1}\,\vert\omega-v_1\vert^{(1-1/q)}\vert\omega-v_2\vert^{(1+1/q)}
= q\,(v_2-v_1)\frac{|y|^{q- 1}}{|y^q-1|^2}\,.
 \ee
We observe that in the asymptotic limit, $\rho\to 0$, the second
expression in eq.~\reef{transform} becomes simply $z=y$. Hence on the
boundary, this transformation reduces to precisely that between the
$\w$-plane and its universal cover given in eq.~\reef{ctrans}.

Now the region that is being cut out by our regularization in
eq.~\reef{regulation} can easily be understood using this coordinate
transformation. The constant factors can be readily absorbed by a
rescaling of the coordinates and will be ignored below. Further we will
set $L=1$ to avoid clutter in the following. We begin by manipulating
the above transformation \reef{transform} by taking ratios of the two
expressions to solve for $\rho e^{-2\hat \phi}$ and then substituting
the result back in to the equation for $z$ to obtain
 \be\labell{disks}
\frac{|z-y|}{\xi^2 + |z-y|^2} = |\partial_y \hat \phi|\,.
 \ee

Now for illustrative purposes, let us focus on $\w\simeq v_1$. In the
analysis of the on-shell action above the integration over the boundary
coordinates was regulated here by cutting off the integral at
$|\w-v_1|=\delta$ --- see eq.~\reef{regulation}. Alternatively, using
eq.~\reef{gamma8}, we can think that this cut-off surface was placed at
$y \simeq \delta^{1/q}/(v_2-v_1)\equiv \hat\delta$. Now from
eq.~\reef{conformalfactz}, in the vicinity of $y=0$, we have
 \be
 e^{\hat \phi}\simeq  q(v_2-v_1)\,|y|^{q-1}\,.
 \ee
Combining these results in eq.~\reef{disks} then yields
 \be
 \labell{ineq}
 \xi^2 + |z- \hat\delta|^2 \simeq 2\frac{\hat\delta}{q-1}|z-
\hat\delta|\,.
  \ee
To simplify the discussion, let us consider the cross-section of this
surface given by fixing $z$ to be real (and positive). Then
eq.~\reef{ineq} describes a semi-circle centered on  the $\xi$-axis at
$z=\hat\delta q/(q-1)$ and with radius $\hat\delta/(q-1)$ --- see
figure \ref{fgplot}. Hence the semi-circle reaches the asymptotic
boundary, \ie $\xi=0$, at $z=\hat\delta$ and $\hat\delta (q+1)/(q-1)$.
However, let us consider general curves of constant $y$ in the vicinity
of $y\simeq0$:
 \be
 \labell{ineq77}
 \xi^2 + (z- y)^2 \simeq 2\frac{y}{q-1}(z-
y)\,,
  \ee
where we are assuming that both $z$ and $y$ are real and positive.
Treating $y$ as a parameter, eq.~\reef{ineq77} describes a semi-circle
in the ($\xi,z$)-plane, which is now centered on the $\xi$-axis at $z=y
q/(q-1)$ and has radius $y/(q-1)$. This may seem slightly confusing
since if we consider some value of $y$ slightly larger than the cut-off
$y= \hat\delta$, we find the corresponding surface is slightly larger
semi-circle which has moved to the right, as illustrated in figure
\ref{fgplot}. In particular, this constant $y$ curve crosses the
cut-off surface described by eq.~\reef{ineq}. This reflects a
degeneracy in the coordinates and we will argue below that we should
only consider the left-hand (blue) portion of any of these
semi-circles.
\FIGURE[!ht]{
\includegraphics[width=0.6\textwidth]{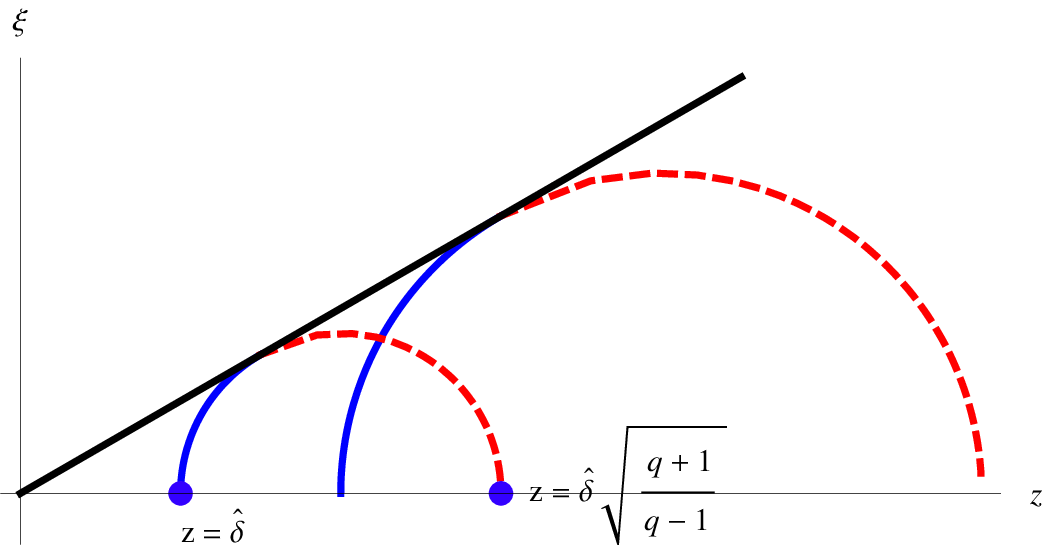}
\caption{(Colour Online) FG coordinate surfaces near the singularity in
the boundary metric at $\w=v$ illustrated in Poincar\'e coordinates.
The semi-circle to the left corresponds to the surface $y=\hat\delta$
given by eq.~\reef{ineq}. The semi-circle to the right again
corresponds to a constant $y$ surface for a slightly larger value of
$y$. The black line corresponds to the IR boundary, \ie
$\rho=\rho^-_{\textrm{crit}}$. As discussed in the main text, the red
dashed portions of the semi-circles are spurious.} \labell{fgplot}}

Recall in our evaluation of the on-shell action above, the integration
over $\rho$ is bounded in the IR by $\rho^-_{\textrm{crit}}$, as given
in eq.~\reef{rhocrit}. Let us examine this bound in the region
$\w\simeq v_1$ or $y\simeq 0$. First we observe that with
$y\ge\hat\delta$ we stay away from the singularity in the boundary
metric and so $R_{(0)}=0$. In this case, eq.~\reef{rhocrit} reduces to
$\rho^-_{\textrm{crit}}=2/|\mathcal{T_{\omega\omega}}|$. Then using
eq.~\reef{Tww}, we find in the vicinity of $y\simeq 0$ that the IR
boundary becomes
 \be
 \rho=\rho^-_{\textrm{crit}}\simeq
\frac{4|y|^{2q}(v_2-v_1)^2}{1-\frac{1}{q^2}}\,.
 \labell{critical}
 \ee
We can express this equation in terms of $\xi$ and $z$ by substituting
the above into eq.~\reef{transform} and then eliminating $y$. The final
result takes the simple form:
 \be
\xi = \frac{z}{\sqrt{q^2-1}}\,.
 \labell{line}
 \ee
That is, the IR boundary is simply a straight line with slope
$1/\sqrt{q^2-1}$ extending out from the origin in the ($\xi,z)$-plane.
Examining this surface more carefully we find that it is precisely
tangent to the constant $y$ semi-circles defined by eq.~\reef{ineq77},
as illustrated in figure \ref{fgplot}. This feature is not a
coincidence as $\rho^-_{\textrm{crit}}$ in eq.~\reef{rhocrit} was
chosen to correspond to the vanishing of $\det[ g]$. We can now infer
that, in the present case, this vanishing corresponds to a degeneracy
of the FG coordinates, \ie $d\rho\propto (d\w+d\bar{\w})$ along the
surface $\rho=\rho^-_{\textrm{crit}}$. In any event, this analysis also
indicates that the part of the semi-circles \reef{ineq77} that bends
downwards as $y$ increases, is actually excluded in the coordinate
patch covered by $\rho$ and $y$. That is, we should only consider the
left-hand (blue) portion of any of the semi-circles illustrated in
figure \ref{fgplot}.

Of course, a similar analysis will reveal analogous behaviour in the
vicinity of the singularity in the boundary metric at $\w\simeq v_2$,
which corresponds to the bulk region near $\xi=0$, $z\to\infty$. In
particular then, the regulator $|\w-v_2|>\delta$ corresponds to a
cut-off surface that extends from the asymptotic boundary at $\xi=0$ to
the IR boundary at $\rho=\rho_{\textrm{crit}}^-$. So far we have only
described the portion of the cut-off surface in the bulk that appears
in the vicinity of the singularities in the universal cover of the
$\w$-plane, \ie near the insertions of the twist operators. The cut-off
surface that completes the regulation of the bulk integrals has a more
conventional form. Namely, it is defined by asymptotic surface at
$\rho_\mtt{min}=\delta^2=(v_2-v_1)^{2q}\hat\delta^{2q}$. Hence the full
cut-off surface can be regarded as having three components, the two
surfaces cutting off the approach to the twist operators at $\w=v_2$
and $\w=v_1$, as well as a conventional UV cut-off surface extending
over the rest of the boundary away from these singular points.

To close, let us comment on the difference between the various
calculations of the on-shell action for $d=2$. In particular, above we
considered a calculation based on choosing FG coordinates, while that
proposed in section \ref{submarine} relies directly on choosing an
unusual regulator surface in the AdS$_3$ vacuum. Similarly as described
in the latter section, calculating the Euclidean action for the
three-dimensional black hole \reef{btzE} has a similar flavour of
choosing an unconventional cut-off surface. However, a feature which
distinguishes the first calculation is the appearance of an `IR
boundary' at $\rho=\rho^-_{\textrm{crit}}$. Again, the motivation for
choosing this upper bound on the $\rho$ integration was to identify the
center of the bulk geometry with the vanishing of det[$g$]. However, as
we can see in the discussion above, \eg in figure \ref{fgplot}, this
interpretation fails in the present calculation. Rather this IR
boundary simply corresponds to a degeneracy in the FG coordinate system
and is not distinguished by any invariant property of the bulk
geometry. A consequence of employing this upper bound on the $\rho$
integration is that a `large' (but finite) part of the AdS$_3$ bulk is
simply excluded in evaluating the on-shell action. On the other hand,
we have explicitly seen that both this calculation and the black hole
calculation reproduce the correct \ren entropy \reef{twod}, expected
for a two-dimensional CFT. In fact, eq.~\reef{twod} corresponds to the
universal contribution to the \ren entropy and in general, we should
expect that there may also be a non-universal constant contribution
\cite{cardy0,cardyCFT}. Hence the difference between the various
holographic calculations must lie in this non-universal contribution.
In particular, the integration over the additional region in the bulk
space beyond $\rho=\rho^-_{\textrm{crit}}$ must only contribute a
finite $\delta$-independent term. Hence this actually provides a
reassuring demonstration that, within the holographic approach,
different regulation schemes lead to the same universal contribution,
as expected.

\end{document}